\documentclass[twocolumn,rmp,aps,amssymb]{revtex4}
\usepackage{epsf}
\usepackage{psfrag}

\usepackage{graphicx}% Include figure files
\usepackage{dcolumn}% Align table columns on decimal point
\usepackage{bm}% bold math

\begin{document}

%\preprint{APS/123-QED}
%\draft

\title{Quantum information with continuous variables}
\thanks{scheduled for publication in Reviews of Modern Physics}
\author{Samuel L.\ Braunstein$^1$ and Peter van Loock$^{2,3}$}
\affiliation{$^1$Computer Science, University of York, York YO10
5DD, United Kingdom,} \affiliation{$^2$National Institute of
Informatics (NII), Tokyo 101-8430, Japan,}
\affiliation{$^3$Institute of Theoretical Physics, Institute of
Optics, Information and Photonics (Max-Planck Forschungsgruppe),
Universit\"{a}t Erlangen-N\"{u}rnberg, 91058 Erlangen, Germany}
\hspace{2cm}
\begin{abstract}
${}$
\\
Quantum information is a rapidly advancing area of interdisciplinary research.
It may lead to real-world applications for communication and computation
unavailable without the exploitation of quantum properties such as
nonorthogonality or entanglement. We review the progress in quantum information
based on continuous quantum variables, with emphasis on quantum optical
implementations in terms of the quadrature amplitudes of the
electromagnetic field.
\end{abstract}

\maketitle
\tableofcontents

%\pacs{Valid PACS appear here}
%\vspace{3ex}
%\begin{multicols}{2}

\section{Introduction}\label{Intro}

{\it Quantum information} is a relatively young branch of physics.
One of its goals is to interpret
the concepts of quantum physics from an information
theoretic point of view.
This may lead to a deeper understanding
of quantum theory.
Conversely, information and computation
are intrinsically physical concepts,
since they rely on physical systems in which
information is stored and by means of which
information is processed or transmitted.
Hence physical concepts, and at a more fundamental level
quantum physical concepts, must be incorporated in a theory
of information and computation.
Furthermore, the exploitation
of quantum effects may even prove beneficial
for various kinds of information processing and
communication.
The most prominent examples for this are quantum
computation and quantum key distribution.
Quantum computation means in particular cases, in principle,
computation faster than any known classical computation.
Quantum key distribution enables one, in principle, unconditionally
secure communication as opposed to communication
based on classical key distribution.

 From a conceptual point of view, it is illuminating
to consider {\it continuous quantum variables} in quantum
information theory. This includes the extension of quantum
communication protocols from discrete to continuous variables and
hence from finite to infinite dimensions. For instance, the
original discrete-variable (dv) quantum teleportation protocol for
qubits and other finite-dimensional systems \cite{Benn} was soon
after its publication translated into the continuous-variable (cv)
setting \cite{Vaid}. The main motivation to deal with continuous
variables in quantum information, however, originated in a more
practical observation: efficient implementation of the essential
steps in quantum communication protocols, namely preparing,
unitarily manipulating, and measuring (entangled) quantum states,
is achievable in quantum optics utilizing continuous quadrature
amplitudes of the quantized electromagnetic field. For example,
the tools for measuring a quadrature with near-unit efficiency or
for displacing an optical mode in phase space are provided by
homodyne detection and feed-forward techniques, respectively.
Continuous-variable entanglement can be efficiently produced using
squeezed light [where the squeezing of a quadrature's quantum
fluctuations is due to a nonlinear optical interaction
\cite{Walls}] and linear optics.

A valuable feature of quantum optical implementations based upon
continuous variables, related to their high efficiency, is their
{\it unconditionalness}. Quantum resources such as entangled
states emerge from the nonlinear optical interaction of a laser
with a crystal (supplemented if necessary by some linear optics)
in an unconditional fashion, i.e., every inverse bandwidth time.
This unconditionalness is hard to obtain in dv qubit-based
implementations based on single-photon states. There, the desired
preparation due to the nonlinear optical interaction depends on
particular (coincidence) measurement results ruling out the
unwanted (in particular, vacuum) contributions in the outgoing
state vector. However, the unconditionalness of the cv
implementations has its price: it is at the expense of the quality
of the entanglement of the prepared states. This entanglement and
hence any entanglement-based quantum protocol is always imperfect,
where the degree of imperfection depends on the amount of
squeezing of the laser light involved. Good quality and
performance require large squeezing which is technologically
demanding, but to a certain extent [about 10 dB \cite{seth16}]
already state of the art. Of course, in cv protocols that do not
rely on entanglement, for instance, coherent-state based quantum
key distribution, these imperfections do not occur.

To summarize at this point:
in the most commonly used optical approaches, the cv
implementations work ``always'' pretty well (and hence efficiently
and unconditionally), but never perfectly.
Their dv counterparts only work ``sometimes''
(conditioned upon rare ``successful'' events), but they succeed,
in principle, perfectly. A similar trade-off occurs when
optical quantum states are sent through noisy channels
(optical fibers), for example,
in a realistic quantum key distribution scenario.
Subject to losses, the cv states accumulate noise
and emerge at the receiver as contaminated versions of the sender's
input states. The dv quantum information encoded in single-photon
states is reliably conveyed for each photon that is not absorbed
during transmission.

Due to the recent results of Knill, Laflamme, and Milburn
[``KLM'', \cite{KLM}], it is known now that ``efficient'' quantum
information processing is possible, in principle, solely by means
of {\it linear optics}. Their scheme is formulated in a dv setting
where the quantum information is encoded in single-photon states.
Apart from entangled auxiliary photon states, generated
``off-line'' without restriction to linear optics, conditional
dynamics (feedforward) is the essential ingredient to make this
approach work. Universal quantum gates such as a controlled-NOT
can be, in principle, built using this scheme without need of any
Kerr-type nonlinear optical interaction (corresponding to an
interaction Hamiltonian quartic in the optical modes' annihilation
and creation operators). This Kerr-type interaction would be hard
to obtain on the level of single photons. However, the
``off-line'' generation of the complicated auxiliary states needed
in the KLM scheme seems impractical too.

Similarly, in the cv setting, when it comes to more advanced
quantum information protocols, such as universal quantum
computation or, in a communication scenario, entanglement
distillation, it turns out that tools more sophisticated than only
{\it Gaussian operations} are needed. In fact, the Gaussian
operations are effectively those described by interaction
Hamiltonians at most quadratic in the optical modes' annihilation
and creation operators, thus leading to {\it linear} input-output
relations as in beam splitter or squeezing transformations.
Gaussian operations, mapping Gaussian states onto Gaussian states,
also include homodyne detections and phase-space displacements. In
contrast, the non-Gaussian operations required for advanced cv
quantum communication (in particular, long-distance communication
based on entanglement distillation and swapping, quantum memory
and teleportation) are either due to at least cubic nonlinear
optical interactions or due to conditional transformations
depending on non-Gaussian measurements such as photon counting. It
seems that, at this very sophisticated level, the difficulties and
requirements of the dv and cv implementations are analogous.

In this Review, our aim is to highlight the strengths
of the cv approaches to quantum information processing.
Therefore, we focus on those protocols which are based
on Gaussian states and their feasible manipulation
through Gaussian operations.
This leads to cv proposals for the implementation
of the ``simplest'' quantum communication protocols
such as quantum teleportation and quantum key distribution,
and includes the efficient generation and detection of
cv entanglement.

Before dealing with quantum communication and computation,
in Sec.~\ref{cvinqopt}, we first introduce continuous quantum
variables within the framework of quantum optics.
The discussion about the quadratures of quantized electromagnetic modes,
about phase-space representations and Gaussian states includes
the notations and conventions that we use throughout this article.
We conclude Sec.~\ref{cvinqopt} with a few remarks on linear
and nonlinear optics, on alternative polarization and spin
representations, and on the necessity of a phase reference
in cv implementations.
The notion of entanglement, indispensable in many quantum protocols,
is described in Sec.~\ref{cventanglement} in the context of
continuous variables. We discuss pure and mixed entangled states,
entanglement between two (bipartite) and between many (multipartite)
parties, and so-called bound (undistillable) entanglement.
The generation, measurement, and verification (both theoretical
and experimental) of cv entanglement are here of particular interest.
As for the properties of the cv entangled states related with
their inseparability, we explain how the nonlocal character
of these states is revealed. This involves, for instance, violations of
Bell-type inequalities imposed by local realism. Such violations,
however, cannot occur when the measurements considered are
exclusively of cv type. This is due to the strict positivity
of the Wigner function of the Gaussian cv entangled states
which allows for a hidden-variable description in terms of
the quadrature observables.

In Sec.~\ref{qcommwcv}, we describe the conceptually and
practically most important quantum communication protocols
formulated in terms of continuous variables and thus utilizing the
cv (entangled) states. These schemes include quantum teleportation
and entanglement swapping (teleportation of entanglement), quantum
(super)dense coding, quantum error correction, quantum
cryptography, and entanglement distillation. Since quantum
teleportation based on non-maximum cv entanglement, using finitely
squeezed two-mode squeezed states, is always imperfect,
teleportation criteria are needed both for the theoretical and for
the experimental verification. As known from classical
communication, light, propagating at high speed and offering a
broad range of different frequencies, is an ideal carrier for the
transmission of information. This applies to quantum communication
as well. However, light is less suited for the storage of
information. In order to store quantum information, for instance,
at the intermediate stations in a quantum repeater, more
appropriate media than light are for example atoms. Significantly,
as another motivation to deal with cv, a feasible light-atom
interface can be built via free-space interaction of light with an
atomic ensemble based on the alternative polarization and
spin-type variables. No strong cavity QED coupling as for single
photons is needed. The concepts of this transfer of quantum
information from light to atoms and vice versa, the essential
ingredient of a quantum memory, are discussed in
Sec.~\ref{qmemory}.

Section \ref{qcloningsec} is devoted to quantum cloning with
continuous variables.
One of the most fundamental (and historically one of the first)
``laws'' of quantum information theory is the so-called
no-cloning theorem \cite{Dieks,Woott}. It forbids the exact
copying of arbitrary quantum states.
However, arbitrary quantum states can be copied approximately,
and the resemblance (in mathematical terms, the overlap
or fidelity) between the clones may attain an optimal value
independent of the original states.
Such optimal cloning can be accomplished ``locally'',
namely by sending the original states (together with some
auxiliary system) through a local unitary quantum circuit.
Optimal cloning of Gaussian cv states appears to be more
interesting than that of general cv states, because
the latter can be mimicked by a simple coin toss.
We describe a non-entanglement based (linear-optics)
implementation for the optimal local cloning of Gaussian cv
states. In addition, for Gaussian cv states, also an
optical implementation of optimal
``cloning at a distance'' (telecloning) exists.
In this case, the optimality requires entanglement.
The corresponding multi-party entanglement
is again producible with nonlinear optics (squeezed light)
and linear optics (beam splitters).

Quantum computation over continuous variables,
discussed in Sec.~\ref{qcompsec},
is a more subtle issue compared to the in some sense
straightforward cv extensions of quantum communication
protocols. At first sight, continuous variables do not appear
well suited for the processing of digital information
in a computation. On the other hand, a cv quantum state
having an infinite-dimensional spectrum of eigenstates
contains a vast amount of quantum information.
Hence it might be promising to adjust the cv states
theoretically to the task of computation (for instance, by
discretization) and yet to exploit their cv character
experimentally in efficient (optical) implementations.
We explain in Sec.~\ref{qcompsec} why universal
quantum computation over continuous variables requires
Hamiltonians at least cubic in the position and momentum
(quadrature) operators.
Similarly, any quantum circuit that consists exclusively
of unitary gates from the ``cv Clifford group'' can be
efficiently simulated by purely classical means.
This is a cv extension of the dv Gottesman-Knill theorem
where the Clifford group elements include gates such
as the Hadamard (in the cv case, Fourier) transform or the
C-NOT (Controlled-Not).
The theorem applies, for example, to
quantum teleportation which is fully describable by
C-NOT's and Hadamard (or Fourier) transforms of some
eigenstates supplemented by measurements in that eigenbasis
and spin/phase flip operations (or phase-space displacements).

Before some concluding remarks in Sec.~\ref{concludremarks},
we present some of the experimental approaches to squeezing
of light and squeezed-state entanglement generation
in Sec.~\ref{expwcvgen}.
Both quadratic and cubic optical nonlinearities are suitable
for this, namely parametric down conversion and the Kerr effect,
respectively. Quantum teleportation experiments that have been
performed already based on cv squeezed-state entanglement
are described in Sec.~\ref{telepexperim}. In Sec.~\ref{expwcv},
we further discuss experiments with long-lived atomic entanglement,
with genuine multipartite entanglement of optical modes,
experimental dense coding,
experimental quantum key distribution, and the demonstration
of a quantum memory effect.

\section{Continuous Variables in Quantum Optics}\label{cvinqopt}

For the transition from classical to quantum mechanics,
the position and momentum observables of the particles turn into
noncommuting Hermitian operators in the Hamiltonian.
In quantum optics, the quantized electromagnetic modes correspond
to quantum harmonic oscillators. The modes' quadratures
play the roles of the oscillators' position and momentum operators
obeying an analogous Heisenberg uncertainty relation.

\subsection{The quadratures of the quantized field}\label{quadratures}

 From the Hamiltonian of a quantum harmonic oscillator expressed in terms
of (dimensionless) creation and annihilation operators and representing
a single mode $k$, $\hat{H}_k=
\hbar\omega_k(\hat{a}_k^{\dagger}\hat{a}_k+\frac{1}{2})$,
we obtain the well-known form written in terms of `position' and
`momentum' operators (unit mass),
\begin{eqnarray}
\hat{H}_k=\frac{1}{2}\,\left(\hat p_k^2+\omega_k^2\hat x_k^2\right),
\end{eqnarray}
with
\begin{eqnarray}\label{pos}
\hat a_k&=&\frac{1}{\sqrt{2\hbar\omega_k}}\,\left(\omega_k\hat x_k+
i\hat{p}_k\right)\;,\\
\hat{a}_k^{\dagger}&=&\frac{1}{\sqrt{2\hbar\omega_k}}\,
\left(\omega_k\hat x_k-i\hat{p}_k\right)\;,
\end{eqnarray}
or, conversely,
\begin{eqnarray}
\hat x_k&=&\sqrt{\frac{\hbar}{2\omega_k}}\,\left(\hat a_k+
\hat{a}_k^{\dagger}\right)\;,\\
\hat p_k&=&-i\sqrt{\frac{\hbar\omega_k}{2}}\,\left(\hat a_k-
\hat{a}_k^{\dagger}\right)\;.
\end{eqnarray}
Here, we have used the well-known commutation relation
for position and momentum,
\begin{eqnarray}
[\hat x_k,\hat p_{k'}]=i\hbar\,\delta_{kk'}\;,
\end{eqnarray}
which is consistent with the bosonic commutation relations
$[\hat{a}_k,\hat{a}_{k'}^{\dagger}]=\delta_{kk'}$,
$[\hat{a}_k,\hat{a}_{k'}]=0$.
In Eq.~(\ref{pos}), we see that up to normalization factors
the position and the momentum are the real and
imaginary parts of the annihilation operator.
Let us now define the {\it dimensionless} pair of conjugate
variables,
\begin{eqnarray}
\hat X_k\equiv\sqrt{\frac{\omega_k}{2\hbar}}\,\hat x_k={\rm Re}\,\hat a_k\,,
\quad
\hat P_k\equiv\frac{1}{\sqrt{2\hbar\omega_k}}\,\hat p_k={\rm Im}\,\hat a_k\,.
\end{eqnarray}
Their commutation relation is then
\begin{eqnarray}\label{quadrcomm}
[\hat X_k,\hat P_{k'}]=\frac{i}{2}\,\delta_{kk'}\;.
\end{eqnarray}
In other words, the dimensionless `position' and `momentum' operators,
$\hat X_k$ and $\hat P_k$,
are defined as if we set $\hbar=1/2$.
These operators represent the quadratures of a single mode $k$,
in classical terms corresponding to the real and imaginary parts
of the oscillator's complex amplitude.
In the following, by using $(\hat X,\hat P)$ or equivalently
$(\hat x,\hat p)$, we will always refer to these dimensionless
quadratures playing the roles of `position' and `momentum'.
Hence also $(\hat x,\hat p)$ shall stand for a conjugate pair of
dimensionless quadratures.

The Heisenberg uncertainty relation, expressed in terms of the variances
of two arbitrary non-commuting observables $\hat A$ and
$\hat B$ in an arbitrary given quantum state,
\begin{eqnarray}
\langle(\Delta\hat A)^2\rangle&\equiv&
\langle(\hat A-\langle\hat A\rangle)^2\rangle
=\langle\hat A^2\rangle - \langle\hat A\rangle^2\;,
\nonumber\\
\langle(\Delta\hat B)^2\rangle&\equiv&
\langle(\hat B-\langle\hat B\rangle)^2\rangle
=\langle\hat B^2\rangle - \langle\hat B\rangle^2\;,
\end{eqnarray}
becomes
\begin{eqnarray}\label{genuncert}
\langle(\Delta\hat A)^2\rangle
\langle(\Delta\hat B)^2\rangle\geq
\frac{1}{4}\,|\langle[\hat A,\hat B]\rangle|^2\;.
\end{eqnarray}
Inserting Eq.~(\ref{quadrcomm}) into Eq.~(\ref{genuncert}) yields
the uncertainty relation for a pair of conjugate quadrature
observables of a single
mode $k$,
\begin{eqnarray}\label{quadraturesdef}
\hat x_k=(\hat a_k+\hat{a}_k^{\dagger})/2\;,\quad\quad
\hat p_k=(\hat a_k-\hat{a}_k^{\dagger})/2i\;,
\end{eqnarray}
namely,
\begin{eqnarray}\label{quadratuncert}
\langle(\Delta\hat x_k)^2\rangle
\langle(\Delta\hat p_k)^2\rangle\geq
\frac{1}{4}\,|\langle[\hat x_k,\hat p_k]\rangle|^2=\frac{1}{16}\;.
\end{eqnarray}
Thus, in our scales, {\it the quadrature variance
for a vacuum or coherent state of a single mode is $1/4$}.
Let us further illuminate the meaning of the quadratures by looking
at a single frequency mode of the electric field
(for a single polarization),
\begin{eqnarray}
\hat{E}_k(\mathbf{r},t)=E_0\,
[\hat{a}_k\,e^{i(\mathbf{k} \cdot \mathbf{r}-\omega_k t)}+
\hat{a}_k^{\dagger}\,e^{-i(\mathbf{k} \cdot \mathbf{r}-\omega_k t)}]\;.
\end{eqnarray}
The constant $E_0$ contains all the dimensional prefactors.
By using Eq.~(\ref{quadraturesdef}), we can rewrite the mode as
\begin{eqnarray}
\hat{E}_k(\mathbf{r},t)&=&2 E_0\,
[\hat{x}_k\,\cos(\omega_k t-\mathbf{k} \cdot \mathbf{r})
\nonumber\\
&&\quad\quad
+\hat{p}_k\,\sin(\omega_k t-\mathbf{k} \cdot \mathbf{r})]\;.
\end{eqnarray}
Apparently, the `position' and `momentum' operators $\hat x_k$
and $\hat p_k$ represent the in-phase and the out-of-phase components
of the electric field amplitude of the single mode $k$ with respect to
a (classical) reference wave $\propto
\cos(\omega_k t-\mathbf{k} \cdot \mathbf{r})$. The choice of the phase of
this wave is arbitrary, of course, and
a more general reference wave would lead us to the single mode description
\begin{eqnarray}
\hat{E}_k(\mathbf{r},t)&=&2 E_0\,
[\hat{x}_k^{(\Theta)}\,\cos(\omega_k t-\mathbf{k} \cdot \mathbf{r}
-\Theta)
\nonumber\\
&&\quad\quad
+\hat{p}_k^{(\Theta)}\,\sin(\omega_k t-\mathbf{k} \cdot \mathbf{r}
-\Theta)]\;,
\end{eqnarray}
with the more general quadratures
\begin{eqnarray}\label{quadraturesdefcontin}
\hat x_k^{(\Theta)}&=&(\hat a_k e^{-i\Theta}+\hat{a}_k^{\dagger} e^{+i\Theta}
)/2\;,\\
\hat p_k^{(\Theta)}&=&(\hat a_k e^{-i\Theta}-\hat{a}_k^{\dagger} e^{+i\Theta}
)/2i\;.
\end{eqnarray}
These ``new'' quadratures can be obtained from $\hat{x}_k$ and
$\hat{p}_k$ via the rotation
\begin{equation}\label{quadrrot}
\left( \begin{array}{c} \hat x_k^{(\Theta)} \\
\hat p_k^{(\Theta)} \end{array} \right)=
\left( \begin{array}{cc} \cos\Theta &
\sin\Theta \\ -\sin\Theta &
\cos\Theta \end{array} \right)
\left( \begin{array}{c} \hat{x}_k \\ \hat{p}_k \end{array} \right) \;.
\end{equation}
Since this is a unitary transformation, we again end up with a pair
of conjugate observables fulfilling the commutation relation
Eq.~(\ref{quadrcomm}).
Furthermore, because $\hat p_k^{(\Theta)}=\hat x_k^{(\Theta+\pi/2)}$,
the whole continuum of quadratures is covered by $\hat x_k^{(\Theta)}$
with $\Theta \in [0,\pi)$. This continuum of observables is indeed
measurable by relatively simple means.
Such a so-called homodyne detection works as follows.

A photodetector measuring an electromagnetic mode converts
the photons into electrons and hence into an electric current,
called the photocurrent $\hat i$.
It is therefore sensible to assume $\hat i \propto
\hat n=\hat{a}^{\dagger}\hat{a}$ or $\hat i=q\,\hat{a}^{\dagger}\hat{a}$
with $q$ a constant \cite{Paul}.
In order to detect a quadrature of the mode $\hat a$, the mode
must be combined
with an intense ``local oscillator'' at a 50:50 beam splitter.
The local oscillator is assumed to be in a coherent state with large
photon number, $|\alpha_{\rm LO}\rangle$.
It is therefore reasonable to describe this oscillator by a classical
complex amplitude $\alpha_{\rm LO}$ rather than by an annihilation operator
$\hat a_{\rm LO}$. The two output modes of the beam splitter,
$(\hat a_{\rm LO}+\hat a)/\sqrt{2}$ and
$(\hat a_{\rm LO}-\hat a)/\sqrt{2}$ (see Sec.~\ref{linopt}),
may then be approximated by
\begin{eqnarray}\label{outputsforcurrent}
\hat a_1=(\alpha_{\rm LO}+\hat a)/\sqrt{2}\;,\quad
\hat a_2=(\alpha_{\rm LO}-\hat a)/\sqrt{2}\;.
\end{eqnarray}
This yields the photocurrents
\begin{eqnarray}\label{currentshere}
\hat i_1 &=&q\,\hat{a}_1^{\dagger}\hat{a}_1=q\,
(\alpha_{\rm LO}^*+\hat a^{\dagger})(\alpha_{\rm LO}+\hat a)/2\;,
\nonumber\\
\hat i_2 &=&q\, \hat{a}_2^{\dagger}\hat{a}_2=q\,
(\alpha_{\rm LO}^*-\hat a^{\dagger})(\alpha_{\rm LO}-\hat a)/2\;.
\end{eqnarray}
The actual quantity to be measured shall be the difference photocurrent
\begin{eqnarray}
\delta\hat i\equiv \hat i_1 -\hat i_2=q\,
(\alpha_{\rm LO}^* \hat a + \alpha_{\rm LO} \hat a^{\dagger})\;.
\end{eqnarray}
By introducing the phase $\Theta$ of the local oscillator,
$\alpha_{\rm LO}=|\alpha_{\rm LO}|\exp(i\Theta)$, we recognize that
the quadrature observable $\hat x^{(\Theta)}$ from
Eq.~(\ref{quadraturesdefcontin}) is measured (without mode index $k$).
Now adjustment of the local oscillator's phase $\Theta \in [0,\pi]$
enables the detection of any quadrature from the whole continuum of
quadratures $\hat x^{(\Theta)}$. A possible way to realize
quantum tomography \cite{Leon},
i.e., the reconstruction of the mode's quantum state
given by its Wigner function, relies on this measurement method,
called (balanced) homodyne detection.
A broadband rather than a single-mode description of homodyne detection
can be found in Ref.~\cite{Samhomodyne}
(in addition, the influence of a quantized local oscillator
is investigated there).

We have seen now that it is not too hard to measure the
quadratures of an electromagnetic mode. Also unitary
transformations such as quadrature displacements (phase-space
displacements) can be relatively easily performed via so-called
feed-forward technique, as opposed to for example ``photon number
displacements''. This simplicity and the high efficiency when
measuring and manipulating the continuous quadratures are the main
reason why continuous-variable schemes appear more attractive than
those based on discrete variables such as the photon number.

In the following, we will mostly refer to the
conjugate pair of quadratures $\hat x_k$ and $\hat p_k$
(`position' and `momentum', i.e., $\Theta=0$ and $\Theta=\pi/2$).
In terms of these quadratures, the number operator becomes
\begin{eqnarray}\label{numberbyquadrat}
\hat n_k=\hat{a}_k^{\dagger}\hat{a}_k=\hat x_k^2 + \hat p_k^2 -\frac{1}{2}\;,
\end{eqnarray}
using Eq.~(\ref{quadrcomm}).

Let us finally review some useful formulas for the single-mode
quadrature eigenstates,
\begin{eqnarray}
\hat x |x\rangle=x |x\rangle\;,\quad\quad
\hat p |p\rangle=p |p\rangle\;,
\end{eqnarray}
where we have now dropped the mode index $k$.
They are orthogonal,
\begin{eqnarray}
\langle x |x'\rangle=\delta(x-x')\;,\quad\quad
\langle p |p'\rangle=\delta(p-p')\;,
\end{eqnarray}
and complete,
\begin{eqnarray}
\int_{-\infty}^{\infty}|x\rangle\langle x|\,dx=
\mbox{1$\!\!${\large 1}}\;,\quad\quad
\int_{-\infty}^{\infty}|p\rangle\langle p|\,dp=
\mbox{1$\!\!${\large 1}}\;.
\end{eqnarray}
As it is known for position and momentum eigenstates, the quadrature
eigenstates are mutually related to each other by Fourier transformation,
\begin{eqnarray}\label{pbasis}
|x\rangle&=&\frac{1}{\sqrt{\pi}}\int_{-\infty}^{\infty}\,e^{-2ixp}|p\rangle
\,dp\;,\\
|p\rangle&=&\frac{1}{\sqrt{\pi}}\int_{-\infty}^{\infty}\,e^{+2ixp}|x\rangle
\,dx\;.
\end{eqnarray}
Despite being unphysical and not square integrable,
the quadrature eigenstates can be very useful in calculations involving
the wave functions $\psi(x)=\langle x|\psi\rangle$ etc. and in idealized
quantum communication protocols based on continuous variables.
For instance, a vacuum state infinitely squeezed in position
may be expressed by a zero-position eigenstate $|x=0\rangle=
\int |p\rangle dp/\sqrt{\pi}$. The physical, finitely
squeezed states are characterized by the quadrature probability distributions
$|\psi(x)|^2$ etc. of which the widths correspond to the quadrature
uncertainties.

\subsection{Phase-space representations}\label{phasespace}

The Wigner function as a ``quantum phase-space
distribution'' is particularly suitable to describe
the effects on the quadrature observables which may arise from quantum
theory and classical statistics. It partly
behaves like a classical probability distribution thus enabling
to calculate measurable quantities such as mean values and variances
of the quadratures in a classical-like fashion.
On the other hand, as opposed to a classical probability distribution,
the Wigner function can become negative.

The Wigner function was originally proposed by Wigner
in his 1932 paper ``On the quantum correction for thermodynamic
equilibrium'' \cite{Wigner}.
There, he gave an expression for the Wigner function
in terms of the position basis which reads (with
$x$ and $p$ being a dimensionless pair of quadratures in our units
with $\hbar=1/2$ as introduced in the previous section) \cite{Wigner}
\begin{eqnarray}\label{Wignerformula}
W(x,p)=\frac{2}{\pi}\,\int\,dy\,e^{+4iyp}\,
\langle x-y|\hat\rho|x+y\rangle\;.
\end{eqnarray}
Here and throughout, unless otherwise specified, the integration
shall be over the entire space of the integration variable (i.e.,
here the integration goes from $-\infty$ to $\infty$). We gave
Wigner's original formula for only one mode or one particle
[Wigner's original equation was in $N$-particle form
\cite{Wigner}], because it simplifies the understanding of the
concept behind the Wigner function approach. The extension to $N$
modes is straightforward.

Why does $W(x,p)$ resemble a classical-like probability distribution?
The most important attributes that explain this are the
proper normalization,
\begin{eqnarray}
\int\,W(\alpha)\,d^2\alpha=1\;,
\end{eqnarray}
the property of yielding the correct marginal distributions,
\begin{eqnarray}
\int\, W(x,p)\,dx=\langle p|\hat\rho|p\rangle\,,\;
\int\, W(x,p)\,dp=\langle x|\hat\rho|x\rangle\,,
\end{eqnarray}
and the equivalence to a probability distribution
in classical averaging when mean values of a certain class of
operators $\hat A$ in a quantum state $\hat\rho$ are to be
calculated,
\begin{eqnarray}\label{meanvaluewithwigner}
\langle\hat A\rangle={\rm Tr}(\hat \rho \hat A)=
\int\,W(\alpha)\,A(\alpha)\,d^2\alpha\;,
\end{eqnarray}
with a function $A(\alpha)$ related to the operator $\hat A$.
The measure of integration is in our case $d^2\alpha=
d({\rm Re}\,\alpha) d({\rm Im}\,\alpha)=dx\,dp$
with $W(\alpha=x+i p)\equiv W(x,p)$, and we will use $d^2\alpha$ and
$dx\,dp$ interchangeably. The operator $\hat A$ represents a particular class
of functions of $\hat a$ and $\hat a^{\dagger}$ or $\hat x$ and $\hat p$.
The marginal distribution for $p$, $\langle p|\hat\rho|p\rangle$,
is obtained by changing the integration  variables ($x-y=u$, $x+y=v$)
and using Eq.~(\ref{pbasis}), that for $x$, $\langle x|\hat\rho|x\rangle$,
by using $\int\exp(+4iyp)dp=(\pi/2)\delta(y)$. The normalization of the
Wigner function then follows from ${\rm Tr}(\hat\rho)=1$.

For any symmetrized operator \cite{Leon}, the so-called Weyl
correspondence \cite{Weyl},
\begin{eqnarray}\label{weylcorres}
{\rm Tr}[\hat\rho\,\mathcal{S}(\hat x^n\hat p^m)]=
\int\,W(x,p)\,x^n p^m\,dx\,dp\;,
\end{eqnarray}
provides a rule how to calculate quantum mechanical expectation
values in a classical-like fashion according to
Eq.~(\ref{meanvaluewithwigner}). Here, $\mathcal{S}(\hat x^n\hat
p^m)$ indicates symmetrization. For example, $\mathcal{S}(\hat
x^2\hat p)=(\hat x^2\hat p + \hat x\hat p\hat x + \hat p\hat
x^2)/3$ corresponds to $x^2 p$ \cite{Leon}.

Such a classical-like
formulation of quantum optics in terms of
quasiprobability distributions is not unique.
In fact, there is a whole family of distributions $P(\alpha,s)$
of which each member corresponds to a particular value of a real
parameter $s$,
\begin{eqnarray}\label{sparameterdistrib}
P(\alpha,s)=\frac{1}{\pi^2}\,
\int\,\chi(\beta,s)\,\exp(i\beta\alpha^*+i\beta^*\alpha)\,
d^2\beta\;,
\end{eqnarray}
with the $s$-parametrized characteristic functions
\begin{eqnarray}\label{sparametercharacterist}
\chi(\beta,s)=
{\rm Tr}[\hat\rho\,\exp(-i\beta\hat a^{\dagger} -i\beta^*\hat a)]
\exp(s|\beta|^2/2)\,.
\end{eqnarray}
The mean values of operators normally and antinormally ordered in
$\hat a$ and $\hat a^{\dagger}$ may be calculated via the
so-called $P$ function ($s=1$) and $Q$ function ($s=-1$),
respectively. The Wigner function ($s=0$) and its characteristic
function $\chi(\beta,0)$ are perfectly suited to provide
expectation values of quantities symmetric in $\hat a$ and $\hat
a^{\dagger}$ such as the quadratures. Hence the Wigner function,
though not always positive definite, appears to be a good
compromise to describe quantum states in terms of quantum
phase-space variables such as the single-mode quadratures. We may
formulate various quantum states relevant to continuous-variable
quantum communication by means of the Wigner representation. These
particular quantum states exhibit extremely nonclassical features
such as entanglement and nonlocality. Yet their Wigner functions
are positive definite, thus belonging to the class of Gaussian
states.

\subsection{Gaussian states}\label{gauss}

Multi-mode Gaussian states may represent
optical quantum states which are potentially useful for quantum
communication or computation purposes. They are efficiently
producible in the laboratory, on demand available in an unconditional fashion.
Their corresponding Wigner functions are normalized Gaussian
distributions of the form (for zero mean)
\begin{eqnarray}\label{Gausswigndef}
W(\xi)=
\frac{1}{(2\pi)^N\sqrt{\det V^{(N)}}}\,\exp\left\{-\frac{1}{2}\,
\xi\left[V^{(N)}\right]^{-1}
\xi^{T}\right\}\;,\nonumber\\
\end{eqnarray}
with the $2N$-dimensional vector $\xi$ having
the quadrature pairs of all $N$ modes as its components,
\begin{eqnarray}
\xi&=&(x_1,p_1,x_2,p_2,...,x_N,p_N)\;,\\
\hat{\xi}&=&(\hat x_1,\hat p_1,\hat x_2,\hat p_2,...,
\hat x_N,\hat p_N)\;,
\end{eqnarray}
and with the $2N\times 2N$ correlation matrix $V^{(N)}$ having
as its elements the second moments symmetrized according to
the Weyl correspondence Eq.~(\ref{weylcorres}),
\begin{eqnarray}\label{corrdef}
{\rm Tr}[\hat\rho\,(\Delta\hat\xi_i\Delta\hat\xi_j+
\Delta\hat\xi_j\Delta\hat\xi_i)/2]
&=&\langle(\hat\xi_i\hat\xi_j+\hat\xi_j\hat\xi_i)/2\rangle\nonumber\\
&=&\int\,W(\xi)\,\xi_i \xi_j\, d^{2N}\xi\nonumber\\
&=&V^{(N)}_{ij}\;,
\end{eqnarray}
where $\Delta\hat\xi_i=\hat\xi_i-\langle\hat\xi_i\rangle=\hat\xi_i$ for
zero mean values.
The last equality defines the correlation matrix for any quantum state.
For Gaussian states of the form Eq.~(\ref{Gausswigndef}),
the Wigner function is completely determined by the second-moment
correlation matrix.

For a {\it classical} probability distribution over the classical
$2N$-dimensional phase space, every physical correlation matrix
is real, symmetric, and positive, and conversely, any
real, symmetric, and positive matrix represents a possible physical
correlation matrix. Apart from reality, symmetry, and positivity,
the Wigner correlation matrix (of any state),
describing the {\it quantum} phase space,
must also comply with the commutation relation
from Eq.~(\ref{quadrcomm}) \cite{Simon,Wolf},
\begin{eqnarray}
[\hat\xi_k,\hat\xi_l]=\frac{i}{2}\,\Lambda_{kl}\;,
\quad\quad k,l=1,2,3,...,2N\;,
\end{eqnarray}
with the $2N\times 2N$ matrix $\Lambda$ having
the $2\times 2$ matrix $J$ as diagonal entry for each
quadrature pair, for example for $N=2$,
\begin{eqnarray}\label{landjmatrices}
\Lambda=
\left( \begin{array}{cc} J & 0 \\
0 & J
\end{array} \right)\;,\quad\quad
J=
\left( \begin{array}{cc} 0 & 1 \\
-1 & 0
\end{array} \right)\;.
\end{eqnarray}
A direct consequence of this commutation relation and the non-negativity
of the density operator $\hat\rho$ is the following $N$-mode
uncertainty relation \cite{Simon,Wolf},
\begin{eqnarray}\label{Nmodeuncert}
V^{(N)}-\frac{i}{4}\,\Lambda\geq 0\;.
\end{eqnarray}
This matrix equation means that the matrix sum on the
left-hand-side has only nonnegative eigenvalues. Note that this
$N$-mode uncertainty relation applies to any state, not only
Gaussian states. Any physical state has to obey it. For Gaussian
states, however, it is not only a necessary condition, but it is
also sufficient to ensure the positivity of $\hat\rho$
\cite{Wolf}. In the simplest case $N=1$, Eq.~(\ref{Nmodeuncert})
is reduced to the statement $\det V^{(1)}\geq 1/16$, which is a
more precise and complete version of the Heisenberg uncertainty
relation in Eq.~(\ref{quadratuncert}). For any $N$,
Eq.~(\ref{Nmodeuncert}) becomes exactly the Heisenberg uncertainty
relation of Eq.~(\ref{quadratuncert}) for each individual mode, if
$V^{(N)}$ is diagonal. The purity condition for an $N$-mode
Gaussian state is given by $\det V^{(N)}=1/16^N$.

\subsection{Linear optics}\label{linopt}

In passive optical devices such as beam splitters and phase
shifters, the photon number
is preserved and the modes' annihilation operators
are transformed only linearly.
This linear-optics toolbox provides essential tools for generating
particular quantum states and for manipulating and measuring them.

A beam splitter can be considered as
a four-port device with the input-output relations in the Heisenberg picture
\begin{equation}
(\hat a'_1 \; \hat a'_2)^T =
U(2)\,
(\hat a_1 \; \hat a_2)^T\;.
\end{equation}
The matrix $U(2)$ must be unitary, $U^{-1}(2)=U^{\dagger}(2)$,
in order to ensure that the commutation relations are preserved,
\begin{eqnarray}
[\hat{a}_i',\hat{a}_j']=[(\hat{a}_i')^{\dagger},(\hat{a}_j')^{\dagger}]
=0\,,\quad [\hat{a}_i',(\hat{a}_j')^{\dagger}]=\delta_{ij}\,.
\end{eqnarray}
This unitarity reflects the fact that the total photon number
remains constant for a lossless beam splitter. Any unitary
transformation acting on two modes can be expressed by the matrix
\cite{Danakas,Bernstein}
\begin{equation}\label{seechapter1}
U(2)=
\left( \begin{array}{cc} e^{-i(\phi+\delta)}\sin\theta &
e^{-i\delta}\cos\theta \\ e^{-i(\phi+\delta')}\cos\theta &
-e^{-i\delta'}\sin\theta \end{array} \right)  \;.
\end{equation}
An ideal phase-free beam splitter operation is then simply
given by the linear transformation
\begin{equation}\label{generalBS1}
\left( \begin{array}{c} \hat a_1' \\ \hat a_2' \end{array} \right)=
\left( \begin{array}{cc} \sin\theta &
\cos\theta \\ \cos\theta &
-\sin\theta \end{array} \right)
\left( \begin{array}{c} \hat a_1 \\ \hat a_2 \end{array} \right) \;,
\end{equation}
with the reflectivity and transmittance parameters $\sin\theta$
and $\cos\theta$.
Thus, the general unitary matrix describes a sequence of phase shifts
and phase-free beam splitter `rotations',
\begin{equation}
U(2)=
\left( \begin{array}{cc} e^{-i\delta} &
0 \\ 0 & e^{-i\delta'} \end{array} \right)
\left( \begin{array}{cc} \sin\theta &
\cos\theta \\ \cos\theta &
-\sin\theta \end{array} \right)
\left( \begin{array}{cc} e^{-i\phi} &
0 \\ 0 & 1 \end{array} \right) \;.
\end{equation}
Not only the above $2\times 2$ matrix
can be decomposed into phase shifting and beam splitting operations.
Any $N\times N$ unitary matrix
as it appears in the linear transformation
\begin{equation}\label{generallinoptmix}
\hat{a}_i'=\sum_{j} U_{ij} \hat{a}_j\;,
\end{equation}
may be expressed by a sequence of phase shifters and beam
splitters \cite{Reck}. This means that any mixing between optical
modes described by a unitary {\it matrix} can be implemented with
linear optics. In general, it does {\it not} mean that any unitary
{\it operator} acting on the Hilbert space of optical modes (or a
subspace of it) is realizable via a fixed network of linear
optics. Conversely, however, any such network can be described by
the {\it linear} transformation in Eq.~(\ref{generallinoptmix}).

The action of an ideal phase-free beam splitter operation on two
modes can be expressed in the Heisenberg picture by
Eq.~(\ref{generalBS1}). The input operators are changed, whereas
the input states remain invariant. The corresponding unitary
operator must satisfy
\begin{equation}\label{BSoperator}
\left( \begin{array}{c} \hat a_1' \\ \hat a_2' \end{array} \right)=
\hat B_{12}^{\dagger}(\theta)\,
\left( \begin{array}{c} \hat a_1 \\ \hat a_2 \end{array} \right) \,
\hat B_{12}(\theta)\;.
\end{equation}
In the Schr\"{o}dinger representation, we have correspondingly
$\hat \rho'= \hat B_{12}(\theta)\hat\rho\hat
B_{12}^{\dagger}(\theta)$ or for pure states, $|\psi'\rangle=\hat
B_{12}(\theta)|\psi\rangle$. Note that $\hat B_{12}(\theta)$ acts
on the position eigenstates as
\begin{eqnarray}\label{BSeigenstates}
\hat{B}_{12}(\theta)|x_1,x_2\rangle&=&
|x_1\sin\theta+x_2\cos\theta,x_1\cos\theta-
x_2\sin\theta\rangle\nonumber\\
&=&|x'_1,x'_2\rangle \;.
\end{eqnarray}
In Eq.~(\ref{BSeigenstates}), $|x_1,x_2\rangle\equiv
|x_1\rangle |x_2\rangle \equiv |x_1\rangle_1\otimes |x_2\rangle_2$
which we will use interchangeably throughout.
The position wave function is transformed according to
\begin{eqnarray}\label{wavefcttransfBS}
\psi(x_1,x_2)&\rightarrow&\psi'(x'_1,x'_2)\\
&&=\psi(x'_1\sin\theta+x'_2\cos\theta,x'_1\cos\theta-
x'_2\sin\theta) \;.\nonumber
\end{eqnarray}
Analogous linear beam-splitter transformation rules apply to the
momentum wave function, the probability densities, and the Wigner
function. Finally, we note that any unitary operator $\hat U$ that
describes a network of passive linear optics acting upon $N$ modes
corresponds to a {\it quadratic} Hamiltonian such that $\hat U =
\exp(-i\vec a^\dagger H \vec a)$, where $\vec a=(\hat a_1,\hat
a_2,...,\hat a_N)^T$, $\vec a^\dagger= (\hat a_1^\dagger,\hat
a_2^\dagger,...,\hat a_N^\dagger)$, and $H$ is an $N\times N$
Hermitian matrix.

\subsection{Nonlinear optics}\label{nonlinopt}

An important tool of many quantum communication protocols is
entanglement, and the essential ingredient in the generation of
continuous-variable entanglement is squeezed light. In order to
squeeze the quantum fluctuations of the electromagnetic field,
nonlinear optical effects are needed. This squeezing of optical
modes is sometimes also referred to as a linear optical process,
because the corresponding interaction Hamiltonian is {\it
quadratic} in $\hat{a}$ and  $\hat{a}^{\dagger}$ which yields a
{\it linear} mixing between annihilation and creation operators in
the input-output relations. In the previous section, we discussed
that a process which ``truly'' originates from linear optics
(based only on passive elements such as beam splitters and phase
shifters) is expressed by Eq.~(\ref{generallinoptmix}). Hence it
is given by linear input-output relations, but it does not involve
mixing between the $\hat{a}$'s and $\hat{a}^{\dagger}$'s. The most
general linear transformation combining elements from passive
linear optics and nonlinear optics is the so-called linear unitary
Bogoliubov (LUBO) transformation \cite{LUBO},
\begin{equation}\label{LUBO}
\hat{a}_i'=\sum_{j} A_{ij} \hat{a}_j + B_{ij} \hat{a}_j^{\dagger}
+\gamma_i\;,
\end{equation}
with the matrices $A$ and $B$ satisfying the conditions
$AB^T=(AB^T)^T$ and $AA^{\dagger}=BB^{\dagger}+
\mbox{1$\!\!${\large 1}}$ due to the bosonic commutation relations
for $\hat{a}_i'$. This input-output relation describes any
combination of linear optical elements (multi-port
interferometers), multi-mode squeezers, and phase-space
displacements or, in other words, any interaction Hamiltonian {\it
quadratic} in $\hat a$ and $\hat a^{\dagger}$. The LUBO
transformations are equivalent to the {\it Gaussian
transformations} that map Gaussian states onto Gaussian states.

In general, squeezing refers to the reduction of quantum
fluctuations in one observable below the standard quantum limit
(the minimal noise level of the vacuum state) at the expense of an
increased uncertainty of the conjugate variable. In the remainder
of this section, we will briefly discuss squeezing schemes
involving a nonlinear-optical $\chi^{(2)}$ interaction,
describable by a quadratic interaction Hamiltonian. Others, based
on a $\chi^{(3)}$ nonlinearity and a quartic Hamiltonian, are
among the topics of Sec.~\ref{expwcv}.

The output state of degenerate parametric amplification, where
the signal and idler frequencies both equal half the pump frequency,
corresponds to a single-mode squeezed state.
This effect of single-mode squeezing can be calculated with
an interaction Hamiltonian quadratic in the creation and annihilation
operators,
\begin{eqnarray}\label{OPOHam}
\hat{H}_{\rm int}=
i\hbar\frac{\kappa}{2}\,
(\hat{a}^{\dagger 2}e^{i\Theta} -\hat{a}^2e^{-i\Theta})\;.
\end{eqnarray}
It describes the amplification of the signal mode $\hat{a}$
at half the pump frequency in an interaction picture
(without explicit time dependence due to the free evolution).
The coherent pump mode is assumed to be classical (the so-called parametric
approximation), its real amplitude $|\alpha_{\rm pump}|$
is absorbed in $\kappa$, and the pump phase is $\Theta$.
The parameter $\kappa$ also contains the susceptibility,
$\kappa\propto \chi^{(2)}|\alpha_{\rm pump}|$.
The fully quantum mechanical Hamiltonian is
$\hat{H}_{\rm int}\propto \hat{a}^{\dagger 2}\hat a_{\rm pump}-
\hat{a}^2 \hat a^{\dagger}_{\rm pump}$, and with the parametric
approximation we assume $\hat a_{\rm pump}\rightarrow \alpha_{\rm pump}
=|\alpha_{\rm pump}|e^{i\Theta}$ \cite{Scully}.
In the interaction picture, we can insert $\hat{H}_{\rm int}$ into
the Heisenberg equation of motion
for the annihilation operator, and obtain
(taking zero pump phase $\Theta=0$)
\begin{eqnarray}\label{diffeqs}
\frac{d}{dt}\hat a(t)=\frac{1}{i\hbar}\,[\hat a(t),\hat H_{\rm int}]=
\kappa\,\hat a^{\dagger}(t)\;.
\end{eqnarray}
This equation is solved by
\begin{eqnarray}
\hat a(t)=\hat a(0)\,\cosh(\kappa t)+\hat a^{\dagger}(0)\,\sinh(\kappa t)\;.
\end{eqnarray}
The quadrature operators evolve correspondingly into
\begin{eqnarray}\label{OPOsqquadratures}
\hat x(t)=e^{+\kappa t}\,\hat x(0)\;,\quad\quad
\hat p(t)=e^{-\kappa t}\,\hat p(0)\;.
\end{eqnarray}
This is in fact the expected result. Due to the unitary evolution,
the uncertainty of the $p$ quadrature
decreases, whereas that of the $x$ quadrature grows:
\begin{eqnarray}
\langle[\Delta\hat x(t)]^2\rangle&=&
e^{+2\kappa t}\,\langle[\Delta\hat x^{(0)}]^2\rangle\;,\nonumber\\
\langle[\Delta\hat p(t)]^2\rangle&=&
e^{-2\kappa t}\,\langle[\Delta\hat p^{(0)}]^2\rangle\;.
\end{eqnarray}
Here we have chosen vacuum-state inputs
and replaced the initial quadratures
by those of the vacuum labeled by a superscript `${(0)}$'.
The evolving states remain minimum uncertainty states, but
they have $p$ fluctuations below and $x$ fluctuations above the vacuum
noise level. They have become quadrature squeezed states.
According to the unitary evolution
\begin{eqnarray}
\hat U(t,t_0)=\exp\left[-\frac{i}{\hbar}\,
\hat H\, (t-t_0)\right]\;,
\end{eqnarray}
with the Hamiltonian from
Eq.~(\ref{OPOHam}) and $t_0=0$,
let us now introduce the unitary squeezing or
squeeze operator $\hat S(\zeta)$ by
defining $\zeta\equiv -r\exp(i\Theta)$ with the squeezing parameter
$r\equiv \kappa t$ (a dimensionless effective interaction time),
\begin{eqnarray}
\hat U(t,0)&=&\exp\left[\frac{\kappa}{2}\,
(\hat{a}^{\dagger 2}e^{i\Theta} -\hat{a}^2e^{-i\Theta})\, t\right]
\nonumber\\
&\equiv& \hat S(\zeta)=
\exp\left(\frac{\zeta^*}{2}\,
\hat{a}^2-\frac{\zeta}{2}\,\hat{a}^{\dagger 2}\right)\;.
\end{eqnarray}
The squeezing operator obviously satisfies
$\hat S^{\dagger}(\zeta)=\hat S^{-1}(\zeta)=\hat S(-\zeta)$.
Applying it to an arbitrary initial mode
$\hat a(0)\equiv \hat a$ yields the transformations
\begin{eqnarray}\label{ONEMODESQSOL}
\hat S^{\dagger}(\zeta)\hat a\hat S(\zeta)&=&
\hat a \cosh r + \hat a^{\dagger} e^{i\Theta} \sinh r \;,\nonumber\\
\hat S^{\dagger}(\zeta)\hat a^{\dagger}\hat S(\zeta)&=&
\hat a^{\dagger} \cosh r + \hat a e^{-i\Theta} \sinh r \;.
\end{eqnarray}
For the rotated mode
\begin{eqnarray}
\hat x^{(\Theta/2)}+i\hat p^{(\Theta/2)}=
(\hat x +i\hat p)\,e^{-i\Theta/2}=\hat a\,e^{-i\Theta/2}\;,
\end{eqnarray}
the squeezing transformation results in
\begin{eqnarray}
\hat S^{\dagger}(\zeta)[\hat x^{(\Theta/2)}+i\hat p^{(\Theta/2)}]
\hat S(\zeta)&=&
\hat a \,e^{-i\Theta/2} \cosh r \\
&&+ \hat a^{\dagger} \,e^{+i\Theta/2} \sinh r \nonumber\\
&=&e^{+r}\,\hat x^{(\Theta/2)}+i\, e^{-r}\,\hat p^{(\Theta/2)}\;.
\nonumber
\end{eqnarray}
Thus, the effect of the squeezing operator on
an arbitrary pair of quadratures,
as generally defined in Eq.~(\ref{quadraturesdefcontin}),
is the attenuation of one quadrature and the amplification of the other.
We have seen that the squeezing operator effectively represents the
unitary evolution due to the OPA Hamiltonian. The corresponding
expressions for the resulting
Heisenberg quadrature operators (with $\Theta=0$ and {\it vacuum inputs}),
\begin{eqnarray}\label{sqquad}
\hat x(r)=e^{+r}\,\hat x^{(0)}\;,\quad\quad
\hat p(r)=e^{-r}\,\hat p^{(0)}\;,
\end{eqnarray}
as in Eq.~(\ref{OPOsqquadratures}) squeezed in $p$ for $t>0$ ($r>0$),
will prove extremely useful for the following investigations.
Note that time reversal ($r<0$) just swaps the squeezed and the
antisqueezed quadrature.
Throughout this article, we may always use $r\geq 0$,
and hence describe a position-squeezed mode via the Heisenberg equations
\begin{eqnarray}\label{sqquad2}
\hat x(r)=e^{-r}\,\hat x^{(0)}\;,\quad\quad
\hat p(r)=e^{+r}\,\hat p^{(0)}\;,
\end{eqnarray}
where $r>0$.
The quadrature squeezing, mathematically defined through the
squeezing operator $\hat S(\zeta)$ and physically associated with
the OPA interaction, is commonly referred to as ``ordinary'' squeezing.
Other kinds of squeezing will be mentioned in Sec.~\ref{expwcv}.

The Heisenberg equations Eq.~(\ref{sqquad}) correspond to
a squeezed vacuum state, in the Schr\"{o}dinger representation
given by the Hilbert vector $\hat S(\zeta)|0\rangle$ (with $\Theta=0$).
More generally, all minimum uncertainty states are displaced squeezed
vacua,
\begin{eqnarray}
|\alpha,\zeta\rangle=\hat D(\alpha)\hat S(\zeta)|0\rangle\;,
\end{eqnarray}
with the unitary displacement operator
\begin{eqnarray}\label{displacementopdefinition}
\hat D(\alpha)=\exp(\alpha\hat{a}^{\dagger}-\alpha^*\hat{a})=
\exp(2ip_{\alpha}\hat x-2ix_{\alpha}\hat p)\;,
\end{eqnarray}
where $\alpha=x_{\alpha}+ip_{\alpha}$ and $\hat a=\hat x +i\hat p$.
The displacement operator acting on $\hat a$ (as a unitary transformation
in the Heisenberg picture) yields a displacement
by the complex number $\alpha$,
\begin{eqnarray}\label{generaldisplacementproperty}
\hat D^{\dagger}(\alpha)\hat a\hat D(\alpha)=
\hat a + \alpha\;.
\end{eqnarray}
The position wave function for the displaced
position-squeezed vacuum is given by
\begin{eqnarray}
\psi(x)&=&\left(\frac{2}{\pi}\right)^{1/4}e^{r/2}
\exp[-e^{2r}(x-x_{\alpha})^2\nonumber\\
&&\quad\quad\quad\quad\quad\quad\quad\quad\;
+2ip_{\alpha} x
-ix_{\alpha} p_{\alpha}]\;.
\end{eqnarray}
The corresponding Wigner function is then
\begin{eqnarray}
W(x,p)=\frac{2}{\pi}\exp[-2e^{+2r}(x-x_{\alpha})^2
-2e^{-2r}(p-p_{\alpha})^2]\;,\nonumber\\
\end{eqnarray}
where the quadrature variances here are $\sigma_x=e^{-2r}/4$ and
$\sigma_p=e^{+2r}/4$.
In the limit of infinite squeezing $r\to\infty$,
the position probability density,
$|\psi(x)|^2=\sqrt{2/\pi}\,e^r\exp[-2e^{2r}(x-x_{\alpha})^2]$
becomes a delta function
$\lim_{\epsilon\to 0}\exp[-(x-x_{\alpha})^2/\epsilon^2]/\epsilon\sqrt{\pi}=
\delta(x-x_{\alpha})$ with $\epsilon=e^{-r}/\sqrt{2}$.
The squeezed vacuum wave function in that limit,
$\psi(x)\propto\delta(x)$, describes a zero position eigenstate,
$\int dx\, \psi(x) |x\rangle\propto |0\rangle$.
The mean photon number of an infinitely squeezed state becomes
infinite, because for the displaced squeezed vacuum we have
\begin{eqnarray}
\langle \hat n \rangle=\langle \hat x^2 \rangle +
\langle \hat p^2 \rangle -\frac{1}{2}=|\alpha|^2+\sinh^2 r\;,
\end{eqnarray}
using Eq.~(\ref{numberbyquadrat}).

Later, we will show that the simplest quantum teleportation protocol
based on continuous variables utilizes two-mode squeezing.
The physical process for producing a two-mode squeezed state via
a nondegenerate optical parametric amplifier (NOPA)
is a generalization of the nonlinear interaction involved in
degenerate optical parametric amplification.
The NOPA interaction relies on the Hamiltonian
\begin{eqnarray}\label{NOPOHam}
\hat{H}_{\rm int}=
i\hbar\kappa\,
(\hat{a}_1^{\dagger}\hat{a}_2^{\dagger}e^{i\Theta}
-\hat{a}_1\hat{a}_2e^{-i\Theta})\;,
\end{eqnarray}
where $\hat a_1$ and $\hat a_2$ refer to the signal and idler modes
emerging at two sidebands around half the pump frequency
and having different polarizations.
Here, we still assume $\kappa\propto \chi^{(2)}|\alpha_{\rm pump}|$.
Mathematically, two-mode squeezing may be defined analogously to
single-mode squeezing by the unitary two-mode squeeze operator
\begin{eqnarray}
\hat U(t,0)&=&\exp\left[\kappa\,
(\hat{a}_1^{\dagger}\hat{a}_2^{\dagger}e^{i\Theta}
-\hat{a}_1\hat{a}_2e^{-i\Theta})\, t\right]
\nonumber\\\label{TWOMODESQoperatordef}
&\equiv& \hat S(\zeta)=
\exp\left(\zeta^*\,\hat{a}_1\hat{a}_2
-\zeta\,\hat{a}_1^{\dagger}\hat{a}_2^{\dagger}\right)\;,
\end{eqnarray}
with the same definitions and conventions as above.
The solution for the output modes, calculated as above for single-mode
squeezing, is (with $\Theta=0$)
\begin{eqnarray}\label{TWOMODESQSOL}
\hat a_1(r)&=&
\hat a_1 \cosh r + \hat a_2^{\dagger} \sinh r \;,\nonumber\\
\hat a_2(r)&=&
\hat a_2 \cosh r + \hat a_1^{\dagger} \sinh r \;.
\end{eqnarray}
These output modes are entangled and exhibit quantum correlations
between the quadratures. More realistically, these correlations
cover a finite range of sideband frequencies. This ``broadband
two-mode squeezing'' will be briefly discussed in
Sec.~\ref{expwcv}. A two-mode squeezed state, produced by the NOPA
interaction, is equivalent to two single-mode squeezed states
(with perpendicular squeezing directions and produced via the
degenerate OPA interaction or alternatively via a $\chi^{(3)}$
interaction, see Sec.~\ref{expwcv}) combined at a beam splitter
\cite{PvLbroad}. This equivalence will be explained in
Sec.~\ref{cventanglement}.

We have discussed the generation of squeezed light within the framework
of nonlinear optics and the manipulation of electromagnetic modes
by linear optics using beam splitters.
Squeezers and beam splitters are the resources and building components
of quantum communication protocols based on continuous variables,
because they represent tools for creating the essential ingredient
of most of these protocols: continuous-variable entanglement.

\subsection{Polarization and spin representations}\label{polspinrep}

The field of quantum information with cv grew out of the analysis
of quadrature-squeezed optical states. However, in order for this
field to mature into one yielding a usable technology, methods for
storing continuous quantum information will be required.
In particular,
it seems clear that continuous quantum variables which can be compatible
with the collective state of a set of atomic systems will be needed to
perform this task.
Here we briefly discuss a useful alternative encoding
for continuous quantum information in terms of
collective spin-like variables.

Optically, we will be interested in encoding the continuous quantum
information onto the collective Stokes (polarization) variables
of an optical field.
Let $\hat a_{+}(t-z/c)$ and $\hat a_{-}(t-z/c)$ be the
annihilation operators for circularly polarized beams of {\it light}
propagating along the positive $z$-axis. Then the Stokes operators
may be defined as
\begin{eqnarray}\label{stokesdef}
\hat S_x &=& \frac{c}{2}\int_0^T d\tau'\,[\hat a_{+}^\dagger(\tau')
\hat a_{-}(\tau') +
\hat a_{-}^\dagger(\tau') \hat a_{+}(\tau')] \;,\nonumber \\
\hat S_y &=& \frac{-ic}{2}\int_0^T d\tau'\,[\hat a_{+}^\dagger(\tau')
\hat a_{-}(\tau') - \hat a_{-}^\dagger(\tau') \hat a_{+}(\tau')]
\;,\nonumber\\
\hat S_z &=& \frac{c}{2}\int_0^T d\tau'\,[\hat a_{+}^\dagger(\tau')
\hat a_{+}(\tau') - \hat a_{-}^\dagger(\tau') \hat a_{-}(\tau')]
\nonumber \;.\\
\end{eqnarray}
Given the usual equal-time commutation relations for the annihilation
operators $\hat a_{\pm}(t,z)$ as $[\hat a_i(t,z),\hat a_j(t,z')] =\delta_{ij}
\delta(z-z')$ where $i,j=\pm$, the commutation relations for
these Stokes operators correspond to those of the usual spin operators,
namely $[\hat S_j,\hat S_k]=i\epsilon_{jkl} \hat S_l$.

Now suppose we restrict our states to those for which
$\langle \hat S_x\rangle$ is near its maximum value. In this case, to
an excellent approximation, we may write
$\hat S_x \simeq \langle \hat S_x\rangle$. For states restricted in this
manner, the resulting commutation relations for $\hat S_y$ and $\hat S_z$
are an excellent approximation (up to rescaling) of the canonical
commutation relations for the usual phase-space variables $\hat x$ and
$\hat p$. Thus, states with near maximum Stokes' polarization
correspond to those on a patch of phase space with $\hat S_y$ and $\hat S_z$
playing the role of $\hat x$ and $\hat p$. Taking the state with
maximum $\langle \hat S_x\rangle$ to represent the phase-space `vacuum'
state, the whole set of usual coherent, squeezed, and cv
entangled states may be constructed via displacements and squeezing
transformations based on this operator translation.

An analogous representation may be constructed for the collective
spin of a set of $N$ spin-$\frac{1}{2}$ {\it atoms}. Defining the
collective spin variables as
$\hat F_i = \frac{1}{N}\sum_{n=1}^{N}\hat F_i^{(n)}$, with the usual
spin commutation relations we find for the collective spin variables
$[\hat F_j, \hat F_k] = i\epsilon_{jkl} \hat F_l$. In a similar manner,
we shall consider states with near maximal polarization along the
negative $z$-axis, i.e., states corresponding to small variations about
$|F,-F\rangle$. Again, for such a subset of states, the variables
$\hat F_x$ and $\hat F_y$ have commutation relations (up to rescaling)
which are excellent approximations to those of the phase-space variables
$\hat x$ and $\hat p$. Thus, within these phase-space patches, we may
encode continuous quantum information as spin-coherent, spin-squeezed
etc. states. A coherent spin state would then be a
minimum uncertainty state that satisfies equality in the
corresponding Heisenberg uncertainty relation, for instance,
according to Eq.~(\ref{genuncert}),
\begin{eqnarray}\label{spinuncert}
\langle(\Delta\hat F_x)^2\rangle
\langle(\Delta\hat F_y)^2\rangle\geq
\frac{1}{4}\,|\langle\hat F_z\rangle|^2\;.
\end{eqnarray}
For a state with mean polarization along the $z$-axis,
$|\langle\hat F_z\rangle | = F$, the ``vacuum variance'' would
correspond to
$\langle(\Delta\hat F_x)^2\rangle=\langle(\Delta\hat F_y)^2\rangle=
F/2$.

\subsection{Necessity of phase reference}\label{phaseref}

The quantum information of continuous variables is stored as phase space
distributions. Almost all such distributions (and hence quantum states)
depend on the orientation of the phase-space coordinate frame.
Physically, this corresponds to a phase of the quantum field
being used to encode the quantum information.

Typically, this phase corresponds to a choice of phase of the
single-mode annihilation operator, via
\begin{equation}
\hat a \rightarrow e^{-i\phi}\hat a \;.
\end{equation}
or equivalently the phase-shift on a single-mode state (or
wave-function) via
\begin{equation}
|\psi\rangle \rightarrow e^{-i\phi\hat a^\dagger \hat a}|\psi\rangle \;.
\end{equation}

When we measure such states to extract some of their quantum information
we will typically use phase sensitive detection schemes such as homodyne
detection. However, this entails picking some phase reference.

In this case, the phase reference comes from a strong local oscillator (LO).
However, the LO itself has phase freedom. So where is this freedom actually
tied down in any given experiment?

The trick that has been used for decades is to both construct and
manipulate the phase-sensitive states from the same phase reference that
one uses to make the measurements. Physically, one splits up the LO into
several pieces each of which controls a different aspect of any given
experiment. However, so long as each process is done within the LO's
dephasing time the common phase cancels out from the response of the
detectors. Thus, quantum information with continuous variables involves
experiments which typically are going to require a phase reference. In
quantum-optics experiments, for example, this LO is just a strong laser
beam that is shared amongst the various `parties' (such as sender and
receiver) in the lab.

Does the phase reference of the LO correspond to a quantum or a
classical channel that must be shared between users? One can expect
that, since multiple copies are being used, the resource must
actually be relying on no more than clonable and hence presumably
classical information.

This whole picture is well appreciated by experimentalists, but
has recently led to some discussion about the validity of this
paradigm. In particular, \textcite{RS01} have recently argued that
since we cannot know the LO's phase $\phi$ \cite{Mol97}, we should
average over it
\begin{eqnarray}
\label{rel_phi}
\hat \rho_{\strut \rm PEF}
&=& \int_0^{2\pi} \frac{d\phi}{2\pi} Pr(\phi) \big|\, |\alpha|
e^{-i\phi}\big\rangle\big\langle |\alpha| e^{-i\phi} \big| \\
&=& \int_0^{2\pi} \frac{d\phi}{2\pi} \big|\, |\alpha|
e^{-i\phi}\big\rangle\big\langle |\alpha| e^{-i\phi} \big|
\label{rel_coherent_s} \\
\label{rel_fock}
&=& e^{-|\alpha|^2}
\sum_{n=0}^{\infty}\frac{|\alpha|^{2n}}{n!}|n\rangle\langle n| \;,
\end{eqnarray}
where  they implicitly took the prior distribution $Pr(\phi)$ to be
uniform. They then argued that the decomposition of
Eq.~(\ref{rel_phi}) into coherent states, as opposed to number states,
was not appropriate in interpreting experiments by invoking
the Partition Ensemble Fallacy \cite{Kok00}.

A number of people have argued against this as counter to common
sense \cite{Gea90,Weis01,Enk02}, however, \textcite{Nem03} noted a
flaw in the argument of Rudolph and Sanders: although choosing
$Pr(\phi)$ as uniform seems eminently reasonable, if $\phi$ is
truly unobservable as they presume and as is generally accepted
\cite{Mol97}, then  Rudolph and Sanders' choice of Prior is
untestable. Indeed, any choice of $Pr(\phi)$ would lead to
completely equivalent predictions for all possible experiments.
The implications of this are that states of the form
Eq.~(\ref{rel_phi}) actually form an equivalence class --- any
member of which may be chosen to represent the class. One of these
members is just a coherent state. Thus, the conventional
experimental interpretation falls out and the long-standing
interpretations do not in fact involve any fallacy.

\section{Continuous-Variable Entanglement}\label{cventanglement}

Historically, the notion of entanglement (``Verschr\"{a}nkung'')
appeared explicitly in the literature first in 1935, long before
the dawn of the relatively young field of quantum information, and
without any reference to dv qubit states. In fact, the entangled
states treated in this paper by Einstein, Podolsky, and Rosen
[``EPR'', \cite{Einst}] were two-particle states quantum
mechanically correlated with respect to their positions and
momenta. Although important milestones in quantum information
theory have been derived and expressed in terms of qubits or dv,
the notion of quantum entanglement itself came to light in a
continuous-variable setting.\footnote{more explicitly, the notion
of entanglement was introduced by \textcite{Schroedingerreply},
inspired by the EPR paper.} Einstein, Podolsky, and Rosen
\cite{Einst} considered the position wave function
$\psi(x_1,x_2)=C\,\delta(x_1 - x_2 - u)$ with a vanishing
normalization constant $C$. Hence the corresponding quantum state,
\begin{eqnarray}\label{originalEPR}
\int dx_1 dx_2\, \psi(x_1,x_2)\, |x_1,x_2\rangle\propto
\int dx\, |x,x-u\rangle\;,
\end{eqnarray}
describes perfectly correlated positions
($x_1 - x_2 = u$) and momenta (total momentum zero,
$p_1 + p_2 = 0$), but this state
is unnormalizable and unphysical.
However, it can be thought of as the limiting case
of a regularized properly normalized version
where the positions and momenta are correlated only to some
finite extent given by a (Gaussian) width.
Such regularized versions are for example two-mode squeezed
states, since the position and momentum
wave functions for the two-mode squeezed vacuum state
are \cite{Leon},
\begin{eqnarray}\label{wavefcttwin}
\psi(x_1,x_2)
&=&\sqrt{\frac{2}{\pi}}
\exp[-e^{-2r}(x_1+x_2)^2/2
\nonumber\\&&\quad\quad\quad\quad-
e^{+2r}(x_1-x_2)^2/2],
\nonumber\\
\bar{\psi}(p_1,p_2)
&=&\sqrt{\frac{2}{\pi}}
\exp[-e^{-2r}(p_1-p_2)^2/2
\nonumber\\&&\quad\quad\quad\quad-
e^{+2r}(p_1+p_2)^2/2],
\end{eqnarray}
approaching $C\,\delta(x_1-x_2)$ and $C\,\delta(p_1+p_2)$
respectively in the limit of infinite squeezing $r\to\infty$.
The corresponding Wigner function of the two-mode squeezed vacuum
state is then \cite{Walls,Bell,SamKimble}
\begin{eqnarray}\label{WEPR}
W(\xi)&=&\frac{4}{\pi^2}
\exp\{-e^{-2r}[(x_1+x_2)^2+(p_1-p_2)^2]\nonumber\\
&&\quad\quad\quad\;\, -e^{+2r}[(x_1-x_2)^2+(p_1+p_2)^2]\} \;,
\nonumber\\
\end{eqnarray}
with $\xi=(x_1,p_1,x_2,p_2)$. This Wigner function
approaches $C\,\delta(x_1-x_2)\delta(p_1+p_2)$
in the limit of infinite squeezing $r\to\infty$,
corresponding to the original (perfectly correlated and maximally
entangled, but unphysical) EPR state \cite{Einst}.
 From the Wigner function, the marginal distributions
for the two positions or the two momenta
are obtained by integration
over the two momenta or the two positions, respectively,
\begin{eqnarray}\label{marginal}
&&\int\,dp_1\,dp_2\, W(\xi)
=|\psi(x_1,x_2)|^2\\
&&=\frac{2}{\pi}\exp[-e^{-2r}(x_1+x_2)^2-e^{+2r}(x_1-x_2)^2],
\nonumber\\
&&\int\,dx_1\,dx_2\, W(\xi)
=|\bar{\psi}(p_1,p_2)|^2\nonumber\\
&&=\frac{2}{\pi}\exp[-e^{-2r}(p_1-p_2)^2-e^{+2r}(p_1+p_2)^2].
\nonumber
\end{eqnarray}
Though having well defined relative position and total momentum
for large squeezing, the two modes of the two-mode squeezed vacuum
state exhibit increasing uncertainties in
their individual positions and momenta as the squeezing grows.
In fact, upon tracing (integrating) out either mode of the
Wigner function in Eq.~(\ref{WEPR}), we obtain the thermal state
\begin{eqnarray}\label{thermal}
\int\,dx_1\,dp_1\, W(\xi)
=\frac{2}{\pi(1+2\bar{n})}\,\exp\left[-\frac{2(x_2^2+p_2^2)}
{1+2\bar{n}}\right],\nonumber\\
\end{eqnarray}
with mean photon number $\bar{n}=\sinh^2 r$.
Instead of the cv position or momentum basis,
the two-mode squeezed vacuum state may also be written in the
discrete (though, of course, still infinite-dimensional)
photon number (Fock) basis.
Applying the two-mode squeeze operator with $\Theta=0$,
as defined in Eq.~(\ref{TWOMODESQoperatordef}), to two vacuum modes,
we obtain the following expression,
\begin{eqnarray}\label{twin}
\hat S(\zeta)|00\rangle&=&
e^{r(\hat a_1^{\dagger}\hat a_2^{\dagger}-\hat a_1\hat a_2)}
|00\rangle\nonumber\\
&=&e^{\tanh r \,\hat a_1^{\dagger}\hat a_2^{\dagger}}
\left(\frac{1}{\cosh r}\right)^{\hat a_1^{\dagger}\hat a_1+
\hat a_2^{\dagger}\hat a_2+1}\nonumber\\
&&\times \;e^{-\tanh r \,\hat a_1\hat a_2}|00\rangle\nonumber\\
&=&\sqrt{1-\lambda}\sum_{n=0}^{\infty}
\lambda^{n/2} |n\rangle |n\rangle \;,
\end{eqnarray}
where $\lambda=\tanh^2 r$. In the second line here, we have used
the disentangling theorem of \textcite{Collett}. The form in
Eq.~(\ref{twin}) reveals that the two modes of the two-mode
squeezed vacuum state are also quantum correlated in photon number
and phase.

The two-mode squeezed vacuum state, as produced
by the unitary two-mode squeeze operator
in Eq.~(\ref{TWOMODESQoperatordef}) corresponding to
the NOPA interaction Hamiltonian in Eq.~(\ref{NOPOHam}),
is equivalent to the two-mode state emerging from
a 50:50 beam splitter with two single-mode squeezed vacuum
states at the input. The simplest way to see this is
in the Heisenberg representation.
A single-mode vacuum state squeezed in $p$ as in
Eq.~(\ref{ONEMODESQSOL}) with $\Theta=0$,
\begin{eqnarray}
\hat a_1&=&
\hat a_1^{(0)} \cosh r + \hat a_1^{(0)\dagger} \sinh r \;,
\end{eqnarray}
and another one squeezed in $x$,
\begin{eqnarray}
\hat a_2&=&
\hat a_2^{(0)} \cosh r - \hat a_2^{(0)\dagger} \sinh r \;,
\end{eqnarray}
are combined at a 50:50 beam splitter,
\begin{eqnarray}\label{TWOMODESQfromBS}
\hat b_1&=&
(\hat a_1 + \hat a_2)/\sqrt{2}\nonumber\\
&=&
\hat b_1^{(0)} \cosh r + \hat b_2^{(0)\dagger} \sinh r \;,\nonumber\\
\hat b_2&=&
(\hat a_1 - \hat a_2)/\sqrt{2}\nonumber\\
&=&
\hat b_2^{(0)} \cosh r + \hat b_1^{(0)\dagger} \sinh r \;,
\end{eqnarray}
where $\hat b_1^{(0)}=(\hat a_1^{(0)}+\hat
a_2^{(0)})/\sqrt{2}$ and $b_2^{(0)}=
(\hat a_1^{(0)}-\hat a_2^{(0)})/\sqrt{2}$ are again
two vacuum modes. The resulting state is a two-mode squeezed
state as in Eq.~(\ref{TWOMODESQSOL}) with vacuum inputs.
The quadrature operators of the
two-mode squeezed vacuum state can be written as
\begin{eqnarray}\label{2modeHeis}
\hat{x}_1&=&(e^{+r} \hat{x}^{(0)}_1+ e^{-r} \hat{x}^{(0)}_2)/\sqrt{2},
\nonumber\\
\hat{p}_1&=&(e^{-r} \hat{p}^{(0)}_1+ e^{+r} \hat{p}^{(0)}_2)/\sqrt{2},
\nonumber\\
\hat{x}_2&=&(e^{+r} \hat{x}^{(0)}_1- e^{-r} \hat{x}^{(0)}_2)/\sqrt{2},
\nonumber\\
\hat{p}_2&=&(e^{-r} \hat{p}^{(0)}_1- e^{+r} \hat{p}^{(0)}_2)/\sqrt{2}
\;,
\end{eqnarray}
where here $\hat b_k=\hat x_k + i\hat p_k$ and
$\hat a_k^{(0)}=\hat x_k^{(0)} + i\hat p_k^{(0)}$.
Whereas the individual quadratures $\hat x_k$ and $\hat p_k$
become very noisy for large squeezing $r$,
the relative position and the total momentum,
\begin{eqnarray}\label{2moderelpostotmom}
\hat x_1 - \hat x_2 &=& \sqrt{2}\,e^{-r}\hat{x}^{(0)}_2,
\nonumber\\
\hat p_1 + \hat p_2 &=& \sqrt{2}\,e^{-r}\hat{p}^{(0)}_1,
\end{eqnarray}
become quiet,
$\langle(\hat x_1 - \hat x_2)^2\rangle = e^{-2r}/2$ and
$\langle(\hat p_1 + \hat p_2)^2\rangle = e^{-2r}/2$.

The two-mode squeezed vacuum state is the quantum optical
representative for bipartite continuous-variable entanglement. In
general, we may refer to cv entanglement whenever the entangled
states are defined in an infinite-dimensional Hilbert space, for
instance, that of two discrete quantized modes having
position-momentum and number-phase quantum correlations. The
Gaussian entangled states are then an important subclass of the cv
entangled states. More general results on the creation of
bipartite Gaussian cv entanglement were presented by
\textcite{WolfEisert03} and \textcite{KrausCirac03}.

Complementary to its occurrence in quantum optical states,
let us illuminate the notion of cv entanglement
and, in particular, the entanglement of Gaussian states
from the perspective and with the tools of
quantum information theory. This leads to a rigorous
{\it definition of entanglement}, necessarily given in
the Schr\"{o}dinger picture as a {\it property of composite
state vectors or, more generally, density operators}.
The link between this definition and the typical
{\it measurable quantities in a cv implementation,
namely the Gaussian moments of the quadratures},
is provided by cv inseparability criteria or EPR-type
nonlocality proofs, which are expressed in terms of
the elements of the second-moment correlation matrix
for the quadrature operators.
We begin with the entanglement shared by only two parties.

\subsection{Bipartite entanglement}\label{bipentanglement}

\subsubsection{Pure states}

Bipartite entanglement, the entanglement of a pair of systems shared
by two parties, is easy to handle for {\it pure states}.
For any pure two-party state, orthonormal bases of each subsystem exist,
$\{|u_n\rangle\}$ and $\{|v_n\rangle\}$, so that the total state vector
can be written in the ``Schmidt decomposition'' \cite{Schmidt} as
\begin{eqnarray}\label{Schmidt}
|\psi\rangle=\sum_n c_n |u_n\rangle |v_n\rangle \;,
\end{eqnarray}
where the summation goes over the smaller of the
dimensionalities of the two subsystems.
The Schmidt coefficients $c_n$ are real and non-negative, and
satisfy $\sum_n c_n^2=1$.
The Schmidt decomposition may be
obtained by transforming the expansion of an arbitrary pure bipartite
state as
\begin{eqnarray}
|\psi\rangle&=&\sum_{mk} a_{mk} |m\rangle |k\rangle
=\sum_{nmk} u_{mn} c_{nn} v_{kn} |m\rangle |k\rangle\nonumber\\
&=&\sum_n c_n |u_n\rangle |v_n\rangle  \;,
\end{eqnarray}
with $c_{nn}\equiv c_n$.
In the first step, the matrix $a$ with complex elements $a_{mk}$ is
diagonalized, $a=u c v^T$, where $u$ and $v$ are unitary matrices and $c$ is
a diagonal matrix with non-negative elements. In the second step,
we defined $|u_n\rangle\equiv \sum_m u_{mn} |m\rangle$ and
$|v_n\rangle\equiv \sum_k v_{kn} |k\rangle$ which form orthonormal sets
due to the unitarity of $u$ and $v$ and the orthonormality of $|m\rangle$
and $|k\rangle$.
A pure state of two $d$-level systems (``qudits'') is now maximally
entangled when the Schmidt coefficients of its total state vector
are all equal. Since the eigenvalues of the reduced density operator
upon tracing out one half of a bipartite state are the Schmidt
coefficients squared,
\begin{eqnarray}
\hat\rho_1={\rm Tr}_2\hat\rho_{12}=
{\rm Tr}_2|\psi\rangle_{12}\langle\psi|=\sum_n c^2_n
|u_n\rangle_1\langle u_n| \;,
\end{eqnarray}
tracing out either qudit of a maximally
entangled state leaves the other half in the maximally mixed state
$\mbox{1$\!\!${\large 1}}/d$.
A pure two-party state is factorizable (not entangled) if and only if
the number of nonzero Schmidt coefficients (``Schmidt rank'') is one.

A unique measure of bipartite entanglement for pure states
is given by the partial von Neumann entropy, the von Neumann entropy
[$S(\hat\rho)=-{\rm Tr}\hat\rho\log\hat\rho$] of the remaining system
after tracing out either subsystem \cite{Benn3}:
$E_{\rm v.N.}=-{\rm Tr}\hat\rho_1\log_d\hat\rho_1=
-{\rm Tr}\hat\rho_2\log_d\hat\rho_2=
-\sum_n c_n^2\log_d c_n^2$, ranging
between zero and one (in units of ``edits''), with
${\rm Tr}_2\hat\rho_{12}=\hat\rho_1$,
${\rm Tr}_1\hat\rho_{12}=\hat\rho_2$.
It corresponds to the number of maximally entangled states
``contained in a given pure state''. For example,
$E_{\rm v.N.}=0.4$ means that asymptotically 1000 copies of the state
can be transformed into 400 maximally entangled states via
deterministic state transformations using local operations
and classical communication (LOCC) \cite{Nielsen}.

According to Eq.~(\ref{twin}),
the Fock basis corresponds to the Schmidt basis of the two-mode
squeezed vacuum state.
In this Schmidt form, we can quantify the
entanglement of the two-mode
squeezed vacuum state via the partial von Neumann entropy
\cite{vanEnk},
\begin{eqnarray}\label{vNeumtwin}
E_{\rm v.N.}&=&
-\log(1-\lambda)-\lambda\log \lambda /(1-\lambda)\\
&=&\cosh^2 r \log(\cosh^2 r)-\sinh^2 r \log(\sinh^2 r)\;.\nonumber
\end{eqnarray}
Note that any pure two-mode Gaussian state can be transformed into
the canonical two-mode squeezed state form via local LUBO
transformations and hence its entanglement can be quantified as in
Eq.~(\ref{vNeumtwin}). More generally, any bipartite pure
multi-mode Gaussian state corresponds to a product of two-mode
squeezed states up to local LUBO transformations
\cite{GiedkeQIC03,Botero}. In addition, the partial von Neumann
entropy of a pure two-mode Gaussian state, corresponding to the
entropy of an arbitrary single-mode Gaussian state, can be also
directly computed \cite{Agarwal}.

In general, an important sign of entanglement is the
violation of inequalities
imposed by local realistic theories \cite{Bell}.
Any pure two-party state is entangled if and only if,
for suitably chosen observables, it yields a violation of
such inequalities. The main features of {\it pure-state}
bipartite entanglement shall be summarized by
\begin{eqnarray}
{\rm entangled}&\Leftarrow\!\!\Rightarrow&
{\rm Schmidt}\,\;{\rm rank}>1,\nonumber\\
{\rm entangled}&\Leftarrow\!\!\Rightarrow&
{\rm partial}\,\;{\rm von}\,\;{\rm Neumann}\,\;{\rm entropy}>0
\;,\nonumber\\
{\rm entangled}&\Leftarrow\!\!\Rightarrow&
{\rm violations}\,\;{\rm of}\,\;{\rm local}\,\;{\rm realism}.\nonumber\\
\end{eqnarray}
All these conditions are necessary and sufficient.

\subsubsection{Mixed states and inseparability criteria}
\label{mixedentanglement}

The definition of pure-state entanglement via the non-factorizability
of the total state vector is generalized to {\it mixed states} through
non-separability (or inseparability) of the total density operator.
A general quantum state of a two-party system is separable if its
total density operator is a mixture (a convex sum) of product states
\cite{Werner},
\begin{eqnarray}\label{PVLconvexsum}
\hat\rho_{12}=\sum_i \eta_i\, \hat\rho_{i,1}\otimes\hat\rho_{i,2}\;.
\end{eqnarray}
Otherwise, it is inseparable.\footnote{Separable states also
exhibit correlations, but those are purely classical. For
instance, compare the separable state
$\hat\rho=\frac{1}{2}(|0\rangle\langle 0|\otimes|0\rangle\langle
0| +|1\rangle\langle 1|\otimes|1\rangle\langle 1|)$ to the pure
maximally entangled ``Bell state''
$|\Phi^+\rangle=\frac{1}{\sqrt{2}}(|0\rangle\otimes|0\rangle
+|1\rangle\otimes|1\rangle)=\frac{1}{\sqrt{2}}(|+\rangle\otimes|+\rangle
+|-\rangle\otimes|-\rangle)$ with the conjugate basis states
$|\pm\rangle=\frac{1}{\sqrt{2}}(|0\rangle\pm |1\rangle)$. The
separable state $\hat\rho$ is classically correlated only with
respect to the predetermined basis $\{|0\rangle,|1\rangle\}$.
However, the Bell state $|\Phi^+\rangle$ is a priori quantum
correlated in both bases $\{|0\rangle,|1\rangle\}$ and
$\{|+\rangle,|-\rangle\}$, and may become a posteriori classically
correlated depending on the particular basis choice in a local
measurement. Similarly, the inseparability criteria for continuous
variables must be expressed in terms of positions and their
conjugate momenta.} In general, it is a non-trivial question
whether a given density operator is separable or inseparable.
Nonetheless, a very convenient method to test for inseparability
is Peres' partial transpose criterion \cite{Peres}. For a
separable state as in Eq.~(\ref{PVLconvexsum}), transposition of
either density matrix yields again a legitimate non-negative
density operator with unit trace,
\begin{eqnarray}
\hat\rho'_{12}=\sum_i \eta_i\, (\hat\rho_{i,1})^T\otimes\hat\rho_{i,2}\;,
\end{eqnarray}
since $(\hat\rho_{i,1})^T=(\hat\rho_{i,1})^*$ corresponds to
a legitimate density matrix.
This is a necessary condition for a separable state, and hence
a single negative eigenvalue of the partially transposed density
matrix is a sufficient condition for inseparability
(transposition is a so-called positive, but not completely
positive map, which means its application to a subsystem
may yield an unphysical state when the subsystem
is entangled to other subsystems).
In general, for states with arbitrary dimension,
negative partial transpose (npt) is only sufficient for inseparability
\cite{Horo}. Similarly, for arbitrary mixed states, the occurrence of
violations of inequalities imposed by local realism is
also only a sufficient, but not a necessary condition for
inseparability \cite{Werner}.
To summarize, for general {\it mixed-state}
bipartite inseparability, the following statements hold,
\begin{eqnarray}
{\rm inseparable}&\Leftarrow&
{\rm npt},
\nonumber\\
{\rm inseparable}&\Leftarrow&
{\rm violations}\,\;{\rm of}\,\;{\rm local}\,\;{\rm realism}.\nonumber\\
\end{eqnarray}
However, there are classes of states where negative partial transpose
becomes both necessary and sufficient, namely
\begin{eqnarray}\label{nptstatements}
2\times 2-{\rm dimensional,}\,\;{\rm inseparable}
&\Leftarrow\!\!\Rightarrow&
{\rm npt},
\nonumber\\
2\times 3-{\rm dimensional,}\,\;{\rm inseparable}
&\Leftarrow\!\!\Rightarrow&
{\rm npt},
\nonumber\\
(1\times N)-{\rm mode}\,\;{\rm Gaussian,}\,\;{\rm inseparable}
&\Leftarrow\!\!\Rightarrow&
{\rm npt}.\nonumber\\
\end{eqnarray}
Other sufficient inseparability criteria include violations of
an entropic inequality
[$E_{\rm v.N.}(\hat\rho_1)> E_{\rm v.N.}(\hat\rho_{12})$,
again with $\hat\rho_1
={\rm Tr}_2\hat\rho_{12}$] \cite{Horo3}, and a condition
based on the theory of majorization \cite{NielsenKempe}.
In the context of bound entanglement and distillability, the
so-called reduction inseparability criterion \cite{Horo4}
proves very useful (see below).

In Eq.~(\ref{nptstatements}), we made a statement about the
inseparability of Gaussian states in terms of the partial
transpose criterion. What does partial transposition applied to
bipartite Gaussian or, more generally, cv states actually mean?
Due to the Hermiticity of a density operator, transposition
corresponds to complex conjugation. Moreover, as for the time
evolution of a quantum system described by the Schr\"{o}dinger
equation, complex conjugation is equivalent to time reversal,
$i\hbar\partial/\partial t\rightarrow -i\hbar\partial/\partial t$.
Hence, intuitively, transposition of a density operator means time
reversal, or, in terms of cv, sign change of the momentum
variables. This observation and its application to the
inseparability problem of cv states is due to \textcite{Simon}.
Thus, in phase space, transposition is described by
$\xi^{T}\rightarrow \Gamma\xi^{T}=
(x_1,-p_1,x_2,-p_2,...,x_N,-p_N)^{T}$, i.e., by transforming the
Wigner function $W(x_1,p_1,x_2,p_2,...,x_N,p_N)\rightarrow
W(x_1,-p_1,x_2,-p_2,...,x_N,-p_N)$. This general transposition
rule for cv is in the case of Gaussian states as in
Eq.~(\ref{Gausswigndef}) reduced to $V^{(N)}\rightarrow\Gamma
V^{(N)}\Gamma$ (where the first moments are not relevant to the
separability properties, since they can be eliminated via LOCC).

Expressing partial transposition of a bipartite
Gaussian system by $\Gamma_a\equiv\Gamma\oplus
\mbox{1$\!\!${\large 1}}$ (where $A\oplus B$
means the block-diagonal matrix with the matrices
$A$ and $B$ as diagonal `entries', and $A$ and $B$
are respectively $2N\times 2N$ and $2M\times 2M$ square
matrices applicable to $N$ modes at $a$'s side
and $M$ modes at $b$'s side),
the condition that the partially transposed
Gaussian state described by $\Gamma_a V^{(N+M)}\Gamma_a$
is unphysical [see Eq.~(\ref{Nmodeuncert})],
\begin{eqnarray}
\Gamma_a V^{(N+M)}\Gamma_a\ngeq \frac{i}{4}\,\Lambda\;,
\end{eqnarray}
is sufficient for the inseparability
between $a$ and $b$ \cite{Simon,Wolf}.
For Gaussian states with $N=M=1$ \cite{Simon} and for those
with $N=1$ and arbitrary $M$ \cite{Wolf},
this condition is necessary and sufficient.
The simplest example where the condition is no longer
necessary for inseparability involves two modes at each
side, $N=M=2$. In that case, states with positive partial
transpose, so-called bound entangled Gaussian states
(see below), exist \cite{Wolf}.
For the general bipartite $N\times M$ case of Gaussian states,
there is also a necessary and sufficient condition:
the correlation matrix $V^{(N+M)}$ corresponds to
a separable state iff a pair of correlation matrices
$V^{(N)}_a$ and $V^{(M)}_b$ exists such that \cite{Wolf}
\begin{eqnarray}
V^{(N+M)}\geq V^{(N)}_a\oplus V^{(M)}_b\;.
\end{eqnarray}
Since it is in general hard to find such a pair of correlation
matrices $V^{(N)}_a$ and $V^{(M)}_b$ for a separable state or to
prove the non-existence of such a pair for an inseparable state,
this criterion in not very practical. A more practical solution
was proposed by \textcite{Giedke01PRL}. The operational criteria
for Gaussian states there, computable and testable via a finite
number of iterations, are entirely independent of the npt
criterion. They rely on a nonlinear map between the correlation
matrices rather than a linear one such as the partial
transposition, and in contrast to the npt criterion, they witness
also the inseparability of bound entangled states. Thus, the
separability problem for bipartite Gaussian states with
arbitrarily many modes at each side is completely solved.

Let us now consider arbitrary bipartite two-mode states.
According to the definition of the $N$-mode
correlation matrix $V^{(N)}$ in Eq.(\ref{corrdef}),
we can write the correlation matrix of an arbitrary
bipartite two-mode system in block form,
\begin{eqnarray}\label{blockform}
V^{(2)}=
\left( \begin{array}{cc} A & C \\
C^{T} & B
\end{array} \right)\;,
\end{eqnarray}
where $A$, $B$, and $C$ are real $2\times 2$ matrices.
Simon's cv version of the Peres-Horodecki partial
transpose criterion reads as follows \cite{Simon},
\begin{eqnarray}\label{generalSimon}
\det A \det B + \left(\frac{1}{16}-|\det C|\right)^2-
{\rm Tr}(AJCJBJC^{T}J)\nonumber\\
\geq \frac{1}{16}(\det A + \det B)\;,\nonumber\\
\end{eqnarray}
where $J$ is the $2\times 2$ matrix
from Eq.~(\ref{landjmatrices}).
Any separable bipartite state satisfies the inequality of
Eq.~(\ref{generalSimon}), so that it
represents a necessary condition for separability, and hence its violation
is a sufficient condition for inseparability.
The inequality Eq.~(\ref{generalSimon}) is a consequence
of the fact that the two-mode uncertainty relation,
Eq.~(\ref{Nmodeuncert}) with $N=2$,
is preserved under partial transpose,
$W(x_1,p_1,x_2,p_2)\rightarrow W(x_1,p_1,x_2,-p_2)$,
provided the state is separable.

We may now define the following two standard forms for the correlation
matrix:
\begin{eqnarray}\label{standardform1}
V_{I}^{(2)}=
\left( \begin{array}{cccc} a & 0 & c & 0 \\
0 & a & 0 & c' \\ c & 0 & b & 0 \\ 0 & c' & 0 & b
\end{array} \right)\;,
\end{eqnarray}
and
\begin{eqnarray}\label{standardform2}
V_{II}^{(2)}=
\left( \begin{array}{cccc} a_1 & 0 & c_1 & 0 \\
0 & a_2 & 0 & c_2 \\ c_1 & 0 & b_1 & 0 \\ 0 & c_2 & 0 & b_2
\end{array} \right)\;,
\end{eqnarray}
where the elements of the second standard form $V_{II}^{(2)}$ satisfy
\begin{eqnarray}\label{standardform2restrictions}
\frac{a_1-1/4}{b_1-1/4}&=&\frac{a_2-1/4}{b_2-1/4},\nonumber\\
|c_1|-|c_2|&=&\sqrt{(a_1-1/4)(b_1-1/4)}\nonumber\\
&&-\sqrt{(a_2-1/4)(b_2-1/4)}\;.
\end{eqnarray}
Any correlation matrix can be transformed into the first standard form
$V_{I}^{(2)}$ via appropriate local canonical transformations \cite{Simon}
[i.e., via local LUBO's with the LUBO given in Eq.~(\ref{LUBO})].
 From the first standard form $V_{I}^{(2)}$, two appropriate local squeezing
operations can always lead to the second standard form $V_{II}^{(2)}$
\cite{Duan}.

For the standard form $V_{I}^{(2)}$, the necessary separability condition
of Eq.~(\ref{generalSimon}) simplifies to
\begin{eqnarray}\label{generalSimonstandard1}
16(ab-c^2)(ab-{c'}^2)\geq (a^2+b^2) + 2|c c'| - \frac{1}{16}.
\end{eqnarray}
Simon's criterion does not rely
on that specific standard form and can, in fact, be applied to an
arbitrary (even non-Gaussian) state using Eq.~(\ref{generalSimon}).
For Gaussian two-mode states, however, Eq.~(\ref{generalSimon}) turns
out to be both a necessary and a sufficient condition for separability
\cite{Simon}.

A similar inseparability criterion, applicable to two-mode
continuous-variable systems and expressed in terms of an
inequality for certain variances involving position and momentum
operators, was derived by \textcite{Duan} using a strategy
independent of the partial transpose. This criterion relies upon
the standard form $V_{II}^{(2)}$ to follow through as a necessary
and sufficient condition for two-mode Gaussian states. Expressed
in terms of the elements of $V_{II}^{(2)}$, the necessary and
sufficient condition for the separability of two-mode Gaussian
states reads
\begin{eqnarray}\label{Duan1}
\langle(\Delta\hat{u})^2\rangle_{\rho}+
\langle(\Delta\hat{v})^2\rangle_{\rho}
\geq \frac{a_0^2}{2}+\frac{1}{2 a_0^2},
\end{eqnarray}
where
\begin{eqnarray}\label{Duan2}
\hat{u}&=&a_0\hat{x}_1-\frac{c_1}{|c_1|a_0}\hat{x}_2,\nonumber\\
\hat{v}&=&a_0\hat{p}_1-\frac{c_2}{|c_2|a_0}\hat{p}_2,\nonumber\\
a_0^2&=&\sqrt{\frac{b_1-1/4}{a_1-1/4}}
=\sqrt{\frac{b_2-1/4}{a_2-1/4}},
\end{eqnarray}
and the bipartite state of interest $\hat\rho$ has been labeled $\rho$.
Without the assumption of Gaussian states, an alternative
approach \cite{Duan}, solely based on the Heisenberg uncertainty
relation of position and momentum and on the Cauchy-Schwarz inequality,
leads to an inequality similar to Eq.~(\ref{Duan1}).
It only represents a necessary condition for
the separability of arbitrary states,
\begin{eqnarray}\label{Duan3}
\langle(\Delta\hat{u})^2\rangle_{\rho}+
\langle(\Delta\hat{v})^2\rangle_{\rho}
&\geq& \bar{a}^2|\langle [\hat{x}_1,\hat{p}_1]
\rangle_{\rho}| + |\langle
[\hat{x}_2,\hat{p}_2]\rangle_{\rho}|/\bar{a}^2\nonumber\\
&=&\frac{\bar{a}^2}{2}+\frac{1}{2 \bar{a}^2},
\end{eqnarray}
with
\begin{eqnarray}\label{Duan4}
\hat{u}&=&|\bar{a}|\hat{x}_1-\frac{1}{\bar{a}}\hat{x}_2,\nonumber\\
\hat{v}&=&|\bar{a}|\hat{p}_1+\frac{1}{\bar{a}}\hat{p}_2.
\end{eqnarray}
Here, $\bar{a}$ is an arbitrary nonzero real parameter. Let us
also mention that a similar (but weaker) inseparability criterion
was derived by \textcite{Tan}, namely the necessary condition for
any separable state,
\begin{eqnarray}\label{Tan}
\langle(\Delta\hat{u})^2\rangle_{\rho}
\langle(\Delta\hat{v})^2\rangle_{\rho}
\geq \frac{1}{4},
\end{eqnarray}
with $\bar{a}=1$ in Eq.~(\ref{Duan4}).
It is simply the product version of the sum condition in Eq.~(\ref{Duan3})
(with $\bar{a}=1$).
A generalization of these two-party separability conditions
can be found in Ref.~\cite{Giovannetti}, including
a discussion on how they are related with each other.
In this respect, defining the general linear combinations
\begin{eqnarray}\label{generalcombtwo}
\hat u\equiv h_1\hat x_1 + h_2\hat x_2 \;,
\quad\hat v\equiv g_1\hat p_1 + g_2\hat p_2 \;,
\end{eqnarray}
let us only note here that for any separable state,
we have
\begin{eqnarray}\label{generalDuan}
\langle(\Delta\hat{u})^2\rangle_{\rho}+
\langle(\Delta\hat{v})^2\rangle_{\rho}\geq
(|h_1 g_1|+|h_2 g_2|)/2\,,
\end{eqnarray}
whereas for a potentially entangled state, this bound
is changed to
\begin{eqnarray}
\langle(\Delta\hat{u})^2\rangle_{\rho}+
\langle(\Delta\hat{v})^2\rangle_{\rho}\geq
(|h_1 g_1 + h_2 g_2|)/2\,.
\end{eqnarray}
The $h_l$ and $g_l$ are arbitrary real parameters.
When choosing, for instance, $h_1=-h_2=g_1=g_2=1$,
the bound for a separable state becomes 1, whereas that
for an entangled state drops to zero.
In fact, with this choice, $\hat u =\hat x_1 - \hat x_2$ and
$\hat v =\hat p_1 + \hat p_2$,
quantum mechanics allows the observables
$\hat u$ and $\hat v$ to simultaneously take on arbitrarily
well defined values because of the vanishing commutator
\begin{eqnarray}
[\hat x_1 - \hat x_2,\hat p_1 + \hat p_2]=0\;.
\end{eqnarray}

All the inseparability criteria discussed above are fulfilled
by the two-mode squeezed vacuum state for any nonzero squeezing.
For example,
according to Eq.~(\ref{WEPR}) and Eq.~(\ref{Gausswigndef}),
its correlation matrix is given by
\begin{eqnarray}\label{corrEPR}
V^{(2)}=\frac{1}{4}
\left( \begin{array}{cccc} \cosh 2r & 0 & \sinh 2r & 0 \\
0 & \cosh 2r & 0 & -\sinh 2r \\  \sinh 2r & 0 & \cosh 2r & 0 \\
0 & -\sinh 2r & 0 & \cosh 2r
\end{array} \right)\;.\nonumber\\
\end{eqnarray}
This matrix is in standard form $V_{I}^{(2)}$. Hence
one can easily verify that Simon's separability condition
Eq.~(\ref{generalSimonstandard1}) is violated for any $r>0$.
Even simpler is the application of Eq.~(\ref{2moderelpostotmom})
for the two-mode squeezed vacuum state to the condition
in Eq.~(\ref{Duan3}) which is also violated for any $r>0$.

Separability conditions similar to those above
can be derived also in terms of the polarization
Stokes operators from Eq.~(\ref{stokesdef})
\cite{BowenTreps02,KorolkovaLoudonRalph02,KorolkovaLoudon03}.
Thereby, in general, one must take into account
the operator-valued commutator of the Stokes operators,
$[\hat S_j, \hat S_k] = i\epsilon_{jkl} \hat S_l$.
Analogously, a possible sum condition in terms of the
collective spin variables of two atomic ensembles
[see Eq.~(\ref{spinuncert})],
always satisfied for separable systems, is \cite{Polzik02}
\begin{eqnarray}\label{Duanforatoms}
&&\langle[\Delta(\hat F_{x1}+\hat F_{x2})]^2\rangle_{\rho}+
\langle[\Delta(\hat F_{y1}+\hat F_{y2})]^2\rangle_{\rho}
\nonumber\\
&&\quad\quad\quad\quad\geq |\langle\hat F_{z1}\rangle_{\rho}| +
|\langle\hat F_{z2}\rangle_{\rho}|\;.
\end{eqnarray}
Note that in contrast to the conditions in Eq.~(\ref{Duan3})
or Eq.~(\ref{generalDuan}), the condition in Eq.~(\ref{Duanforatoms})
has, in general, a state-dependent
bound due to the operator-valued commutator
$[\hat F_j, \hat F_k] = i\epsilon_{jkl} \hat F_l$.
However, as discussed in Sec.~\ref{polspinrep},
within the subset of states
with a large classical mean polarization along the $z$-axis,
the commutators of $\hat F_x$ and $\hat F_y$ resemble those
of $\hat x$ and $\hat p$. One may choose two spins with
opposite classical orientation,
$\langle\hat F_{z1}\rangle = -\langle\hat F_{z2}\rangle=
F$, yielding
\begin{eqnarray}\label{Duanforatoms2}
\langle[\Delta(\hat F_{x1}+\hat F_{x2})]^2\rangle_{\rho}+
\langle[\Delta(\hat F_{y1}+\hat F_{y2})]^2\rangle_{\rho}
\geq 2\,F\;.\nonumber\\
\end{eqnarray}
With this choice, and using
$\hat F_{zj} \simeq \langle \hat F_{zj}\rangle$, $j=1,2$,
the vanishing commutator
\begin{eqnarray}
[\hat F_{x1} + \hat F_{x2},\hat F_{y1} + \hat F_{y2}]=
i(\hat F_{z1} + \hat F_{z2})=0\;,
\end{eqnarray}
permits an arbitrarily large violation of
Eq.~(\ref{Duanforatoms2}) for entangled states.
Similar to Eq.~(\ref{Duan3}), where (for $\bar{a}=1$)
the bound is four vacuum units corresponding to
uncorrelated quadratures of the two modes,
entanglement is present according to Eq.~(\ref{Duanforatoms2})
when the total spin variances are smaller than those of uncorrelated
atoms in two collective ``vacuum states'', each with
$\langle(\Delta\hat F_x)^2\rangle=\langle(\Delta\hat F_y)^2\rangle=
F/2$ [see Eq.~(\ref{spinuncert})].

As for the quantification of bipartite mixed-state entanglement,
there are various measures available such as the entanglement of
formation (EoF) and distillation \cite{Benn5}. Only for pure
states do these measures coincide and equal the partial von
Neumann entropy. In general, the EoF is hard to compute. However,
apart from the qubit case \cite{WoottersEoF}, also for symmetric
two-mode Gaussian states given by a correlation matrix in
Eq.~(\ref{standardform1}) with $a=b$, the EoF can be calculated
via the total variances in Eq.~(\ref{Duan3}) \cite{GiedkeEoF}. A
Gaussian version of the EoF was proposed by \textcite{WolfEoF}.
Another computable measure of entanglement for any mixed state of
an arbitrary bipartite system, including bipartite Gaussian
states, is the ``logarithmic negativity'' based on the negativity
of the partial transpose \cite{VidalWerner02}.

\subsection{Multipartite entanglement}\label{multipentanglement}

Multipartite entanglement, the entanglement shared by more
than two parties, is a subtle issue even for pure states.
In that case, for {\it pure} multi-party states,
a Schmidt decomposition does not exist in general.
The total state vector then cannot be written as a single sum over
orthonormal basis states.
Let us first consider dv multipartite entanglement.

\subsubsection{Discrete variables}

There is one very important representative of multipartite
entanglement which does have the form of a multi-party Schmidt decomposition,
namely the Greenberger-Horne-Zeilinger (GHZ) state \cite{GHZ}
\begin{equation}\label{PVLGHZdef}
|{\rm GHZ}\rangle=\frac{1}{\sqrt{2}}\left(|000\rangle
+|111\rangle\right)\;,
\end{equation}
here given as a three-qubit state.
Although there is no rigorous definition
of maximally entangled multi-party states due to the lack of a general
Schmidt decomposition, the form of the GHZ state with
all ``Schmidt coefficients'' equal suggests
that it exhibits maximum multipartite entanglement.
In fact, there are various reasons for assigning the attribute ``maximally
entangled'' to the $N$-party GHZ states,
$(|000\cdots 000\rangle+|111\cdots 111\rangle)/\sqrt{2}$.
For example, they yield the maximum violations
of multi-party inequalities imposed by local realistic theories
\cite{Mermin,Klyshko,GisinBech}.
Further, their entanglement heavily relies on all parties,
and, if examined pairwise,
they do not contain simple bipartite entanglement
(see below).

For the case of three qubits, any pure and fully entangled state can be
transformed to either the GHZ state or the so-called W state \cite{DUER},
\begin{equation}\label{PVLWdef}
|{\rm W}\rangle=\frac{1}{\sqrt{3}}\left(|100\rangle
+|010\rangle+|001\rangle\right)\;,
\end{equation}
via stochastic local operations and classical communication
(``SLOCC'', where stochastic means that the state is transformed
with non-zero probability). Thus,
with respect to SLOCC, there are two inequivalent classes of genuine
tripartite entanglement, represented by the GHZ and the W state.
Genuinely or fully tripartite entangled here means that the
entanglement of the three-qubit state is not just present between two
parties while the remaining party can be separated by a tensor product.
Though genuinely tripartite, the entanglement of the W state is also
``readily bipartite''. This means that the remaining two-party state after
tracing out one party,
\begin{eqnarray}
{\rm Tr}_1|{\rm W}\rangle\langle{\rm W}|&=&
\frac{1}{3}\big(
|00\rangle\langle 00| + |10\rangle\langle 10| + |01\rangle\langle 01|
\nonumber\\
&&\quad+ |01\rangle\langle 10| + |10\rangle\langle 01| \big)\;,
\end{eqnarray}
is inseparable which can be verified by taking the partial transpose
[the eigenvalues are $1/3$, $1/3$, $(1\pm\sqrt{5})/6$].
This is in contrast to the GHZ state where tracing out one party
yields the separable two-qubit state
\begin{eqnarray}
{\rm Tr}_1|{\rm GHZ}\rangle\langle {\rm GHZ}|= \frac{1}{2}\left(
|00\rangle\langle 00| + |11\rangle\langle 11|
\right)\;.\nonumber\\
\end{eqnarray}
Maximum bipartite entanglement is available from the GHZ state
through a local measurement of one party in the
conjugate basis $\{|\pm\rangle=
(|0\rangle\pm |1\rangle)/\sqrt{2}\}$
(plus classical communication about the result),
\begin{eqnarray}\label{PVLGHZmeasonconjbasis}
\frac{ |\pm\rangle_{1\,1}\langle \pm |{\rm GHZ}\rangle }{|\!|
|\pm\rangle_{1\,1}\langle \pm |{\rm GHZ}\rangle |\!|}=
|\pm\rangle_1 \otimes |\Phi^{\pm}\rangle \;.\nonumber\\
\end{eqnarray}
Here, $|\Phi^{\pm}\rangle$ are two of the four Bell states,
$|\Phi^{\pm}\rangle=(|00\rangle\pm |11\rangle)/\sqrt{2}$,
$|\Psi^{\pm}\rangle=(|01\rangle\pm |10\rangle)/\sqrt{2}$.

What can be said about arbitrary {\it mixed} entangled states
of more than two parties?
There is of course an immense variety of inequivalent classes
of multi-party mixed states,
e.g., five classes of three-qubit states of which
the extreme cases are the fully separable,
$\sum_i \eta_i\, \hat\rho_{i,1}\otimes\hat\rho_{i,2}
\otimes\hat\rho_{i,3}$, and the genuinely tripartite inseparable states
\cite{DuerCiracTarr}.

\subsubsection{Genuine multipartite entanglement}

By using the term {\it genuine} multipartite entanglement,
we refer to states where none of the parties
can be separated from any other party in a mixture
of product states. In general,
multi-party inseparability criteria
cannot be formulated in such a compact form as for two parties.
In order to verify genuine $N$-party
entanglement, one has to rule out
any possible partially separable form.
In principle, this can be done by considering all possible
bipartite splittings (or groupings) and, for instance, applying
the npt criterion.
Moreover, the quantification of multipartite entanglement,
even for pure states, is still subject of current research.
Violations of multi-party inequalities imposed by local
realism do not necessarily imply genuine multi-party inseparability.
The following statements hold for pure or mixed
{\it multipartite} entangled states (both dv and cv),
\begin{eqnarray}
{\rm partially}\,\;{\rm entangled}&\Leftarrow&
{\rm violations}\,\;{\rm of}\,\;N-{\rm party}\nonumber\\
&&{\rm Bell-type}\,\;{\rm inequalities},\nonumber\\
{\rm genuinely}\quad\quad\;\;\;&\Leftarrow\!\!\!\!/\!\!\!\!\Rightarrow&
\nonumber\\
{\rm multipartite}
\,\;{\rm entangled}&&\quad\quad ''\quad\quad''\quad\quad''
\end{eqnarray}
An example for the last statement is the
pure genuinely $N$-party entangled state
%\cite{Gisintalk}
\begin{eqnarray}
|\psi\rangle = \cos\alpha\, |00\cdots 0\rangle +
\sin\alpha\, |11\cdots 1\rangle \;,
\end{eqnarray}
which for $\sin 2\alpha\leq 1/\sqrt{2^{N-1}}$
does not violate any $N$-party Bell inequality \cite{Bell},
if $N$ odd, and does not violate Mermin-Klyshko inequalities
\cite{Mermin,Klyshko,GisinBech} for any $N$.
Genuine multipartite entanglement can be verified only
via sufficiently large violations \cite{Seevinck} of
Mermin-Klyshko inequalities.
Another sufficient condition for the genuine $N$-party
entanglement of an $N$-qubit state $\hat\rho$ exists,
namely $\langle{\rm GHZ}|\hat\rho|{\rm GHZ}\rangle >1/2$
\cite{Seevinck}.

\subsubsection{Separability properties of Gaussian states}
\label{seppropgaussian}

As for the cv case, the criteria by \textcite{Giedke01} determine
to which of five possible classes of fully and partially
separable, and fully inseparable states a three-party three-mode
Gaussian state belongs. Hence genuine tripartite entanglement if
present can be unambiguously identified. The classification is
mainly based on the npt criterion for cv states. For three-party
three-mode Gaussian states, the only partially separable forms are
those with a bipartite splitting of $1 \times 2$ modes. Hence
already the npt criterion is necessary and sufficient.

The classification of tripartite three-mode Gaussian states
\cite{Giedke01},
\begin{eqnarray}\label{classification}
&&\quad{\rm class}\;1\,:\quad\quad
\bar V^{(3)}_1\ngeq \frac{i}{4}\,\Lambda\,,
\bar V^{(3)}_2\ngeq \frac{i}{4}\,\Lambda\,,
\bar V^{(3)}_3\ngeq \frac{i}{4}\,\Lambda\,,
\nonumber\\
&&\quad{\rm class}\;2\,:\quad\quad
\bar V^{(3)}_k\geq \frac{i}{4}\,\Lambda\,,
\bar V^{(3)}_m\ngeq \frac{i}{4}\,\Lambda\,,
\bar V^{(3)}_n\ngeq \frac{i}{4}\,\Lambda\,,
\nonumber\\
&&\quad{\rm class}\;3\,:\quad\quad
\bar V^{(3)}_k\geq \frac{i}{4}\,\Lambda\,,
\bar V^{(3)}_m\geq \frac{i}{4}\,\Lambda\,,
\bar V^{(3)}_n\ngeq \frac{i}{4}\,\Lambda\,,
\nonumber\\
&&{\rm class}\;4\;{\rm or}\;5\,:\quad
\bar V^{(3)}_1\geq \frac{i}{4}\,\Lambda\,,
\bar V^{(3)}_2\geq \frac{i}{4}\,\Lambda\,,
\bar V^{(3)}_3\geq \frac{i}{4}\,\Lambda\,,
\nonumber\\
\end{eqnarray}
is solely based on the npt criterion, where
$\bar V^{(3)}_j\equiv\Gamma_j V^{(3)}\Gamma_j$ denotes
the partial transposition with respect to one mode $j$.
In classes 2 and 3, any permutation of modes ($k,m,n$)
must be considered.
Class 1 corresponds to the fully inseparable states.
Class 5 shall contain the fully separable states.
A Gaussian state described by $V^{(3)}$ is fully separable
iff one-mode correlation matrices
$V^{(1)}_1$, $V^{(1)}_2$, and $V^{(1)}_3$ exist such that
$V^{(3)}\geq V^{(1)}_1\oplus V^{(1)}_2\oplus V^{(1)}_3$.
In general, fully separable quantum states can be written as
a mixture of tripartite product states,
$\sum_i \eta_i\, \hat\rho_{i,1}\otimes\hat\rho_{i,2}
\otimes\hat\rho_{i,3}$.
In class 2, we have the one-mode biseparable states,
where only one particular mode is separable from
the remaining pair of modes.
This means in the Gaussian case that
only for one particular mode $k$,
$V^{(3)}\geq V^{(1)}_k\oplus V^{(2)}_{mn}$
with some two-mode correlation matrix $V^{(2)}_{mn}$
and one-mode correlation matrix $V^{(1)}_k$.
In general, such a state can be written as
$\sum_i \eta_i\, \hat\rho_{i,k}\otimes\hat\rho_{i,mn}$
for one mode $k$.
Class 3 contains those states where two but not three
bipartite splittings are possible, i.e.,
two different modes $k$ and $m$
are separable from the remaining
pair of modes (two-mode biseparable states).
The states of class 4 (three-mode biseparable states)
can be written as a mixture of products
between any mode 1, 2, or 3
and the remaining pair of modes,
but not as a mixture of three-mode product states.
Obviously, classes 4 and 5 are not distinguishable
via the npt criterion.
An additional criterion for this distinction
of class 4 and 5 Gaussian states
is given in Ref.~\cite{Giedke01},
deciding whether one-mode correlation matrices
$V^{(1)}_1$, $V^{(1)}_2$, and $V^{(1)}_3$ exist such that
$V^{(3)}\geq V^{(1)}_1\oplus V^{(1)}_2\oplus V^{(1)}_3$.
For the identification of genuinely tripartite entangled
Gaussian states, only class 1 has to be distinguished
from the rest. Hence the npt criterion alone suffices.

What about more than three parties and modes? Even for only four
parties and modes, the separability issue becomes more subtle. The
one-mode bipartite splittings, $\sum_i \eta_i\,
\hat\rho_{i,klm}\otimes\hat\rho_{i,n}$, can be tested and possibly
ruled out via the npt criterion with respect to any mode $n$. In
the Gaussian language, if $\bar
V^{(4)}_n\ngeq\frac{i}{4}\,\Lambda$ for any $n$, the state cannot
be written in the above form. Since we consider here the bipartite
splitting of $1\times 3$ modes, the npt condition is necessary and
sufficient for Gaussian states. However, also a state of the form
$\sum_i \eta_i\, \hat\rho_{i,kl}\otimes\hat\rho_{i,mn}$ leads to
negative partial transpose with respect to any of the four modes
when the two pairs ($k,l$) and ($m,n$) are each entangled. Thus,
npt with respect to any individual mode is necessary but not
sufficient for genuine four-party entanglement. One has to
consider also the partial transposition with respect to any pair
of modes. For this $2\times 2$ mode case, however, we know that
entangled Gaussian states with positive partial transpose exist
\cite{Wolf}. But the npt criterion is still sufficient for the
inseparability between any two pairs. As for a necessary and
sufficient condition, one can use those from
\textcite{Giedke01PRL}. In any case, in order to confirm genuine
four-party or even $N$-party entanglement, one has to rule out any
possible partially separable form. In principle, this can be done
by considering all possible bipartite splittings (or groupings)
and applying either the npt criterion or the stronger operational
criteria from \textcite{Giedke01PRL}. Although a full theoretical
characterization including criteria for entanglement
classification has not been considered yet for more than three
parties and modes, the presence of genuine multipartite
entanglement can be confirmed, once the complete $2N\times 2N$
correlation matrix is given.

\subsubsection{Generating entanglement}\label{genentsection}

In this section, we discuss how to {\it generate} genuine multipartite
cv entanglement of arbitrarily many modes
from finitely squeezed sources.
The resulting states are nonmaximally entangled due to the
finite squeezing. How {\it measurements} onto the
maximally entangled basis of arbitrarily many modes can be performed
with linear optics and homodyne detections will be shown in the
next section.

Let us consider a family of genuinely $N$-party entangled cv
states. The members of this family are those states that emerge
from a particular sequence of $N-1$ phase-free beam splitters
(``$N$-splitter'') with $N$ squeezed-state inputs \cite{PvLPRL00}.
The recipe for the generation of these states stems from the
quantum circuit for creating qubit GHZ states.

\begin{figure}[tb]
\begin{center}
%\begin{psfrags}
    % \psfrag{H}{\large $H$}
    % \psfrag{C-NOT}{\large C-NOT}
    % \psfrag{0}{\large $|0\rangle$}
    % \psfrag{GHZ}{
%\large $|{\rm GHZ}\rangle$}
\epsfxsize=3.3in \epsfbox[0 370 550 540]{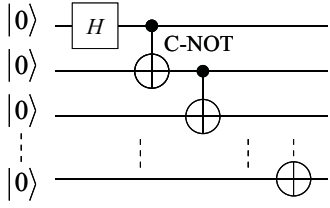}
%\end{psfrags}
\end{center}
\caption{Quantum circuit for generating the $N$-qubit GHZ state.
The gates (unitary transformations) are a
Hadamard gate (``$H$'') and
pairwise acting C-NOT gates.}
\label{PVLcircuit}
\end{figure}

Let us consider the generation of entanglement between arbitrarily
many qubits. The quantum circuit shall turn $N$ independent
qubits into an $N$-partite entangled state. Initially, the $N$
qubits shall be in the eigenstate $|0\rangle$. All we need
is a circuit with the following two elementary gates:
the Hadamard gate, acting on a single
qubit as
\begin{eqnarray}
|0\rangle \longrightarrow \frac{1}{\sqrt{2}}
(|0\rangle + |1\rangle )\,,\quad
|1\rangle \longrightarrow \frac{1}{\sqrt{2}}
(|0\rangle - |1\rangle )\,,
\end{eqnarray}
and the controlled-NOT (C-NOT) gate, a two-qubit operation
acting as
\begin{eqnarray}
|00\rangle \longrightarrow |00\rangle\,,\quad
|01\rangle \longrightarrow |01\rangle\,,\nonumber\\
|10\rangle \longrightarrow |11\rangle\,,\quad
|11\rangle \longrightarrow |10\rangle\,.
\end{eqnarray}
The first qubit (control qubit) remains unchanged under the C-NOT.
The second qubit (target qubit) is flipped if the control qubit is
set to 1, and is left unchanged otherwise.
Equivalently, we can describe the action of the C-NOT gate
by $|y_1,y_2\rangle\rightarrow |y_1,y_1\oplus y_2\rangle$ with
$y_1,y_2 = 0,1$ and the addition modulo two $\oplus$.
The $N$-partite entangled output state of the circuit (see
Fig.~\ref{PVLcircuit}) is the $N$-qubit GHZ state.

Let us translate the qubit quantum circuit to continuous
variables \cite{PvLPRL00}.
For this purpose, it is convenient to consider
position and momentum eigenstates.
We may replace the Hadamard by a Fourier transform,
\begin{eqnarray}\label{fourier}
\hat\mathcal{F}|x\rangle_{\rm position}&=&
\frac{1}{\sqrt{\pi}}\int_{-\infty}^{\infty}\, dy\, e^{2ixy}
|y\rangle_{\rm position}\nonumber\\
&=&|p=x\rangle_{\rm momentum}\;,
\end{eqnarray}
and the C-NOT gates by appropriate beam splitter
operations.\footnote{A possible continuous-variable generalization
of the C-NOT gate is $|x_1,x_2\rangle\rightarrow |x_1,x_1+
x_2\rangle$, where the addition modulo two of the qubit C-NOT,
$|y_1,y_2\rangle\rightarrow |y_1,y_1\oplus y_2\rangle$ with
$y_1,y_2 = 0,1$, has been replaced by the normal addition.
However, for the quantum circuit here, a beam splitter operation
as described by Eq.~(\ref{BSeigenstates}) is a suitable substitute
for the generalized C-NOT gate.} The input states are taken to be
zero-position eigenstates $|x=0\rangle$. The sequence of beam
splitter operations $\hat B_{jk}(\theta)$ is provided by a network
of ideal phase-free beam splitters (with typically asymmetric
transmittance and reflectivity) acting on the position eigenstates
of two modes as in Eq.~(\ref{BSeigenstates}).

Now we apply this sequence of beam splitters (making an ``$N$-splitter''),
\begin{eqnarray}\label{PVLNsplitt}
\hat B_{N-1\,N}(\pi/4)\hat B_{N-2\,N-1}
\left(\sin^{-1}1/\sqrt{3}\right)\nonumber\\
\times\cdots\times\hat B_{12}\left(\sin^{-1}1/\sqrt{N}\right),
\end{eqnarray}
to a zero-momentum eigenstate
$|p=0\rangle\propto\int dx\,|x\rangle$ of mode 1
(the Fourier transformed zero-position eigenstate)
and $N-1$ zero-position eigenstates $|x=0\rangle$ in modes $2$ through $N$.
We obtain the entangled $N$-mode state $\int dx\,|x,x,\ldots ,x\rangle$.
This state is an eigenstate with total momentum zero and all relative
positions $x_i-x_j=0$ $(i,j=1,2,\ldots ,N)$. It is clearly an
analogue to the qubit GHZ state with perfect correlations among the
quadratures. However, it is an unphysical and unnormalizable state.
Rather than sending infinitely squeezed position eigenstates
through the entanglement-generating circuit, we will now use
finitely squeezed states.

In the Heisenberg representation, an ideal phase-free beam splitter
operation acting on two modes is described by
Eq.~(\ref{generalBS1}).
Let us now define a matrix $B_{kl}(\theta)$
which is an $N$-dimensional identity matrix with the entries
$I_{kk}$, $I_{kl}$, $I_{lk}$, and $I_{ll}$ replaced by the corresponding
entries of the beam splitter matrix in Eq.~(\ref{generalBS1}).
Thus, the matrix for the $N$-splitter becomes
\begin{eqnarray}
{\mathcal U}(N)&\equiv&
B_{N-1\,N}\left(\sin^{-1}\frac{1}{\sqrt{2}}\right)B_{N-2\,N-1}
\left(\sin^{-1}\frac{1}{\sqrt{3}}\right) \nonumber\\
\label{PVLnsplit}
&&\times\cdots\times
B_{12}\left(\sin^{-1}\frac{1}{\sqrt{N}}\right) \;.
\end{eqnarray}
The entanglement-generating circuit is now
applied to $N$ position-squeezed vacuum modes.
In other words, one momentum-squeezed and $N-1$ position-squeezed
vacuum modes are coupled by an $N$-splitter,
\begin{equation}\label{PVLNsplcircuit}
\left(\begin{array}{cccc} \hat a'_1 & \hat a'_2
& \cdots & \hat a'_N \end{array}\right)^T =
{\mathcal U}(N)
 \left(\begin{array}{cccc} \hat a_1 & \hat a_2
& \cdots & \hat a_N \end{array}\right)^T ,
\end{equation}
where the input modes are squeezed according to
\begin{eqnarray}
\hat{a}_1&=&\cosh r_1 \hat{a}_1^{(0)} +
\sinh r_1
\hat{a}_1^{(0)\dagger},\nonumber\\
\label{PVLNsplinputs}
\hat{a}_i&=&\cosh r_2 \hat{a}_i^{(0)} -
\sinh r_2
\hat{a}_i^{(0)\dagger}\;,
\end{eqnarray}
with $i=2,3,...,N$.
In terms of the input quadratures, we have
\begin{eqnarray}
\hat{x}_1=e^{+r_1} \hat{x}_1^{(0)},&\quad&
\hat{p}_1=e^{-r_1} \hat{p}_1^{(0)},\nonumber\\
\label{PVLinputsquadr}
\hat{x}_i=e^{-r_2} \hat{x}_i^{(0)},&\quad&
\hat{p}_i=e^{+r_2} \hat{p}_i^{(0)}\;,
\end{eqnarray}
for $\hat{a}_j=\hat{x}_j+i\hat{p}_j$ ($j=1,2,...,N$).
The correlations between the output quadratures
are revealed by the arbitrarily small noise in the
relative positions and the total momentum for sufficiently
large squeezing $r_1$ and $r_2$,
\begin{eqnarray}\label{PVLcorrfamily}
\langle(\hat{x}'_k-\hat{x}'_l)^2\rangle&=&e^{-2r_2}/2\;,
\nonumber\\
\langle(\hat{p}'_1+\hat{p}'_2+\cdots+\hat{p}'_N)^2\rangle&=&
N e^{-2r_1}/4 \;,
\end{eqnarray}
for $k\neq l$ ($k,l=1,2,...,N$) and
$\hat{a}'_k=\hat{x}'_k+i\hat{p}'_k$.
Note that all modes involved have zero mean values, thus the
variances and the second moments are identical.

% FIG circuit

The output states from Eq.~(\ref{PVLNsplcircuit})
are pure $N$-mode states,
totally symmetric under interchange of modes,
and they retain the Gaussian character of the input states.
Hence they are entirely described by their second-moment
correlation matrix,
\begin{eqnarray}\label{GHZcorr}
V^{(N)}=\frac{1}{4}
\left( \begin{array}{ccccccc}
a & 0 & c & 0 & c & 0
& \cdots \\
0 & b & 0 & d & 0 & d
& \cdots \\
c & 0 & a & 0 & c & 0
& \cdots \\
0 & d & 0 & b & 0 & d
& \cdots \\
c & 0 & c & 0 & a & 0
& \cdots \\
0 & d & 0 & d & 0 & b
& \cdots \\
\vdots & \vdots & \vdots & \vdots & \vdots & \vdots & \vdots \\
\end{array} \right)\;,
\end{eqnarray}
where
\begin{eqnarray}\label{GHZcorrelements}
a&=&\frac{1}{N}e^{+2r_1}+\frac{N-1}{N}e^{-2r_2}\;,\nonumber\\
b&=&\frac{1}{N}e^{-2r_1}+\frac{N-1}{N}e^{+2r_2}\;,\nonumber\\
c&=&\frac{1}{N} (e^{+2r_1} - e^{-2r_2})\;,\nonumber\\
d&=&\frac{1}{N} (e^{-2r_1} - e^{+2r_2})\;.
\end{eqnarray}
For squeezed vacuum inputs, the multi-mode output states have zero
mean and their Wigner function is of the form
Eq.~(\ref{Gausswigndef}). The particularly simple form of the
correlation matrix in Eq.~(\ref{GHZcorr}) is, in addition to the
general correlation matrix properties, symmetric with respect to
all modes and contains no intermode or intramode $x$-$p$
correlations (hence only the four parameters $a$, $b$, $c$, and
$d$ appear in the matrix). However, the states of this form are in
general biased with respect to $x$ and $p$ ($a\neq b$). Only for a
particular relation between the squeezing values $(r_1,r_2)$
\cite{vanloockFdP02,vanloockcvbook02},
\begin{eqnarray}\label{Bowenrelation}
e^{\pm 2 r_1}&=&(N-1)\\
&&\times \sinh 2r_2 \,\left[
\sqrt{1+\frac{1}{(N-1)^2\sinh^2 2r_2}} \pm 1 \right],
\nonumber
\end{eqnarray}
the states are unbiased (all diagonal entries of the correlation
matrix equal), thus having minimum energy at a given degree of
entanglement or, in other words, maximum entanglement for a given
mean photon number \cite{Bowen01}. The other $N$-mode states of
the family \cite{vanloockFdP02,vanloockcvbook02} can be converted
into the minimum-energy state via local squeezing operations
\cite{Bowen01}. Only for $N=2$, we obtain $r=r_1=r_2$. In this
case, the matrix $V^{(N)}$ reduces to that of a two-mode squeezed
state which is the maximally entangled state of two modes at a
given mean energy with the correlation matrix given in
Eq.~(\ref{corrEPR}). For general $N$, the first squeezer with
$r_1$ and the $N-1$ remaining squeezers with $r_2$ have different
squeezing. In the limit of large squeezing ($\sinh 2r_2\approx
e^{+2r_2}/2$), we obtain approximately
\cite{vanloockFdP02,vanloockcvbook02}
\begin{eqnarray}\label{Bowenrelationlargesq}
e^{+ 2 r_1}\approx (N-1) e^{+ 2 r_2}  \;.
\end{eqnarray}
We see that in order to produce the minimum-energy $N$-mode state,
the single $r_1$-squeezer is, in terms of the squeezing factor,
$N-1$ times as much squeezed as each $r_2$-squeezer.
However, also in this general $N$-mode case, the other $N$-mode states
of the family can be converted into the minimum-energy state via
local squeezing operations.

Why is the $N$-mode state described by the correlation matrix
in Eq.~(\ref{GHZcorr}) genuinely $N$-partite entangled?
Simple arguments suffice already to confirm this.
One such argument is that the Wigner function of the
$N$-mode state is not even partially factorizable.
Neither the Wigner function of a single mode nor that
of a group of modes can be factored out from the total Wigner
function. This argument depends on the purity of the
$N$-mode state, since the Wigner function of a mixed and only
classically correlated state is not factorizable either.
The $N$-mode state described by Eq.~(\ref{GHZcorr})
is indeed pure, because it is built from $N$ pure single-mode
states via linear optics.
Apart from its purity, taking into account also its total
symmetry, the presence of any kind of (partial) entanglement proves
the genuine $N$-partite entanglement of the state \cite{vanloockFdP02}.
For example, according to Eq.~(\ref{GHZcorr}),
upon tracing out all modes except one,
the remaining single-mode correlation matrix satisfies
$\det V^{(1)}>1/16$ for any $N$ and any $r_1>0$ or $r_2>0$.
The remaining mode is then in a mixed state, thus
proving its entanglement with at least some of the other modes.
Hence due to the purity and symmetry, the total state is
genuinely $N$-partite entangled.
Note that this holds true even for $r_1>0$ and $r_2=0$.
Thus, only one squeezed state suffices to make
genuine $N$-partite entanglement via linear optics \cite{PvLPRL00}.
For three parties and modes, $N=3$, checking the npt criterion
is simple too. One has to apply transposition only with respect to
each mode 1, 2, and 3, in order to rule out
any partially separable form
$\sum_i \eta_i\, \hat\rho_{i,k}\otimes\hat\rho_{i,mn}$.
Due to the symmetry, npt with respect to mode 1 is sufficient
(and necessary) for the genuine tripartite entanglement.
The resulting three-mode state belongs to the fully inseparable
class 1 in Eq.~(\ref{classification}).

The separability properties
of mixed versions of the three-mode state
in Eq.~(\ref{GHZcorr}) with $N=3$, having the same
correlation matrix contaminated by some noise,
$V_{\rm noisy}^{(3)}=V^{(3)}+\mu\mbox{1$\!\!${\large 1}}/4$
are more subtle.
In that case, for given squeezing $r=r_1=r_2>0$,
the state becomes three-mode biseparable [class 4 in
Eq.~(\ref{classification})] above some threshold value $\mu_0$,
$\mu\geq\mu_0>0$, and fully separable [class 5 in
Eq.~(\ref{classification})] above some greater
value $\mu_1$, $\mu\geq\mu_1>\mu_0$ \cite{Giedke01}.
Note that due to symmetry, the state described by
$V_{\rm noisy}^{(3)}$ can only belong to the classes 1, 4, and 5.
Classes 4 and 5 are not distinguishable via partial transpose,
but the full separability (class 5) is proven
iff one-mode correlation matrices
$V^{(1)}_1$, $V^{(1)}_2$, and $V^{(1)}_3$ exist such that
$V_{\rm noisy}^{(3)}\geq V^{(1)}_1\oplus V^{(1)}_2\oplus V^{(1)}_3$
\cite{Giedke01}.
In particular, for $\mu\geq 1$, we have
$V^{(1)}_1=V^{(1)}_2= V^{(1)}_3=\mbox{1$\!\!${\large 1}}/4$,
thus confirming the full separability of the state in that case.

How do the ``GHZ-like cv states'' described by Eq.~(\ref{GHZcorr})
behave compared to the qubit GHZ states? For finite squeezing,
they actually behave more like the qubit W state in
Eq.~(\ref{PVLWdef}) rather than the maximally entangled qubit GHZ
state in Eq.~(\ref{PVLGHZdef}) \cite{vanloockcvbook02}. For three
parties, for instance, bipartite two-mode mixed-state entanglement
is readily available upon tracing out one mode
\cite{vanloockcvbook02,vanloockFdP02}[see also the very recent
results on Gaussian multipartite entanglement by
\textcite{Adesso1} and \textcite{Adesso2}].

\subsubsection{Measuring entanglement}

Rather than generating entangled states, another important task is
the measurement of multi-party entanglement, i.e., the projection
onto the basis of maximally entangled multi-party states. For
qubits, it is well-known that this can be achieved simply by
inverting the above entanglement-generating circuit
(Fig.~\ref{PVLcircuit}). A similar strategy also works for
$d$-level systems \cite{Dusek}.

For creating continuous-variable entanglement, we replaced the
C-NOT gates in Fig.~\ref{PVLcircuit} by appropriate beam splitter
operations. The same strategy, after inverting the circuit in
Fig.~\ref{PVLcircuit}, also enables one to measure
continuous-variable entanglement. In other words, a projection
onto the ``continuous-variable GHZ basis'' can be performed by
applying an inverse $N$-splitter followed by a Fourier transform
of one mode and by subsequently measuring the positions of all
modes \cite{vanloockFdP02}. The simplest example is the cv Bell
measurement, needed, for instance, in cv quantum teleportation
(see Sec.~\ref{qtelep}). It can be accomplished by using a
symmetric beam splitter and detecting the position of one output
mode and the momentum of the other output mode.

We see that the requirements of a ``Bell state analyzer'' and,
more generally, a ``GHZ state analyzer'' for continuous variables
are easily met by current experimental capabilities. This is in
contrast to the Bell and GHZ state analyzer for photonic qubits
\cite{Luetkenhaus99,VaidmYoram,PvLLutkenhaus04}]. Although
arbitrarily high efficiencies can be approached, in principle,
using linear optics, photon number detectors, and feedforward, one
would need sufficiently many, highly entangled auxiliary photons
and detectors resolving correspondingly large photon numbers
\cite{KLM,Dusek}. Neither of these requirements is met by current
technology. Of course, the C-NOT gates of a qubit Bell and GHZ
state measurement device can, in principle, be implemented via the
cross Kerr effect using nonlinear optics. However, on the
single-photon level, the required optical nonlinearities are hard
to obtain.

The efficient and unconditional generation of (multipartite)
entanglement, though nonmaximum for finite squeezing, and the
simple and feasible linear-optics schemes for measuring ``maximum
(multipartite) cv entanglement'' demonstrate the power of quantum
optical entanglement manipulation based on cv. However, the
capabilities of such purely cv based schemes, which rely
exclusively on Gaussian operations such as beam splitting,
homodyne detection, and squeezing, are not unlimited. This is
revealed, in particular, by the No-Go results for cv entanglement
distillation (see Sec.~\ref{entdistillation}).

\subsection{Bound entanglement}\label{boundent}

There are two big issues related with composite mixed quantum
states: the separability and the distillability issue.
The former was subject of the previous sections.
For cv, the separability problem is, in general, not
completely solved yet. In the special case of Gaussian states,
however, we mentioned that necessary and sufficient
inseparability criteria (different from the npt criterion)
for bipartite states of arbitrarily many modes
exist. The situation is similar with regard to distillability.
The distillability of general cv states is an open question,
whereas that of bipartite Gaussian states with arbitrarily many modes
is completely characterized by the partial transpose criterion:
any $N\times M$ Gaussian state is distillable iff it is npt
\cite{Giedke01PRL,GezaFdP}.
A state is distillable if a sufficiently large number of copies
of the state can be converted into a pure
maximally entangled state (or arbitrarily close to it)
via local operations and classical communication.
Entanglement distillation (or ``purification'' \cite{Benn2}) is essential
for quantum communication when the two halves of a supply
of entangled states are distributed through noisy channels,
distilled, and subsequently used, for instance, for high-fidelity
quantum teleportation.

In general, any inseparable state with positive partial transpose
cannot be distilled to a maximally entangled state, thus representing a
so-called bound entangled state \cite{Horo2}. In other words, npt is
necessary for distillability. Is it in general sufficient too?
There are conjectures that this is not the case and
that undistillable (bound) npt states exist
\cite{diVicenzoGezaFdP11a,diVicenzoGezaFdP11b}.
On the other hand, any state $\hat\rho_{12}$ that satisfies the
so-called reduction criterion, $\hat\rho_1\otimes\mbox{1$\!\!${\large
1}}-\hat\rho_{12}\ngeq 0$ or $\mbox{1$\!\!${\large
1}}\otimes\hat\rho_2-\hat\rho_{12}\ngeq 0$, where
$\hat\rho_1={\rm Tr}_2\hat\rho_{12}$ etc., is both inseparable and
distillable \cite{Horo4}. The reduction criterion is sufficient
for distillability, but it was shown not to be a necessary condition
\cite{GezaFdPRef12}.
The known criteria for distillability are summarized by the
following statements,
\begin{eqnarray}
{\rm general}\,\;{\rm states,}\,\;{\rm distillable}
&\Rightarrow& {\rm npt}\nonumber\\
%?\!\!\!\!&\Leftarrow\!\!\!\!\!\!/&\!\!?\nonumber\\
\quad\quad''\quad\quad''\quad\quad''\quad\quad\quad
&\stackrel{?}{\!\!\!\!\Leftarrow\!\!\!\!\!\!/\!\!}&
\;''\nonumber\\
{\rm general}\,\;{\rm states,}\,\;{\rm distillable}
&\Leftarrow& \hat\rho_1\otimes\mbox{1$\!\!${\large
1}}-\hat\rho_{12}\ngeq 0\nonumber\\
\quad\quad''\quad\quad''\quad\quad''\quad\quad\quad
&\Rightarrow\!\!\!\!\!\!\!/&
\quad\quad''\quad\quad''\nonumber\\
{\rm Gaussian}\,\;{\rm states,}\,\;{\rm distillable}
&\Leftarrow\!\!\Rightarrow& {\rm npt}
\end{eqnarray}
Bound entangled npt Gaussian states do definitely not exist
\cite{Giedke01PRL,GezaFdP}. Hence the set of Gaussian states is
fully explored, consisting only of npt distillable, ppt entangled
(undistillable), and separable states. The simplest bound
entangled Gaussian states are those with two modes at each side,
$N=M=2$. Explicit examples were constructed by \textcite{Wolf}. An
example for tripartite bound entangled states are the Gaussian
three-mode states of class 4 in Eq.~(\ref{classification}). These
states are ppt with respect to any of the three modes, but
nonetheless entangled. Unfortunately, the distillation of npt
Gaussian states to maximally entangled finite-dimensional states,
though possible in principle, is not very feasible with current
technology. It relies upon non-Gaussian operations (see
Sec.~\ref{entdistillation}). As for the existence of generic bound
entanglement of non-Gaussian cv states, examples were discussed by
\textcite{HoroLewenstein} and \textcite{HoroLewensteinCirac}.

\subsection{Nonlocality}\label{nonlocalsec}

We mentioned earlier that the notion of entanglement
was introduced in 1935 by Schr\"{o}dinger in his reply
\cite{Schroedingerreply} to the EPR paper \cite{Einst}.
The EPR argument itself, based on the cv entangled state in
Eq.~(\ref{originalEPR}), already contained
as an essential ingredient the notion of {\it nonlocality}.
More precisely, the reasoning behind the EPR paradox
relies upon two major assumptions: firstly, there is something
like an objective reality, and secondly, there is no action
at a distance. Objective reality becomes manifest
``if without in any way disturbing the system, we can predict
with certainty the value of a physical quantity, then there
exists an element of physical reality corresponding to this
quantity.'' Now two particles sharing the entangled state of
Eq.~(\ref{originalEPR}) are perfectly correlated in
their positions {\it and} momenta.
Measuring say the position of one particle means that
the result obtainable in a subsequent position measurement
of the other particle can be predicted with certainty.
If there is no action at a distance, this prediction is made
without disturbing the second particle. Hence, due to
EPR's realism, there must be a definite predetermined
position of that particle. The same arguments apply to the
momenta, leading also to a definite predetermined momentum
for the second particle.
Since quantum theory does not allow for
such states of definite position {\it and} momentum, EPR conclude
that the quantum mechanical description is incomplete.
Similar to the inseparability criteria for continuous-variable
states, which need to be expressed in terms of position
{\it and} momentum, also EPR's conclusion (local realism implies
incompleteness of quantum theory) crucially
depends on the presence of correlations in {\it both}
conjugate variables. Hence, as discussed earlier, the nature of these
correlations must be quantum rather than classical.

Later, in 1964, by extending the EPR program
John Bell showed that nonlocality can
be revealed via the constraints that local realism imposes on the
statistics of two physically separated
systems \cite{Bell}. These constraints, expressed in
terms of the Bell inequalities, can be violated by quantum mechanics.
There are then three possible conclusions that can be drawn when
inequalities imposed by local realism are violated:
the correlations of the relevant quantum state contradict locality
or realism or both.
What is today loosely called ``nonlocality'' includes these three
alternatives.

As for a demonstration of this nonlocality,
several quantum optical experiments have been performed.
The first detection of violations of Bell-type inequalities
was based on dv two-photon states
\cite{OuMandel,Aspect}.
These states are analogous to the spin-entangled states
used by Bohm in his dv version of the EPR paradox \cite{Bohm}.
Such single-photon based dv experiments rely on photon counting.

\subsubsection{Traditional EPR-type approach}

A quantum optical cv experiment more reminiscent to the original
EPR paradox and distinct from tests of Bell inequalities was that
by Ou {\it et al.} based on the quantum correlations of position
and momentum in a two-mode squeezed state
\cite{Ou92,Ou92applphys}. There, the experimentally determined
quantities were the quadrature variances of one mode conditioned
upon the results obtainable in quadrature measurements of the
other mode (``inferred variances''). This quantum optical
demonstration of the original EPR paradox was based on a proposal
by Reid who extended the EPR scenario to the case of finite
quantum correlations using the {\it inferred} quadrature variances
\cite{Reid89},
%notation: $V$ is corr. matrix -> use Var for variance
\begin{eqnarray}\label{inferred}
{\rm Var}_{\rm inf}^{\hat{x}}&\equiv&
{\rm Var}\left(\hat x_1 - \hat x_1^{\rm est}\right)=
{\rm Var}\left(\hat x_1 - g_x\hat x_2\right)\nonumber\\
&=&\left\langle\left(\hat x_1 - g_x\hat x_2\right)^2\right\rangle -
\left\langle\hat x_1 - g_x\hat x_2\right\rangle^2\;,
\end{eqnarray}
where $\hat x_1^{\rm est}=g_x\hat x_2$ is the inferred estimate of
mode's 1 position $\hat x_1$ based on the scaled readout
$\hat x_1^{\rm est}$ of mode's 2 position $\hat x_2$.
The scaling parameter $g_x$ may then be chosen optimally
in order to ensure the most accurate inference. The smaller
the deviation of $\hat x_1^{\rm est}$ from the true
values $\hat x_1$, the better $\hat x_1$ may be
determined at a distance by detecting $\hat x_2$.
On average, this deviation is quantified by
${\rm Var}_{\rm inf}^{\hat{x}}$.
Similarly, one can define ${\rm Var}_{\rm inf}^{\hat{p}}$,
the inferred variance for the momentum, with
$\hat{x}\rightarrow\hat{p}$ throughout.
By calculating $\partial {\rm Var}_{\rm inf}^{\hat{x}}/
\partial g_x =0$, we obtain the optimal scaling factor
\begin{eqnarray}\label{optimalscaling}
g_x=\frac{\langle\hat x_1\hat x_2 \rangle-
\langle\hat x_1\rangle\langle\hat x_2 \rangle}{{\rm Var}(\hat x_2)}
=\frac{\langle\Delta\hat x_1
\Delta\hat x_2 \rangle}{{\rm Var}(\hat x_2)}\;,
\end{eqnarray}
with $\Delta\hat x_i\equiv\hat x_i-\langle \hat x_i\rangle$.
For the momentum, we have correspondingly
$g_p=\langle\Delta\hat p_1
\Delta\hat p_2 \rangle/{\rm Var}(\hat p_2)$.
With these scaling factors,
the optimal (minimal) value for the inferred variance
becomes
\begin{eqnarray}\label{optimalinfvar}
[{\rm Var}_{\rm inf}^{\hat{x}}]_{\rm min}&=&
{\rm Var}(\hat x_1)\left(1-\frac{\langle\Delta\hat x_1
\Delta\hat x_2 \rangle^2}{{\rm Var}(\hat x_1)
{\rm Var}(\hat x_2)}\right)\nonumber\\
&\equiv&{\rm Var}_{\rm cond}^{\hat{x}}\;,
\end{eqnarray}
and similarly for
$[{\rm Var}_{\rm inf}^{\hat{p}}]_{\rm min}\equiv
{\rm Var}_{\rm cond}^{\hat{p}}$
with
$\hat{x}\rightarrow\hat{p}$ throughout in
Eq.~(\ref{optimalinfvar}).
The minimal inferred variances are also referred to as
conditional variances
${\rm Var}_{\rm cond}^{\hat{x}}$ and
${\rm Var}_{\rm cond}^{\hat{p}}$, a measure
for the noise degrading the otherwise perfect correlations
between the two modes.

Following the EPR program and
assuming now that the two modes 1 and 2 are spatially separated,
but also that there is no action at a distance,
one would have to assign to mode 1 predetermined values
for $\hat x_1$ and $\hat p_1$ up to, on average, some noise
${\rm Var}_{\rm inf}^{\hat{x}}$ and
${\rm Var}_{\rm inf}^{\hat{p}}$ respectively.
Thus, if this leads to a state where $\hat x_1$ and $\hat p_1$
are defined to accuracy of
\begin{eqnarray}\label{EPRcondition}
{\rm Var}_{\rm inf}^{\hat{x}}{\rm Var}_{\rm inf}^{\hat{p}}
<\frac{1}{16}\;,
\end{eqnarray}
a contradiction to the Heisenberg uncertainty relation
Eq.~(\ref{quadratuncert}) would occur.
In fact, the ``EPR condition'' in Eq.~(\ref{EPRcondition})
is satisfied with the two-mode squeezed (vacuum) state
for any nonzero squeezing $r>0$, since
its correlation matrix in Eq.~(\ref{corrEPR})
using Eq.~(\ref{optimalinfvar})
yields the conditional variances
\begin{eqnarray}\label{2msqstcondvar}
{\rm Var}_{\rm cond}^{\hat{x}}={\rm Var}_{\rm cond}^{\hat{p}}=
\frac{1}{4\cosh 2r}\;.
\end{eqnarray}
The optimal scaling factors to ensure the best inference
and hence to fulfill Eq.~(\ref{EPRcondition})
for any nonzero squeezing are
$g_x=\tanh 2r$ and $g_p=-\tanh 2r$.

Apparently, ``EPR nonlocality'' as given by
Eq.~(\ref{EPRcondition}) and inseparability as indicated by a
violation of Eq.~(\ref{Tan}) are equivalent for pure Gaussian
states such as two-mode squeezed states. In general, however, EPR
nonlocality is only a sufficient \cite{Reid2,Kim02} and not a
necessary condition for inseparability. This is not different from
the relation between inseparability and the nonlocality as
expressed by violations of Bell inequalities. As mentioned
earlier, except for pure states, entanglement does not
automatically imply ``Bell nonlocality'', but the converse holds
true in general. For instance, the qubit Werner states are mixed
states which can be inseparable without violating any
(non-collective) Bell inequality \cite{Werner}. Hence the two
formally distinct approaches of EPR and Bell nonlocality lead both
to criteria generally stricter than those for inseparability. Due
to this similarity and the fact that the EPR concept, initially
designed for cv \cite{Einst}, was later translated into the dv
domain \cite{Bohm}, one may ask whether the concept of Bell
nonlocality, originally derived in terms of dv \cite{Bell}, is
completely describable in terms of cv too. Bell argued that the
original EPR state directly reveals a local hidden-variable
description in terms of position and momentum, since its Wigner
function is positive everywhere and hence serves as a classical
probability distribution for the hidden variables \cite{Bell2}.
Thus, attempts to derive homodyne-based cv violations of Bell
inequalities for the two-mode squeezed state with its positive
Gaussian Wigner function must fail. However, whether the (for
pure-state entanglement always present) nonlocality is uncovered
depends on the observables and the measurements considered in a
specific Bell inequality and not only on the quantum state itself.
In fact, it was shown by \textcite{Bana} how to demonstrate the
nonlocality of the two-mode squeezed vacuum state: it violates a
Clauser-Horne-Shimony-Holt (CHSH) inequality \cite{CHSH} when
measurements of photon number parity are considered.

\subsubsection{Phase-space approach}

Following Bell \cite{Bell2}, an always positive Wigner function
can serve as the hidden-variable probability distribution with
respect to measurements corresponding to any linear combination of
$\hat x$ and $\hat p$. In this sense, one will not obtain a
violation of the CHSH inequality for the two-mode squeezed state
Wigner function of Eq.~(\ref{WEPR}) when restricted to such
measurements \cite{Bana}. The same applies to the always positive
Wigner function of the Gaussian $N$-party entangled $N$-mode state
with correlation matrix given by Eq.~(\ref{GHZcorr}). Thus, in
order to reveal the nonlocality of these Gaussian states,
non-Gaussian measurements, which are not only based upon homodyne
detection, must be considered.

For their analysis using photon number parity measurements,
Banaszek and Wodkiewicz exploited the fact that the Wigner
function is proportional to the quantum expectation value of a
displaced parity operator \cite{Royer,Bana}. Extending this
observation from two to an arbitrary number of $N$ modes, one
obtains
\begin{eqnarray}\label{PVLparity1}
W({\boldsymbol{\alpha}})=\left(\frac{2}{\pi}\right)^N\left\langle
\hat{\Pi}({\boldsymbol{\alpha}})\right\rangle=
\left(\frac{2}{\pi}\right)^N\Pi({\boldsymbol{\alpha}}) \;,
\end{eqnarray}
where ${\boldsymbol{\alpha}}={\bf x}+i{\bf p}=
(\alpha_1,\alpha_2,...,\alpha_N)$
and $\Pi({\boldsymbol{\alpha}})$ is the quantum expectation value of the
operator
\begin{eqnarray}\label{PVLparity2}
\hat{\Pi}({\boldsymbol{\alpha}})=\bigotimes_{i=1}^N\hat{\Pi}_i(\alpha_i)=
\bigotimes_{i=1}^N\hat{D}_i(\alpha_i)
(-1)^{\hat{n}_i}\hat{D}_i^{\dagger}(\alpha_i) \;.
\end{eqnarray}
The operator $\hat{D}_i(\alpha_i)$ is the displacement operator of
Eq.~(\ref{displacementopdefinition}) acting on mode $i$. Thus,
$\hat{\Pi}({\boldsymbol{\alpha}})$ is a product of displaced
parity operators corresponding to the measurement of an even
(parity $+1$) or an odd (parity $-1$) number of photons in mode
$i$. Each mode is then characterized by a {\it dichotomic}
variable similar to the spin of a spin-1/2 particle or the
single-photon polarization. Different spin or polarizer
orientations from the original qubit-based Bell inequality are
replaced by different displacements in phase space. The
nonlocality test then simply relies on this set of two-valued
measurements for each different setting.

In order to expose the nonlocal two-party correlations of the
two-mode squeezed state, one may then consider the combination
\cite{Bana}
\begin{eqnarray}\label{PVLB2}
{\mathcal{B}}_2=\Pi(0,0)+\Pi(0,\beta)+\Pi(\alpha,0)-\Pi(\alpha,\beta) \;,
\end{eqnarray}
which satisfies $|{\mathcal{B}}_2|\leq 2$ for local realistic
theories according to the CHSH inequality \cite{CHSH}
\begin{eqnarray}\label{PVLineq2}
|C(a_1,a_2)+C(a_1,a_2')+C(a_1',a_2)-C(a_1',a_2')| \leq 2.
\nonumber\\
\end{eqnarray}
Here, $C(a_1,a_2)$ are the correlation functions of measurements
on particle 1 and 2 for two possible measurement settings (denoted
by $a_i$ and $a_i'$ for each particle $i$) .

By writing the two-mode squeezed state according to
Eq.~(\ref{PVLparity1}) for $N=2$ as $\Pi(\alpha_1,\alpha_2)$ and
inserting it into Eq.~(\ref{PVLB2}) with, for example,
$\alpha=\beta=i\sqrt{\mathcal{J}}$ (where ${\mathcal{J}}\geq 0$ is
a real displacement parameter), one obtains
${\mathcal{B}}_2=1+2\exp(-2{\mathcal{J}}\cosh 2r)
-\exp(-4{\mathcal{J}}e^{+2r})$. In the limit of large $r$ (so
$\cosh 2r\approx e^{+2r}/2$) and small ${\mathcal{J}}$,
${\mathcal{B}}_2$ is maximized for ${\mathcal{J}}e^{+2r}=(\ln
2)/3$, yielding ${\mathcal{B}}_2^{\rm max}\approx 2.19$
\cite{Bana}, which is a clear violation of the inequality
$|{\mathcal{B}}_2|\leq 2$. A similar analysis, using
Eq.~(\ref{PVLparity1}), reveals the nonlocal $N$-party
correlations of the multipartite entangled Gaussian $N$-mode state
with correlation matrix in Eq.~(\ref{GHZcorr}) \cite{PvLnonlocal}.
In this case, the nonlocality test is possible using $N$-particle
generalizations of the two-particle Bell-CHSH inequality
\cite{Klyshko,GisinBech}. For growing number of parties and modes
of the $N$-mode state in Eq.~(\ref{GHZcorr}), its nonlocality
represented by the maximum violation of the corresponding
Mermin-Klyshko-type inequality seems to increase nonexponentially
\cite{PvLnonlocal}. This is different from the qubit GHZ states
which show an exponential increase of the violations as the number
of parties grows \cite{Mermin,Klyshko,GisinBech}. Correspondingly,
an exponential increase can be also obtained by considering the
maximally entangled $N$-party states defined via the
two-dimensional parity-spin (``pseudospin'') subspace of the
infinite-dimensional Hilbert space for each electromagnetic mode
\cite{Chen2}.

\subsubsection{Pseudospin approach}

Alternatively, different from the phase-space approach of
\textcite{Bana}, one can also reveal the nonlocality of the cv
states by introducing a ``pseudospin operator'' \cite{Chen1},
$\vec s=(\hat s_x,\hat s_y,\hat s_z)^T$, in analogy to the spin
operator of spin-$\frac{1}{2}$ systems,
$\vec\sigma=(\sigma_1,\sigma_2,\sigma_3)^T$ with the Pauli
matrices $\sigma_i$ \cite{Preskillectures}. The components of the
pseudospin operator are then defined in the photon number basis as
\cite{Chen1,Halvorson}
\begin{eqnarray}
\hat s_z = (-1)^{\hat n}\;,\quad
\hat s_+=\hat s_-^\dagger = \sum_{n=0}^{\infty}
|2n\rangle\langle 2n+1|\;,
\end{eqnarray}
where $\hat s_x \pm i\hat s_y = 2\hat s_\pm$.
Hence $\hat s_z$ and $\hat s_\pm$ are the photon number
parity operator and the ``parity flip'' operators, respectively.
They obey the commutation relations
\begin{eqnarray}
[\hat s_z,\hat s_\pm]=\pm 2\hat s_\pm\;,\quad
[\hat s_+,\hat s_-]=\hat s_z\;,
\end{eqnarray}
equivalent to those of spin-$\frac{1}{2}$ systems which satisfy
the Pauli matrix algebra, $[\hat s_i,\hat
s_j]=2i\epsilon_{ijk}\hat s_k$ and $\hat s_i^2=1$, with
$(i,j,k)\leftarrow\!\!\!\rightarrow (x,y,z)$. The two-valued
measurements for the nonlocality test are now represented by the
Hermitian operator $\vec a\cdot\vec s$ with eigenvalues $\pm 1$.
The unit vector $\vec a$ describes the direction along which the
parity spin $\vec s$ is measured. Like in the well-known qubit
context, different measurement settings correspond to different
(parity) spin orientations. In the CHSH inequality
Eq.~(\ref{PVLineq2}), for instance, substituting the setting
parameters $a_1$, $a_2$, $a_1'$, and $a_2'$ by the vectors $\vec
a_1$, $\vec a_2$, $\vec a_1'$, and $\vec a_2'$, respectively, the
correlation functions now become $\langle(\vec a_1\cdot\vec
s_1)\otimes (\vec a_2\cdot\vec s_2)\rangle\equiv C(\vec a_1,\vec
a_2)$ \cite{Chen1}. By parametrizing the unit vectors in terms of
spherical coordinates and choosing the right angles, it can be
shown \cite{Chen1} that the two-mode squeezed vacuum state of
Eq.~(\ref{twin}) maximally violates the CHSH inequality in the
limit of infinite squeezing. For this maximum violation, the
l.h.s. of Eq.~(\ref{PVLineq2}) with correlation functions $C(\vec
a_1,\vec a_2)$ takes on a value of $2\sqrt{2}$, the upper
(Cirel'son) bound for any cv quantum state \cite{Chen1,Cirelson}.
Recently, a comparison was made between the different formalisms
for revealing the nonlocality of the cv states \cite{Jeong03}, in
particular, using a generalized version \cite{Wilson} of the
phase-space formalism of \textcite{Bana}. It turns out that after
generalizing the formalism of Banaszek and Wodkiewicz
\cite{Wilson,Bana}, the two-mode squeezed vacuum state even in the
limit of infinite squeezing, though yielding larger violations
than before the generalization, cannot maximally violate the
Bell-CHSH inequality. Thus, in order to reveal the maximum
violation of the original EPR state as the limiting case of the
two-mode squeezed vacuum state, the parity spin formalism of
\textcite{Chen1} must be employed. Similarly, the nonlocality of a
two-mode squeezed state is more robust against a dissipative
environment such as an absorbing optical fiber when it is based on
the parity spin formalism \cite{Filip} rather than the phase-space
formalism \cite{Jeong00}.

As for an experimental implementation of these nonlocality tests,
both the measurement of the photon number parity alone and that of
the entire parity spin operator are difficult. Already the former
requires detectors capable of resolving large photon numbers.
However, there are reports on an experimental nonlocality test of
the optical EPR state (made via a nonlinear $\chi^{(2)}$
interaction with ultrashort pump pulses) utilizing homodyne-type
measurements with weak local oscillators
\cite{Kuzmich00PRA,Kuzmich00PRL}. Since the measured observables
in this experiment are not reducible to the field quadratures
[like in the experiment of \textcite{Ou92,Ou92applphys} where a
strong local oscillator was used], a local hidden-variable model
for the measurements cannot be simply constructed from the Wigner
function of the state. Hence Bell-type violations of local realism
are detected, similar to those proposed by \textcite{Grangier88}.
The nonlocality of the kind of \textcite{Bana} becomes manifest in
this experiment too.

The advantage of the nonlocality tests in the ``cv domain'' is
that the entangled Gaussian states are easy to build from squeezed
light, though the required measurements are difficult to perform
(as they are not truly cv, but rather of discretized dichotomic
form). Conversely, other ``cv approaches'' to quantum nonlocality
are based on feasible measurements of states for which no
generation scheme is yet known. An example for this is the
Hardy-type \cite{Hardy92} nonlocality proof in the cv domain
involving simple position and momentum, i.e., quadrature
measurements \cite{Yurke99}. Yet another approach is the proposal
of \textcite{RalphMunro00} which relies on states built from
optical parametric amplification in the low-squeezing
(polarization-based dv) limit and utilizes efficient homodyne
detections. Further theoretical work on quantum nonlocality tests
using homodyne-type cv measurements were published recently by
\textcite{BanaszekDragan02} and \textcite{Wenger03}.

As for proposals for revealing the nonlocality of multi-party
entangled cv states, there is a similar trade-off between the
feasibility of the state generation and that of the measurements.
For instance, as discussed previously, applying the phase-space
approach of \textcite{Bana} to the $N$-party Mermin-Klyshko
inequalities using the Gaussian $N$-mode state with correlation
matrix in Eq.~(\ref{GHZcorr}) \cite{PvLnonlocal} requires
measurements of the photon number parity. In contrast, more
feasible position and momentum measurements can be used to
construct a GHZ paradox \cite{GHZ} for cv \cite{MassarPironio},
but the states involved in this scheme are not simply producible
from squeezers and beam splitters. Similarly, one may consider the
$N$-party $N$-mode eigenstates of the parity spin operator for
different orientations $\vec a\cdot\vec s$. These states take on a
form identical to that in Eq.~(\ref{PVLGHZdef}), now for each mode
defined in the two-dimensional parity-spin subspace of the
infinite-dimensional Hilbert space of the electromagnetic mode
\cite{Chen2}. They are, being completely analogous to the qubit
GHZ states, maximally entangled $N$-party states, thus leading to
an exponential increase of violations of the $N$-party
Mermin-Klyshko inequalities as the number of parties grows
\cite{Chen2}. Hence, the nonexponential increase of violations for
the Gaussian $N$-mode states with correlation matrix in
Eq.~(\ref{GHZcorr}) may be due to the ``limited'' phase-space
formalism of \textcite{Bana} rather than the nonmaximum W-type
entanglement of these states for finite squeezing. However, it has
not been shown yet whether the maximally entangled parity-spin GHZ
states of \textcite{Chen2} are obtainable as the
infinite-squeezing limit of the Gaussian $N$-mode states with
correlation matrix in Eq.~(\ref{GHZcorr}). Only in the two-party
case, it is known that the state which maximally violates the
parity-spin CHSH inequality \cite{Chen1} is the infinite-squeezing
limit of the two-mode squeezed vacuum state.

\subsection{Verifying entanglement experimentally}\label{verentanglement}

So far, we have discussed the notion of entanglement
primarily from a theoretical point of view.
How may one verify the presence of entanglement experimentally?
In general, theoretical tests might be as well applicable to the experimental
verification. For instance, measuring
a violation of inequalities imposed by local realism
confirms the presence of entanglement.
In general, any theoretical test is applicable when the experimentalist has
full information about the quantum state after measurements on an ensemble
of identically prepared states [e.g. by quantum tomography
\cite{Leon}]. In the cv setting,
for the special case of Gaussian states,
the presence of (even genuine multipartite) entanglement
can be confirmed, once the complete correlation matrix is given.
In this case one can apply, for instance,
the npt criterion, as discussed in
Sec.~\ref{mixedentanglement} and Sec.~\ref{seppropgaussian}.

The complete measurement of an $N$-mode Gaussian state is
accomplished by determining the $2N\times 2N$ second-moment
correlation matrix. This corresponds to $N(1+2N)$ independent
entries taking into account the symmetry of the correlation
matrix. \textcite{Kim02} recently demonstrated how to determine
all these entries in the two-party two-mode case using beam
splitters and homodyne detectors. Joint homodyne detections of the
two modes yield the intermode correlations such as $\langle\hat
x_1\hat x_2\rangle - \langle\hat x_1\rangle\langle\hat
x_2\rangle$, $\langle\hat x_1\hat p_2\rangle - \langle\hat
x_1\rangle\langle\hat p_2\rangle$, etc. Determining the local
intramode correlations such as $\langle\hat x_1\hat p_1+\hat
p_1\hat x_1\rangle/2 - \langle\hat x_1\rangle\langle\hat
p_1\rangle$ is more subtle and requires additional beam splitters
and homodyne detections (or, alternatively, heterodyne
detections). Once the $4\times 4$ two-mode correlation matrix is
known, the npt criterion can be applied as a necessary and
sufficient condition for bipartite Gaussian two-mode
inseparability (see Sec.~\ref{mixedentanglement}). In fact, the
entanglement can also be quantified then for a given correlation
matrix \cite{Kim02,VidalWerner02}. For three-party three-mode
Gaussian states, one may pursue a similar strategy. After
measuring the 21 independent entries of the correlation matrix
[for example, by extending Kim {\it et al.'s} scheme \cite{Kim02}
to the three-mode case], the necessary and sufficient criteria by
\textcite{Giedke01} can be applied (see
Sec.~\ref{seppropgaussian}).

However, even for Gaussian states,
such a verification of entanglement via a complete state
determination is very demanding to the experimentalist,
in particular, when the state to be determined
is a potentially multi-party entangled multi-mode state.
Alternatively,
rather than detecting all the entries of the correlation matrix,
one may measure only the variances of appropriate
linear combinations of the quadratures of all modes involved.
This may still be sufficient to unambiguously
verify the presence of (genuine multipartite) entanglement.
For example, the sufficient inseparability criteria from
Sec.~\ref{mixedentanglement}, expressed by violations of
Eq.~(\ref{Tan}) or Eq.~(\ref{generalDuan}),
can be used for witnessing entanglement experimentally.
The indirect experimental confirmation of the presence of entanglement
then relies, for example, on the detection of the quadrature variances
$\langle[\Delta(\hat{x}_1-\hat{x}_2)]^2\rangle$ and
$\langle[\Delta(\hat{p}_1+\hat{p}_2)]^2\rangle$
after combining the two relevant modes at a beam splitter
\cite{Tan}. As for a more direct verification, the measured quadratures
of the relevant state can also be combined electronically
\cite{Furucvbook}.

Similarly, for three parties and modes,
one may attempt to detect violations of inequalities of the form
\cite{PvLAkira03}
\begin{eqnarray}\label{3partycritgenansatz}
\langle(\Delta\hat{u})^2\rangle_{\rho}+
\langle(\Delta\hat{v})^2\rangle_{\rho}\geq
f(h_1,h_2,h_3,g_1,g_2,g_3)\;,
\end{eqnarray}
where
\begin{eqnarray}\label{threemodecombin}
\hat u\equiv h_1\hat x_1 + h_2\hat x_2 + h_3\hat x_3 \;,
\hat v\equiv g_1\hat p_1 + g_2\hat p_2 + g_3\hat p_3 \;.
\nonumber\\
\end{eqnarray}
The $h_l$ and $g_l$ are again arbitrary real parameters.
For (at least partially) separable states,
the following statements hold \cite{PvLAkira03},
\begin{eqnarray}\label{3partyassumption}
\hat\rho&=&\sum_i \eta_i\, \hat\rho_{i,km}\otimes\hat\rho_{i,n}
\nonumber\\
&& \Rightarrow \;
f(h_1,h_2,h_3,g_1,g_2,g_3)=\nonumber\\
\label{3partystatements}
&&\quad\quad(|h_n g_n|+|h_k g_k + h_m g_m|)/2 \,.
\end{eqnarray}
Here, $\hat\rho_{i,km}\otimes\hat\rho_{i,n}$ indicates
that the three-party density operator is a mixture of states $i$
where parties (modes) $k$ and $m$ may be entangled or not, but party $n$
is not entangled with the rest, and where $(k,m,n)$ is any triple
of $(1,2,3)$.
Hence also the fully separable state
is included in the above statements.
In fact, for the fully separable state, we have \cite{PvLAkira03}
\begin{eqnarray}\label{3partyfullysep}
\hat\rho&=&\sum_i \eta_i\, \hat\rho_{i,1}\otimes\hat\rho_{i,2}
\otimes\hat\rho_{i,3}
\nonumber\\
&& \Rightarrow \;
f(h_1,h_2,h_3,g_1,g_2,g_3)=\nonumber\\
\label{3partyfullysep2}
&&\quad\quad(|h_1 g_1| + |h_2 g_2| + |h_3 g_3|)/2\,,
\end{eqnarray}
which is always greater or equal than any of the
boundaries in Eq.~(\ref{3partystatements}).
By detecting violations of the conditions in
Eq.~(\ref{3partycritgenansatz}) with Eq.~(\ref{3partyassumption}),
one can rule out any partially separable (biseparable) form
of the state in question and hence verify genuine
tripartite entanglement.
An experiment in which this verification was achieved
will be briefly described in Sec.~\ref{expwcv}.

The advantage of all these cv inseparability criteria is that,
though still relying upon the rigorous definition of entanglement
in terms of states as given in Sec.~\ref{bipentanglement}
and Sec.~\ref{multipentanglement},
they can be easily checked via efficient homodyne detections
of the quadrature operator statistics.

\section{Quantum Communication with Continuous Variables}
\label{qcommwcv}

When using the term quantum communication, we refer to any
protocol in which the participants' capabilities to communicate
are enhanced due to the exploitation of quantum features such as
nonorthogonality or entanglement. Such an enhancement includes,
for instance, the secure transmission of classical information
from a sender (``Alice'') to a receiver (``Bob'') based on the
transfer of nonorthogonal quantum states. This quantum key
distribution \cite{BB84,Ekert,B92} is the prime example of a full
quantum solution to an otherwise unsolvable classical problem,
namely that of unconditionally secure communication. The, in
principle, unconditional security of quantum cryptography is
provided by the fact that an eavesdropper is revealed when she
(``Eve'') is trying to extract the classical information from the
quantum system being transmitted. She would have to perform
measurements, and thereby (at least for some measurements)
inevitably disturb and alter the quantum system. In general, a
solution to a classical communication problem provided by a
quantum-based protocol should be ``complete'' \cite{Norbert},
connecting the classical world of Alice with that of Bob. For
example, a secure quantum cryptographic communication scheme also
relies on the classical one-time pad. By contrast, there are
``quantum subroutines'' \cite{Norbert} that run entirely on the
quantum level. A prime example for this is quantum teleportation,
whereby the quantum information encoded in nonorthogonal
(``arbitrary'' or ``unknown'') quantum states is reliably
transferred from Alice to Bob using shared entanglement and
classical communication. In order to teleport a qubit, two bits of
classical information must be sent. As an application of quantum
teleportation, one may imagine the reliable connection of the
nodes in a network of quantum computers.

Nonorthogonal quantum states when sent directly through a quantum
communication channel are in any realistic situation subject to
environmental-induced noise, i.e., the quantum channel is noisy.
The coherent superposition of the signal then turns into an
incoherent mixture, a process called decoherence. There are
various methods to circumvent the effect of decoherence, all of
which were originally proposed for dv systems. These methods are
quantum teleportation, combined with a purification of the
distributed entanglement \cite{Benn2}, or quantum error correction
[originally proposed for reducing decoherence in a quantum
computer \cite{Shor2} rather than in a quantum communication
channel]. They enable, in principle, completely reliable
transmission of quantum information.

Apart from the above-mentioned quantum communication scenarios in
which Alice and Bob {\it benefit} from using quantum resources,
there are also fundamental results of ``quantum communication'' on
the {\it restrictions} imposed by quantum theory on classical
communication via quantum states. A very famous result in this
context is that from \textcite{ASH98}, sometimes referred to as
the ``fundamental law of quantum communication''
\cite{CavesDrummondRMP}. It places an upper bound (``Holevo
bound'') on the mutual information of Alice and Bob,
\begin{equation}\label{Holevobound}
I(A:B)\le S(\hat\rho) - \sum_ap_a S(\hat \rho_a)\le S(\hat\rho) \;,
\end{equation}
where $S(\hat \rho)$ is the von Neumann entropy,
$\hat\rho$ is the mean channel state, and $\hat\rho_a$ are the signal
states with a priori probabilities $p_a$. In this relation, equality is
achievable if Alice sends pure orthogonal signal states.

Even assuming an ideal (noiseless) channel, any attempt by Bob to retrieve
the classical information sent from Alice introduces ``noise'' when the
signal states are nonorthogonal. In fact, there is an optimum of
``accessible information'', depending on the measurement strategy
that Bob employs. The most general measurement strategy is described
by so-called generalized measurements or positive operator-valued
measures (POVM's).
Such POVM's $\hat E_b$ are generalizations of projection operators
and satisfy
\begin{equation}\label{POVMdef}
\hat E_b = \hat E_b^\dagger \ge 0\;, \qquad
\sum_b \hat E_b = \hat {\openone} \;.
\end{equation}
When Bob is presented with a state $\hat\rho_a$ representing letter $a$
from Alice's alphabet, he will find instead letter $b$ from his own alphabet
with a conditional probability given by
\begin{equation}\label{POVMcondprobdef}
p_{b|a} = {\rm Tr}\, \hat E_b \hat\rho_a \;.
\end{equation}
 From this, one may compute the mutual information $I(A:B)$,
as we will show in Sec.~\ref{qdensecod} in the context of dense
coding. In this protocol, the roles of the classical and quantum
channels are interchanged relative to those in quantum
teleportation. Instead of reliably transferring quantum
information through a classical channel using entanglement as in
teleportation, in a dense coding scheme, the amount of classical
information transmitted from Alice to Bob is increased when Alice
sends quantum information (her half of an entangled state shared
with Bob) through a quantum channel to Bob. For instance, two bits
of classical information can be conveyed by sending just one
qubit. Like quantum teleportation, dense coding also relies on
preshared entanglement. Thus, dense coding is still in agreement
with Holevo's rule that at most one classical bit can be
transmitted by sending one qubit, because, taking into account
Bob's half of the entangled state transmitted to him prior to the
actual communication (``off-peak''), in total two qubits must be
sent to Bob. This entanglement-based dense coding is sometimes
referred to as superdense coding, as opposed to the dense coding
or ``quantum coding'' schemes introduced by
\textcite{Schumacher95}. The latter enable Alice and Bob to
approach the Holevo bound even for nonorthogonal or mixed signal
states via appropriate encoding of the classical information into
these states. These issues of quantum coding, including the
results of Holevo, may be considered as an extension of Shannon's
classical information theory (applied to communication) to the
quantum realm \cite{Preskillectures}. A brief introduction to the
mathematical description of classical information \'{a} la Shannon
\cite{Shannon} will be given in Sec.~\ref{qdensecod}.

The following review of quantum communication protocols with cv
mainly contains entanglement-based schemes.
 From the preceding sections, we know
that in the cv domain, not only the generation of cv entanglement,
but also its {\it manipulation} via (local) measurements and
unitary operations turns out to be very easy. For instance,
distinguishing exactly between maximally entangled states through
a suitable measurement, as needed for quantum teleportation, is
not possible with photonic qubit Bell states using only linear
optics \cite{Luetkenhaus99,VaidmYoram}. In contrast, such a
``complete Bell detection'' for cv only requires a beam splitter
and homodyne detections. Similarly, unitary transformations such
as phase-space displacements can be easily performed for the
continuous quadrature amplitudes using feed-forward techniques.
Homodyne-based Bell detection and feed-forward are the main tools
both in cv quantum teleportation and cv dense coding.

The cv quantum communication schemes presented below
include the entanglement-based protocols for quantum teleportation
(Sec.~\ref{qtelep}) and (super)dense coding (Sec.~\ref{qdensecod}).
A potentially important application of
quantum teleportation, the teleportation of one half
of an entangled state (entanglement swapping), will be also discussed
in the cv setting. Entanglement swapping is an essential ingredient
of potential long-distance implementations of quantum communication:
the two remote ends of a noisy quantum channel are then provided
with good entanglement after purifying the noisy entanglement
in different segments of the channel and combining the segments
via entanglement swapping. Several ``cv approaches''
to entanglement distillation (including the purification of
mixed entangled states and the concentration of pure nonmaximally
entangled states) will be discussed in Sec.~\ref{entdistillation}.
Apart from quantum teleportation combined with entanglement
purification, alternatively, quantum error correction codes
can be used to protect quantum information from decoherence
when being sent through a noisy quantum communication channel.
A possible cv implementation of quantum error correction
for communication based on linear optics and squeezed light
will be presented in Sec.~\ref{qerrorcorr}.
As for quantum communication schemes not necessarily
based on entanglement, we will discuss some cv approaches to secure
communication (cv quantum key distribution)
in Sec.~\ref{qcryptsec}. In Sec.~\ref{qmemory},
we will also briefly mention the proposals for
potential atom-light interfaces for the storage of cv quantum
information (quantum memory) in a quantum repeater.

\subsection{Quantum teleportation}\label{qtelep}

Quantum teleportation, in general, is the reliable transfer of
quantum information through a classical communication channel
using shared entanglement. The teleportation of continuous quantum
variables such as position and momentum of a particle, as first
proposed by \textcite{Vaid}, relies on the entanglement of the
states in the original Einstein, Podolsky, and Rosen (EPR) paradox
\cite{Einst}. In quantum optical terms, the observables analogous
to the two conjugate variables position and momentum of a particle
are the quadratures of a single mode of the electromagnetic field,
as we have discussed in Sec.~\ref{cvinqopt}. By considering the
finite quantum correlations between these quadratures in a
two-mode squeezed state, a realistic implementation for the
teleportation of continuous quantum variables was proposed by
\textcite{SamKimble}. Based on this proposal, in fact, quantum
teleportation of arbitrary coherent states has been achieved with
a fidelity $F=0.58\pm 0.02$ \cite{Fur98}. Without using
entanglement, by purely classical communication, an average
fidelity of 0.5 is the best that can be achieved if the alphabet
of input states includes potentially all coherent states with even
weight \cite{Fuchs}. We will discuss the issue of delineating a
boundary between classical and quantum domains for teleportation
in more detail later. The scheme based on the continuous
quadrature amplitudes enables the realization of quantum
teleportation in an ``unconditional'' fashion with high efficiency
\cite{SamKimble}, as reported in Refs.~\cite{Sam4,Fur98}. In this
experiment, the following three criteria necessary for quantum
teleportation were achieved:

\noindent
1.) An ``unknown'' quantum state enters the sending station
for teleportation.

\noindent
2.) A teleported state emerges from the receiving station for subsequent
evaluation or exploitation.

\noindent
3.) The degree of overlap between the input and the teleported states is
higher than that which could be achieved if the sending and the receiving
stations were linked only by a classical channel.

There are several aspects of quantum teleportation that are worth
pointing out:

\noindent
1.) The arbitrary input state can even be unknown to
both Alice and Bob. If Alice knew the state
she could send her knowledge classically to Bob
and Bob could prepare the state.
Hence, in quantum teleportation,
the state remains completely unknown
to both Alice and Bob throughout the entire teleportation
process.

\noindent
2.) The input system does not remain in its
initial state because of the Bell measurement.
This fact ensures that no-cloning is not violated.

\noindent
3.) A contradiction to special relativity is avoided, because the
classical communication required between Alice and Bob is restricted by
the speed of light.

Let us now describe the teleportation of continuous quantum
variables in the simplest way, considering just (discrete) single
modes of the electromagnetic field. The generalization to a
realistic broadband description, in particular, with respect to
the mentioned teleportation experiment, is straightforward and
will be briefly discussed in Sec.~\ref{expwcv}. In the
teleportation scheme of a single mode of the electromagnetic
field, the shared entanglement resource is a two-mode squeezed
state [e.g. as in Eq.~(\ref{2modeHeis}); in the original proposal
\cite{SamKimble}, equivalently, the Wigner function of the
finitely squeezed EPR state of Eq.~(\ref{WEPR}) was used]. The
entangled state is sent in two halves: one to ``Alice'' (the
teleporter or sender) and the other one to ``Bob'' (the receiver),
as illustrated in Fig.~\ref{telep}. In order to perform the
teleportation, Alice has to couple the input mode she wants to
teleport with her ``EPR mode'' at a beam splitter. The ``Bell
detection'' of the $x$ quadrature at one beam splitter output, and
of the $p$ quadrature at the other output, yields the classical
results to be sent to Bob via a classical communication channel.
In the limit of an infinitely squeezed EPR source, these classical
results contain no information about the mode to be teleported.
This is analogous to the Bell measurement of the
spin-$\frac{1}{2}$-particle pair by Alice for the teleportation of
a spin-$\frac{1}{2}$-particle state. The measured Bell state of
the spin-$\frac{1}{2}$-particle pair determines, for instance,
whether the particles have equal or different spin projections.
The spin projection of the individual particles, i.e., Alice's
``EPR particle'' and her unknown input particle, remains
completely unknown. According to this analogy, we call Alice's
quadrature measurements for the teleportation of the state of a
single mode ``Bell detection''. It corresponds to the projection
onto the maximally entangled cv basis of two modes. Due to the
Bell detection and the entanglement between Alice's ``EPR mode''
and Bob's ``EPR mode'', suitable phase-space displacements of
Bob's mode convert it into a replica of Alice's unknown input mode
(a perfect replica for infinite squeezing). In order to perform
the right displacements, Bob needs the classical results of
Alice's Bell measurement.

Quantum teleportation is a conceptually remarkable phenomenon.
However, as a potential application, it becomes only significant
when combined with entanglement purification protocols
(Sec.~\ref{entdistillation}). In this case, instead of sending quantum
information directly through a noisy channel,
a more reliable transfer can be achieved by first distributing
entanglement through the same channel, purifying it, and eventually
exploiting it in a quantum teleportation protocol.
We will now turn to the cv protocol for quantum teleportation
in more detail.

\subsubsection{Teleportation protocol}\label{telepprotocol}

\begin{figure}[t]
\begin{psfrags}
     \psfrag{1}{\large ``in''}
     \psfrag{2}{\large 1}
     \psfrag{3}{\large 2}
     \psfrag{EPR}{\large{\bf EPR}}
     \psfrag{tel}{\large ``tel''}
     \psfrag{Bob}{\large{\bf Bob}}
     \psfrag{Alice}{\large{\bf Alice}}
     \psfrag{x}{\large ${\rm D}^x$}
     \psfrag{z}{\large ${\rm D}^p$}
     \psfrag{a}{\large u}
     \psfrag{b}{\large v}
     \epsfxsize=3.5in
%     \epsfbox[0 170 450 540]{fig5a.eps}
    \epsfbox[-40 5 450 420]{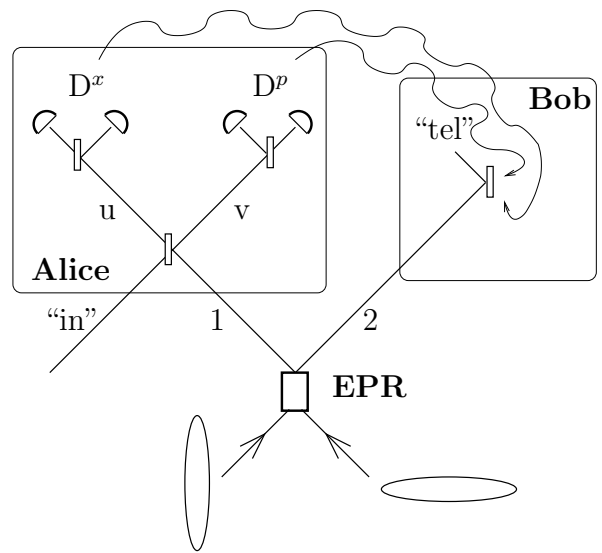}
\end{psfrags}
\caption{Teleportation of a single mode of the electromagnetic field.
Alice and Bob share the entangled state of
modes 1 and 2. Alice combines the mode ``in'' to be teleported with
her half of the EPR state at a beam splitter. The homodyne
detectors ${\rm D}^x$ and ${\rm D}^p$ yield classical photocurrents
for the quadratures $x_{\rm u}$ and $p_{\rm v}$, respectively.
Bob performs phase-space displacements of his half of the EPR state
depending on Alice's classical results.
The entangled state ``EPR'', eventually shared by Alice and Bob,
may be produced by combining two single-mode squeezed states at a
beam splitter or directly via a nonlinear two-mode squeezing interaction.}
\label{telep}
\end{figure}

The simplest formalism to describe cv quantum teleportation
is based on the {\it Heisenberg representation}.
In Eq.~(\ref{2modeHeis}), modes 1 and 2 are entangled to a finite degree,
corresponding to a (pure) nonmaximally entangled state.
In the limit of infinite squeezing,
$r\to\infty$, the individual modes become infinitely noisy, but also
the EPR correlations between them become ideal:
$(\hat{x}_1-\hat{x}_2)\to 0$, $(\hat{p}_1+\hat{p}_2)\to 0$.
Now mode 1 is sent to Alice and mode 2 is sent to Bob
(Fig.~\ref{telep}). Alice's mode is
then combined at a (phase-free) 50:50 beam splitter with the input mode
``in'':
\begin{eqnarray}\label{1.3}
&&\hat{x}_{\rm u}=\frac{1}{\sqrt{2}}\hat{x}_{\rm in}-\frac{1}{\sqrt{2}}
\hat{x}_1,\;\;\;
\hat{p}_{\rm u}=\frac{1}{\sqrt{2}}\hat{p}_{\rm in}-\frac{1}{\sqrt{2}}
\hat{p}_1,\nonumber\\
&&\hat{x}_{\rm v}=\frac{1}{\sqrt{2}}\hat{x}_{\rm in}+\frac{1}{\sqrt{2}}
\hat{x}_1,\;\;\;
\hat{p}_{\rm v}=\frac{1}{\sqrt{2}}\hat{p}_{\rm in}+\frac{1}{\sqrt{2}}
\hat{p}_1.\nonumber\\
\end{eqnarray}
% FIG 5
Using Eqs.~(\ref{1.3}) and (\ref{2modeHeis}), we may write Bob's mode 2 as
\begin{eqnarray}\label{mode2}
\hat{x}_2&=&\hat{x}_{\rm in}-(\hat{x}_1-\hat{x}_2)
-\sqrt{2}\hat{x}_{\rm u}\nonumber\\
&=&\hat{x}_{\rm in}-\sqrt{2}e^{-r} \hat{x}^{(0)}_2
-\sqrt{2}\hat{x}_{\rm u},\nonumber\\
\hat{p}_2&=&\hat{p}_{\rm in}+(\hat{p}_1+\hat{p}_2)
-\sqrt{2}\hat{p}_{\rm v}\nonumber\\
&=&\hat{p}_{\rm in}+\sqrt{2}e^{-r} \hat{p}^{(0)}_1
-\sqrt{2}\hat{p}_{\rm v}.
\end{eqnarray}
Alice's Bell detection yields certain classical values
$x_{\rm u}$ and $p_{\rm v}$ for $\hat{x}_{\rm u}$ and $\hat{p}_{\rm v}$.
The quantum variables $\hat{x}_{\rm u}$ and $\hat{p}_{\rm v}$
become classically determined, random variables
$x_{\rm u}$ and $p_{\rm v}$.
Now, due to the entanglement, Bob's mode 2 collapses into
states that for $r\to\infty$ differ from Alice's input state
only in (random) classical phase-space displacements. After receiving
Alice's classical results $x_{\rm u}$ and $p_{\rm v}$, Bob displaces
his mode,
\begin{eqnarray}\label{1.5}
\hat{x}_2\longrightarrow\hat{x}_{\rm tel}&=&\hat{x}_2+g\sqrt{2}
\hat x_{\rm u},\nonumber\\
\hat{p}_2\longrightarrow\hat{p}_{\rm tel}&=&\hat{p}_2+g\sqrt{2}
\hat p_{\rm v},
\end{eqnarray}
thus accomplishing the teleportation. The parameter
$g$ describes a gain for the transformation from classical
photocurrent to complex field amplitude. For $g=1$,
Bob's displacements eliminate $\hat x_{\rm u}$ and $\hat p_{\rm v}$
in Eq.~(\ref{mode2}).
The teleported mode then becomes
\begin{eqnarray}\label{1.6}
\hat{x}_{\rm tel}&=&\hat{x}_{\rm in}-\sqrt{2}e^{-r}
\hat{x}^{(0)}_2,
\nonumber\\
\hat{p}_{\rm tel}&=&\hat{p}_{\rm in}+\sqrt{2}e^{-r}
\hat{p}^{(0)}_1.
\end{eqnarray}
For an arbitrary gain $g$, we obtain
\begin{eqnarray}\label{gain}
\hat{x}_{\rm tel}&=&g\hat{x}_{\rm in}-\frac{g-1}{\sqrt{2}}
e^{+r}\hat{x}^{(0)}_1-\frac{g+1}{\sqrt{2}}e^{-r}
\hat{x}^{(0)}_2,\nonumber\\
\hat{p}_{\rm tel}&=&g\hat{p}_{\rm in}+\frac{g-1}{\sqrt{2}}
e^{+r}\hat{p}^{(0)}_2+\frac{g+1}{\sqrt{2}}e^{-r}
\hat{p}^{(0)}_1.
\end{eqnarray}
Note that these equations do not take into account
Bell detector inefficiencies.

Consider the case $g=1$. For infinite squeezing $r\to\infty$,
Eqs.~(\ref{1.6}) describe perfect teleportation of the quantum
state of the input mode. On the other hand, for the classical case
of $r=0$, i.e., no squeezing and hence no entanglement, each of
the teleported quadratures has {\it two} additional units of
vacuum noise compared to the original input quadratures. These two
units are so-called quantum duties or ``quduties'' which have to
be paid when crossing the border between quantum and classical
domains \cite{SamKimble}. The two quduties represent the minimal
tariff for every ``classical teleportation'' scheme \cite{Fuchs}.
One quduty, the unit of vacuum noise due to Alice's detection,
arises from her attempt to ``simultaneously measure'' the two
conjugate variables $x_{\rm in}$ and $p_{\rm in}$ in an
Arthurs-Kelly measurement \cite{Arth}. This is the standard
quantum limit for the detection of both quadratures when
attempting to gain as much information as possible about the
quantum state. The standard quantum limit yields a product of the
measurement accuracies which is twice as large as the Heisenberg
minimum uncertainty product. This product of the measurement
accuracies contains the intrinsic quantum limit, the Heisenberg
uncertainty of the mode to be detected, plus an additional unit of
vacuum noise due to the detection. In other words, when measuring
the input Wigner function the resulting distribution is the Wigner
function convoluted with one unit of vacuum, i.e., the $Q$
function [see Eqs.~(\ref{sparameterdistrib}) and
(\ref{sparametercharacterist})]. The second quduty arises when Bob
uses the information of Alice's detection to generate the state at
amplitude $\sqrt{2}x_{\rm u}+i\sqrt{2}p_{\rm v}$ \cite{SamKimble}.
It can be interpreted as the standard quantum limit imposed on
state broadcasting.

The original proposal for the quantum teleportation of
continuous variables with a finite degree of entanglement based on
two-mode squeezed states
used the {\it Wigner representation} and its convolution
formalism \cite{SamKimble}. With the EPR-state Wigner function from
Eq.~(\ref{WEPR}), $W(\xi)\equiv
W_{\rm EPR}(\alpha_1,\alpha_2)$, the whole system after combining
mode ``in'' [which is in an unknown arbitrary quantum state
described by $W_{\rm in}(x_{\rm in},p_{\rm in})$] with mode 1
at a phase-free 50:50 beam splitter (having the two outgoing modes
$\alpha_{\rm u}=x_{\rm u}+ip_{\rm u}$ and
$\alpha_{\rm v}=x_{\rm v}+ip_{\rm v}$) can be written, according to
the transformation rules for Wigner functions
under linear optics, as
\begin{eqnarray}\label{telepafterBS}
&&W(\alpha_{\rm u},\alpha_{\rm v},\alpha_2)=
\int dx_{\rm in}dp_{\rm in} W_{\rm in}(x_{\rm in},p_{\rm in})\nonumber\\
&\times&W_{\rm EPR}\left[\alpha_1=
\frac{1}{\sqrt{2}}(\alpha_{\rm v}-\alpha_{\rm u}),\alpha_2\right]
\nonumber\\
&\times&\delta\left[\frac{1}{\sqrt{2}}(x_{\rm u}+x_{\rm v})
-x_{\rm in}\right]
\,\delta\left[\frac{1}{\sqrt{2}}
(p_{\rm u}+p_{\rm v})-p_{\rm in}\right].\nonumber\\
\end{eqnarray}
Alice's Bell detection on the maximally entangled basis,
i.e., homodyne detections of
$x_{\rm u}=(x_{\rm in}-x_{\rm 1})/\sqrt{2}$ and
$p_{\rm v}=(p_{\rm in}+p_{\rm 1})/\sqrt{2}$,
is described
via integration over $x_{\rm v}$ and $p_{\rm u}$:
\begin{eqnarray}\label{telepreduced}
\int dx_{\rm v}\, dp_{\rm u}\,W(\alpha_{\rm u},\alpha_{\rm v},\alpha_2)
=\int dx\, dp\, W_{\rm in}(x,p)\nonumber\\
\times W_{\rm EPR}\left[x-\sqrt{2}x_{\rm u}+i(\sqrt{2}p_{\rm v}-p),
\alpha_2\right].\nonumber\\
\end{eqnarray}
Bob's displacements are now incorporated by the substitution
$\alpha_2=x_2'-\sqrt{2}x_{\rm u}+i(p_2'-\sqrt{2}p_{\rm v})$
in $W_{\rm EPR}$ in Eq.~(\ref{telepreduced}).
Finally, integration over $x_{\rm u}$ and $p_{\rm v}$
yields the teleported ensemble state (for an ensemble of input states),
\begin{eqnarray}\label{teleportensemble}
W_{\rm tel}(\alpha_2')&=&\frac{1}{\pi\,e^{-2r}}
\int d^2\alpha\, W_{\rm in}(\alpha)\,\exp\left(-\frac{|\alpha_2'
-\alpha|^2}{e^{-2r}}\right)\nonumber\\
&\equiv&W_{\rm in}\circ G_{\sigma}\;.
\end{eqnarray}
The teleported state is a convolution of the input state with
the complex Gaussian $G_{\sigma}(\alpha)\equiv [1/(\pi\sigma)]
\exp(-|\alpha|^2/\sigma)$ with the complex variance $\sigma=e^{-2r}$.
This convolution adds the excess noise variance $e^{-2r}/2$
to each quadrature of the input state.

As for a description of cv quantum teleportation in the {\it
Schr\"{o}dinger representation}, there are several works. In
Ref.~\cite{MilburnSam99}, two different teleportation protocols
were considered, referring to two different kinds of Bell
measurements made by Alice. Depending on this choice, measuring
either relative position and total momentum or photon-number
difference and phase sum, the entire protocol, including the
two-mode squeezed vacuum resource, is written in the position or
in the Fock basis, respectively \cite{MilburnSam99}. The protocol
based on number-difference and phase-sum measurements was later
modified and extended by \textcite{Clausen00} and
\textcite{Cochrane00,Cochrane02}. The extended cv quantum
teleportation protocol of \textcite{Opatr} is based on quadrature
Bell measurements and leads to an enhancement of the teleportation
fidelity (see the next section) via subtraction of single photons.
This protocol is also formulated in the Schr\"{o}dinger
representation. Using the approach of \textcite{Opatr}, i.e.,
making conditional measurements on two-mode squeezed states, it is
demonstrated in Ref.~\cite{Cochrane02} that the teleportation
fidelity can be improved in both the quadrature-measurement based
and the number-difference and phase-sum measurement based scheme.
In particular, for quantifying the performance of cv quantum
teleportation of dv (non-Gaussian) states such as Fock states, the
transfer-operator formalism in the Schr\"{o}dinger picture by
\textcite{HofmannIde00} and \textcite{IdeHofmann01} is very
useful. In this formalism, the first step is writing the ``cv Bell
states'' in the Fock basis, $\hat D(\beta)\otimes
\mbox{1$\!\!${\large 1}} \frac{1}{\sqrt{\pi}}\sum_{n=0}^\infty
|n\rangle |n\rangle$, where $\hat D(\beta)$ is the displacement
operator and $\beta=u+iv$ is the measurement result. After
projecting Alice's input state $|\phi\rangle_{\rm in}$ and her
half of the two-mode squeezed vacuum state in Eq.~(\ref{twin})
onto this Bell basis and reversing the measured displacement in
Bob's half, $|\phi\rangle$ is transferred to Bob's location in the
form of \cite{HofmannIde00}
\begin{eqnarray}\label{cvqtelepintransfer}
|\phi_{\rm tel}(\beta)\rangle=
\hat T(\beta)|\phi\rangle\;,
\end{eqnarray}
with the transfer operator
\begin{eqnarray}\label{transferoperator}
\hat T(\beta)=
\hat D(\beta)\mathcal{D}(\lambda)\hat D(-\beta)\;,
\end{eqnarray}
and the ``distortion operator''
\begin{eqnarray}\label{distortionoperator}
\mathcal{D}(\lambda)=\sqrt{\frac{1-\lambda}{\pi}}
\sum_{n=0}^{\infty}\lambda^{n/2}|n\rangle\langle n|\;.
\end{eqnarray}
Here, the teleported state $|\phi_{\rm tel}(\beta)\rangle$
is unnormalized and
$\langle\phi_{\rm tel}(\beta)|\phi_{\rm tel}(\beta)\rangle$
is the probability for obtaining result $\beta$.
The imperfection of the entanglement resource
is expressed by the distortion operator, where the factors
$\sqrt{1-\lambda}\;\lambda^{n/2}$
are the Schmidt coefficients of the finitely squeezed,
only nonmaximally entangled two-mode squeezed vacuum state
in Eq.~(\ref{twin}).
Note that for infinite squeezing, $\lambda\to 1$,
the distortion operator, and hence the transfer
operator too, becomes proportional
to the identity operator.
The teleported state $|\phi_{\rm tel}(\beta)\rangle$
corresponds to a single shot (a single teleportation event).
Thus, the teleported ensemble state, averaged over
all measurement results for an ensemble of input states
becomes
\begin{eqnarray}\label{teleportensembleschroed}
\hat\rho_{\rm tel}=
\int d^2\beta\;|\phi_{\rm tel}(\beta)\rangle
\langle\phi_{\rm tel}(\beta)|\;,
\end{eqnarray}
corresponding to $W_{\rm tel}(\alpha_2')$ in
Eq.~(\ref{teleportensemble}). The distortion operator is also used
in Ref.~\cite{twist} to describe the effect of the nonmaximally
entangled EPR channel. Moreover, using a universal formalism in
the Schr\"{o}dinger picture, it is shown in Ref.~\cite{twist} that
in both dv and cv quantum teleportation, Bob's local unitary
operation means twisting the shared entanglement relative to the
entangled state measured in the Bell detection. The
transfer-operator description of cv quantum teleportation in
Eq.~(\ref{cvqtelepintransfer}) corresponds to a completely
positive map and $\hat T(\beta)$ is just a Kraus operator
\cite{Kraus}. This map projects the input state onto the
conditional teleported state. Upon averaging over all results
$\beta$, the map becomes completely positive {\it and}
trace-preserving (CPTP), yielding the normalized teleported state
$\hat\rho_{\rm tel}$ in Eq.~(\ref{teleportensembleschroed}). A
position-momentum basis description of this transfer-operator or
CP-map formalism was given by \textcite{Takeoka}. The CP map
derived in this paper is also more general, including mixed
entangled states as a resource for quantum teleportation. Another
alternative formulation of nonideal cv quantum teleportation was
proposed by \textcite{Vukics02}, utilizing the coherent-state
basis.

\subsubsection{Teleportation criteria}\label{telepcritsec}

\begin{figure}[tb]
\begin{psfrags}
     \psfrag{1}{\large ``in''$~~$}
     \psfrag{2}{\large 1}
     \psfrag{3}{\large 2}
     \psfrag{EPR}{\large{\bf EPR}}
     \psfrag{tel}{\large ``tel''$~~~$}
     \psfrag{Bob}{\large{\bf Bob$~~~$}}
     \psfrag{Alice}{\large{\bf Alice}}
     \psfrag{Victor}{\large{\bf Victor}}
     \psfrag{x}{\large ${\rm D}^x$}
     \psfrag{p}{\large ${\rm D}^p$}
     \psfrag{a}{\large u}
     \psfrag{b}{\large v}
     \epsfxsize=3.0in
%     \epsfbox[-80 5 450 490]{fig6a.eps}
     \epsfbox[-40 5 450 490]{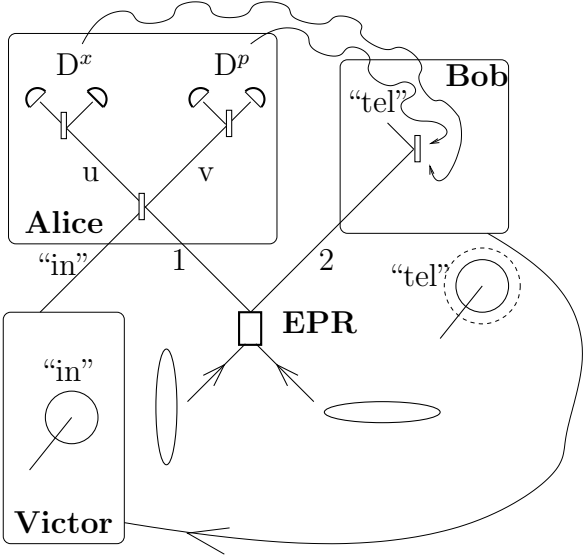}
\end{psfrags}
\caption{Verification of quantum teleportation. The verifier ``Victor''
is independent of Alice and Bob. Victor prepares the input states
which are known to him, but unknown to Alice and Bob. After a supposed
quantum teleportation from Alice to Bob, the teleported states are given
back to Victor. Due to his knowledge of the input states,
Victor can compare the teleported states with the input states.}
\label{telepverif}
\end{figure}

The teleportation scheme with Alice and Bob is complete
without any further measurement. The teleported state remains
unknown to both Alice and Bob
and need not be demolished in a detection by Bob as a final step.
However, maybe Alice and Bob are cheating. Suppose that
instead of using an EPR channel,
they try to get away without entanglement and use only a
classical channel. In particular, for the realistic experimental
situation with finite squeezing and inefficient detectors where perfect
teleportation is unattainable, how may we verify that successful
quantum teleportation has taken place?
To make this verification we shall introduce a third party,
``Victor'' (the verifier), who is independent of Alice and Bob
(Fig.~\ref{telepverif}).
We assume that he prepares the initial input state
(drawn from a fixed set of states) and passes it on to
Alice. After accomplishing the supposed teleportation, Bob sends the
resulting teleported state back to Victor.
Victor's knowledge about the input state
and detection of the teleported state enable him to verify
whether quantum teleportation has really occurred.
For this purpose, however, Victor needs some measure that helps him to
assess when the similarity between the teleported state and the input state
exceeds a boundary that is only exceedable with entanglement.

% FIG 6

One such measure is the so-called fidelity $F$, for an arbitrary
input state $|\phi_{\rm in}\rangle$ defined by \footnote{apart
from the fidelity, we also used the symbol $F$ for the mean value
of a collective spin. We stick with this notation in both cases to
be consistent with the common notation in the literature.}
\cite{Fuchs}
\begin{eqnarray}\label{fid1}
F\equiv\langle\phi_{\rm in}|\hat{\rho}_{\rm tel}|
\phi_{\rm in}\rangle.
\end{eqnarray}
It equals one only if
$\hat{\rho}_{\rm tel}=|\phi_{\rm in}\rangle\langle\phi_{\rm in}|$.
Now Alice and Bob know that Victor draws his states
$|\phi_{\rm in}\rangle$ from a fixed set, but they do not know
which particular state is drawn in a single trial.
Therefore, an average fidelity should be considered \cite{Fuchs},
\begin{eqnarray}\label{fidaveragegeneral}
F_{\rm av}=\int P(|\phi_{\rm in}\rangle)
\langle\phi_{\rm in}|\hat{\rho}_{\rm tel}|
\phi_{\rm in}\rangle d|\phi_{\rm in}\rangle,
\end{eqnarray}
where $P(|\phi_{\rm in}\rangle)$ is the probability of drawing
a particular state $|\phi_{\rm in}\rangle$, and the integral runs
over the entire set of input states.
If the set of input states contains all possible quantum states
in an infinite-dimensional Hilbert space (i.e., the input state is
completely unknown apart from its infinite Hilbert-space dimension), the
best average fidelity $F_{\rm av}$ achievable by Alice and Bob
without using entanglement is zero.
The corresponding best average fidelity if the set of input
states contains all possible quantum states in a $d$-dimensional Hilbert
space is $F_{\rm av}=2/(1+d)$ \cite{Barnum}. Thus, one obtains
$F_{\rm av}=0$ for $d\to\infty$, and the qubit boundary
$F_{\rm av}=2/3$ for $d=2$.
If the input alphabet is restricted to coherent states of amplitude
$\alpha_{\rm in}=x_{\rm in}+ip_{\rm in}$ and
$F=\langle\alpha_{\rm in}|\hat{\rho}_{\rm tel}|\alpha_{\rm in}\rangle$,
on average, the fidelity achievable in a purely classical scheme
(when averaged across the entire complex plane for arbitrary
coherent-state inputs) is bounded by
\cite{Fuchs}
\begin{eqnarray}\label{fid3}
F_{\rm av}\leq\frac{1}{2}\,\,\,.
\end{eqnarray}

Let us apply the fidelity criterion to the single-mode
teleportation equations Eq.~(\ref{gain}) and assume an
input alphabet containing all coherent states
with equal probability.
Up to a factor $\pi$, the fidelity
$F=\langle\alpha_{\rm in}|\hat{\rho}_{\rm tel}|\alpha_{\rm in}\rangle$
is the $Q$ function of the teleported mode evaluated for
$\alpha_{\rm in}$. This Q function is, in general, a bivariate
Gaussian with mean value $g (x_{\rm in}+i p_{\rm in})$,
\begin{eqnarray}\label{fid4}
F&=&\pi Q_{\rm tel}(\alpha_{\rm in})\\
&=&\frac{1}{2\sqrt{\sigma_x\sigma_p}}
\exp\left[-(1-g)^2\left(\frac{x_{\rm in}^2}{2\sigma_x}
+\frac{p_{\rm in}^2}{2\sigma_p}\right)\right],\nonumber
\end{eqnarray}
where $g$ is the gain and $\sigma_x$
and $\sigma_p$ are the variances of the
Q function of the teleported mode for the corresponding quadratures.
The Q function is a convolution of the Wigner function
with a Gaussian of one unit of vacuum
[Eqs.~(\ref{sparameterdistrib}) and (\ref{sparametercharacterist})],
i.e., we have to add this unit
to the actual variances of the teleported quadratures.
According to Eq.~(\ref{gain}), for a coherent-state
input, the variances of the Q function are then given by
\begin{eqnarray}\label{fid5}
\sigma_x=\sigma_p=
\frac{1}{4}(1+g^2)+\frac{e^{+2r}}{8}(g-1)^2+
\frac{e^{-2r}}{8}(g+1)^2\;.\nonumber\\
\end{eqnarray}
Teleporting states with a coherent amplitude
as reliably as possible requires unit-gain teleportation (unit gain
in Bob's final displacements). Only in this case do the coherent
amplitudes of the teleported states always match those of the input
states provided by Victor.
For classical teleportation ($r=0$) and $g=1$, we obtain
$\sigma_x=\sigma_p=1$ and indeed $F=F_{\rm av}=1/2$.
In order to obtain a better fidelity, entanglement is needed.
Then, if $g=1$, we obtain $F=F_{\rm av}>1/2$ for any $r>0$.
For this unit-gain teleportation, we have seen that the
teleported state $W_{\rm tel}$
is a convolution of the input $W_{\rm in}$ with a complex Gaussian of
variance $e^{-2r}$ [Eq.~(\ref{teleportensemble})].
Classical teleportation with $r=0$ then means
the teleported mode has an excess noise of two complex units of vacuum,
$1/2+1/2$, relative to the input.
Any $r>0$ beats this classical scheme. Hence,
if the input state is always recreated with the right amplitude and
less than two units of vacuum excess noise, Alice and Bob must
have employed entanglement as a resource.

Let us also write the fidelity in terms of the transfer operator
of Eq.~(\ref{transferoperator}). Using
Eqs.~(\ref{teleportensembleschroed}), (\ref{cvqtelepintransfer}),
and (\ref{fid1}),
with a coherent-state input, leads to
\begin{eqnarray}\label{fidintransfer}
F=\int d^2\beta\;|\langle\alpha_{\rm in}|\hat T(\beta)|\alpha_{\rm in}
\rangle|^2
=\frac{1 + \sqrt{\lambda}}{2}\;.
\end{eqnarray}
Here we used $\hat D(-\beta)|\alpha\rangle=
|\alpha-\beta\rangle$ and
$|\langle n|\alpha\rangle|^2=
|\alpha|^{2n}e^{-|\alpha|^2}/n!$.
This fidelity becomes independent of $\alpha_{\rm in}$,
because the transfer operator of Eq.~(\ref{transferoperator})
corresponds to unit-gain teleportation.
Hence, with $\lambda=\tanh^2 r$, we obtain $F=F_{\rm av}=1/(1+e^{-2r})$,
identical to the result for $g=1$ using Eqs.~(\ref{fid4})
and (\ref{fid5}).

In Sec.~\ref{multipentanglement}, we found that one squeezed state
is a sufficient resource for generating entanglement between an
arbitrary number of parties. In fact, applied to the two-party
teleportation scenario, the entanglement from only one squeezed
state enables quantum teleportation with $F_{\rm av}>1/2$ for any
nonzero squeezing \cite{PvLPRL00}. Unless Alice and Bob have
access to additional local squeezers,\footnote{using additional
local squeezers, Alice and Bob can transform the shared entangled
state built from one squeezed state into the ``canonical''
two-mode squeezed state
\cite{Bowen01,vanloockFdP02,vanloockcvbook02}. Note that such
local squeezing operations do not change the amount of the
entanglement. The resulting two-mode squeezed state obviously can
approach unit fidelity when used for quantum teleportation. Though
conceptually interesting (the amount of entanglement inherent in
an entangled state built with one squeezer is arbitrarily large
for sufficiently large squeezing and hence there is no fidelity
limit), this would not be the most practical way to achieve
high-fidelity quantum teleportation. The entire teleportation
process would require three squeezers with squeezing $2r$, $r$,
and $r$, instead of only two $r$-squeezers needed to produce the
``canonical'' two-mode squeezed state
\cite{Bowen01,vanloockFdP02,vanloockcvbook02}.} the maximum
fidelity of coherent-state teleportation achievable with one
single-mode squeezed state is $F_{\rm av}=1/\sqrt{2}$ in the limit
of infinite squeezing \cite{PvLPRL00}.

Alternative criteria for quantum teleportation were proposed by
\textcite{Ralph}. Reminiscent of the criteria \cite{Holland} for
quantum non-demolition (QND) measurements \cite{Brag}, these are
expressed in terms of two inequalities for the conditional
variances [Eq.~(\ref{optimalinfvar})] and the so-called signal
transfer coefficients of both conjugate quadratures. The boundary
between classical and quantum teleportation defined by the
criteria of \textcite{Ralph} differs from that in Eq.~(\ref{fid3})
in terms of fidelity. According to Ralph and Lam, the best
classical protocol permits output states completely different from
the input states, corresponding to zero fidelity. This can be
achieved, for instance, via an asymmetric detection scheme, where
the lack of information in one quadrature leads, on average, to
output states with amplitudes completely different from those of
the input states. However, certain correlations between the input
and the teleported quadratures can still attain the optimal value
allowed without using entanglement. By contrast, the best
classical protocol in terms of fidelity always achieves output
states pretty similar to the input states. The fidelity boundary
in Eq.~(\ref{fid3}) is exceeded for any squeezing in the EPR
channel, whereas fulfillment of the teleportation criteria of
\textcite{Ralph} requires more than 3 dB squeezing.

Finally, \textcite{GG3} advertise $F_{\rm av}\leq 2/3$ as the
fidelity boundary between classical and quantum teleportation of
arbitrary coherent states. Exceeding this bound would also require
an EPR channel with more than 3 dB squeezing. The reasoning by
\textcite{GG3} is that only when Bob receives a state with $F_{\rm
av}>2/3$, it is guaranteed that nobody else (neither Alice nor an
eavesdropper ``Eve'') can have an equally good or better copy.
Otherwise, two copies of the unknown input state with $F_{\rm
av}>2/3$ would exist which contradicts the no-cloning boundary for
coherent-state duplication (see Sec.~\ref{qcloningsec}). On the
other hand, when Bob receives for example a state with $1/2<F_{\rm
av}< 2/3$, Alice might have locally made two asymmetric copies,
one with $F_{\rm av}>2/3$ and one with $1/2<F_{\rm av}< 2/3$. She
might have sent the worse copy to Bob via a perfectly entangled
(or sufficiently entangled) EPR channel and kept the better copy.
Thus, according to Grangier and Grosshans, the ``quantum faxing''
region \cite{GG3}, $1/2<F_{\rm av}\leq 2/3$, does not indicate
true quantum teleportation of coherent states. Similarly, one
would have to give the region $2/3<F_{\rm av}\leq 5/6$ (where
$5/6$ is the qubit duplication limit, see Sec.~\ref{qcloningsec})
an analogous status of only quantum faxing when teleportation of
arbitrary {\it qubits} is considered. By contrast, the fidelity
boundary between classical and quantum teleportation of arbitrary
qubit states, analogous to that of arbitrary coherent states in
Eq.~(\ref{fid3}), is $F_{\rm av}\leq 2/3$ \cite{Barnum}. A
detailed discussion about the different fidelity boundaries for
coherent-state teleportation can be found in
Ref.~\cite{Kimblecrit}.

What are the ``right'' criteria for cv quantum teleportation
among those discussed above?
What the most appropriate criteria in a specific scenario are
certainly depends on the particular task that
is to be fulfilled by quantum teleportation.
For example, using quantum teleportation
of coherent states as a subroutine
for quantum cryptography, the security
is (to some extent) ensured by $F_{\rm av}>2/3$ \cite{GG3}.
However, for an input alphabet of coherent states to be
transferred, a rigorous boundary that unambiguously
separates entanglement-based quantum teleportation schemes
from schemes solely based on classical communication
is given by $F_{\rm av}\leq\frac{1}{2}$, Eq.~(\ref{fid3}).

Another possible way to assess whether a cv teleportation scheme
is truly quantum is to check to what extent nonclassical
properties such as squeezing or photon antibunching can be
preserved in the teleported field \cite{Li02}. Once Alice and Bob
do not share entanglement [for instance, when the pure two-mode
squeezed vacuum state becomes mixed in a thermal environment
\cite{LeeKimJeong00}], nonclassical properties can no longer be
transferred from Alice to Bob. Let us now consider the
teleportation of a truly quantum mechanical system, namely an
electromagnetic mode entangled with another mode. This
entanglement shall be transferred to a third mode via quantum
teleportation.

\subsubsection{Entanglement swapping}\label{entswappsec}

In the three optical teleportation experiments
in Innsbruck \cite{Bou}, in Rome \cite{PVLBoschi},
and in Pasadena \cite{Fur98},
the nonorthogonal input states to be teleported were
single-photon polarization states \cite{Bou,PVLBoschi} and
coherent states \cite{Fur98}.
 From a true quantum teleportation device, however, we would also require
the capability of teleporting the entanglement source itself. This
teleportation of one half of an entangled state (``entanglement
swapping'') was first introduced for single-photon polarization
states \cite{Zuk}. In general, it allows for the entanglement of
two quantum systems that have never directly interacted with each
other. A demonstration of entanglement swapping with single
photons was reported by \textcite{Pan}. Practical uses of
entanglement swapping have been suggested
\cite{Bose1,Bose2,Briegel,Duer2} and it has also been generalized
to multiparticle systems \cite{Bose2}. All these investigations
have referred exclusively to dv systems.

Involving continuous variables, there have been several
theoretical proposals for an entanglement swapping experiment.
\textcite{Polk} suggested teleporting polarization-entangled
states of single photons using squeezed-state entanglement (in the
limit of small squeezing) where the output correlations are
verified via Bell inequalities. \textcite{Tan} and \textcite{PvL}
considered the unconditional teleportation (without post-selection
of ``successful'' events by photon detections) of one half of a
two-mode squeezed state using quadrature Bell measurements and
different verification schemes. In Ref.~\cite{PvL}, entanglement
swapping is verified through a second quantum teleportation
process utilizing the entangled output state. In Ref.~\cite{Tan},
the output entanglement is verified via Eq.~(\ref{Tan}) using
unit-gain displacements and hence requiring more than 3 dB
squeezing. The gain, however, can be optimized, enabling a
verification of entanglement swapping for any squeezing
\cite{PvL,vanloockFdP02}.

Choosing the right gain is essential in entanglement swapping, for
instance, in order to optimize the fidelity in a second round of
teleportation \cite{PvL} or to maximize the violations of Bell
inequalities at the output in the low-squeezing (single-photon)
limit \cite{Polk}. The explanation for this is as follows: if the
conditional states are displaced with the right gain, they no
longer depend on the Bell measurement results, always being
transformed to the same canonical two-mode squeezed state. The
optimal scheme then leads to a {\it pure ensemble output state}.
Even after averaging upon all incoming states and measurement
results, the output remains a pure two-mode squeezed vacuum state
with a new squeezing parameter $R$ modified by
\cite{vanloockFdP02},
\begin{eqnarray}\label{newsqueez}
\tanh R=\tanh r\, \tanh r'\;.
\end{eqnarray}
Up to a phase-space displacement, the resulting ensemble state is
then the same as the projected displaced two-mode squeezed state
for a single shot of the cv Bell measurement. In
Eq.~(\ref{newsqueez}), $r$ and $r'$ are the squeezing parameters
of the two initial entangled two-mode squeezed states. For any
nonzero squeezing and hence entanglement in {\it both} input
states, $r>0$ {\it and} $r'>0$, entanglement swapping occurs,
i.e., $R>0$. However, the quality of the entanglement always
deteriorates, $R<r$ and $R<r'$, unless either of the input states
approaches a maximally entangled state, $r\to\infty$ and hence
$R=r'$, or $r'\to\infty$, thus $R=r$.

As a consequence, in a quantum repeater \cite{Briegel}, connecting
many segments of a quantum channel via cv entanglement swapping,
after purifying the mixed entangled states in each segment to pure
ones (two-mode squeezed states), will produce never vanishing, but
increasingly small entanglement between the ends of the channel.
This statement, however, is restricted to an entanglement swapping
protocol based on cv (quadrature) Bell measurements.

\subsection{Dense coding}\label{qdensecod}

Dense coding aims to use shared entanglement to increase the
capacity of a communication channel \cite{CHB97}. Relative to
quantum teleportation, in dense coding, the roles played by the
quantum and classical channels are interchanged. Dense coding was
translated to continuous quantum variables by \textcite{Ban} and
\textcite{Samdense}. It was shown that by utilizing the
entanglement of a two-mode squeezed state, coherent communication
(based on coherent states) can always be beaten \cite{Samdense}.
The cv scheme attains a capacity approaching (in the limit of
large squeezing) twice that theoretically achievable in the
absence of entanglement \cite{Samdense}. Before we discuss how
dense coding can be implemented with continuous variables we shall
review the ideas behind quantifying information for communication.

\subsubsection{Information: a measure}

In classical information theory one constructs a measure of information
which tries to capture the `surprise' attached to receiving a particular
message. Thus messages which are common occurrences, are assumed to
contain very little useful information, whereas rare messages are deemed
valuable and so containing more information. This concept suggests
that the underlying symbols or letters or alphabet used to transmit
the message are themselves unimportant, but only the probabilities of
these symbols or messages.

In addition to this conceptual framework, it turns out that the measure
of information is essentially unique if one takes it to be additive
for independent messages.

To formalize the idea of an alphabet we define it to be a set
$A=\{a:a\in A\}$ of symbols $a$ each of which has an associated
probability $p_a$. The information content of such an alphabet
will be denoted $I(A)$.  The average information content per letter
in alphabet $A$ is given by
\begin{equation}
I(A) = -\sum_a p_a \log_b p_a \;.
\end{equation}
This result is the unique measure of average information per letter.

\subsubsection{Mutual information}

In order to quantify the information in a communication channel we must
introduce a measure of information corresponding to the amount of
information accessible to the receiver which contains information about
the message sent. This suggests we attempt to quantify the information
mutual (or common) to a pair of alphabets $A$ and $B$. If these two
alphabets have letters with probabilities given by $p_a$ and $p_b$
respectively, and if the joint alphabet $AB$ has letters with probabilities
$p_{ab}$ (we no longer assume these alphabets are independent), then
the natural information theoretic measures for these three quantities
are given by
\begin{eqnarray}
I(A) &=& -\sum_a p_a \log p_a \nonumber \\
I(B) &=& -\sum_b p_b \log p_b \nonumber \\
I(A,B) &=& -\sum_{ab} p_{ab} \log p_{ab} \;.
\end{eqnarray}

The so-called mutual information $I(A:B)$ in a pair of alphabets is given by
\begin{eqnarray}\label{mutualinfodef}
I(A:B) &=& I(A) + I(B) - I(A,B) \nonumber \\
&=&\sum_{ab}p_{ab} \log\Bigl(\frac{p_{ab}}{p_a p_b}\Bigr)\;.
\label{mut_info}
\end{eqnarray}
The idea behind this equation is that the sum $ I(A) + I(B)$ accounts
for the joint information in both alphabets, but double counts that
part which is mutual to both alphabets. By subtracting the correct
expression for the joint information $I(A,B)$ we are left solely with
the information that is common or mutual.

\subsubsection{Classical communication}

We are now in a position to apply our expression for the mutual information
to quantify the information received through a communication channel
that contains information or that is mutual or common to the information
actually sent.

Let us suppose the sender, called Alice, has a source alphabet $A$ with
probabilities $p_a$. For each letter $a$ sent the receiver, called Bob,
tries to determine its value through observation at his end of the channel.
However, because of whatever source of noise or
imprecision in general Bob will see an alphabet $B$ which is not identical
to Alice's.

For simplicity, we shall suppose that the channel has no memory so that
each signal sent is independent of earlier or later channel usage.
In this case, we may characterize the channel by the conditional
probabilities $p_{b|a}$, for the probability of observing letter $b$
in Bob's alphabet, given that Alice sent letter $a$. The joint
probability is therefore given by
\begin{equation}
p_{ab} = p_{b|a} p_a \;.
\end{equation}
Given these expressions the mutual information content, per usage of
the channel, in Bob's received data about Alice's messages is
\begin{equation}
I(A:B)=\sum_{ab} p_{b|a}p_a \log\Bigl(\frac{p_{b|a}}{p_b}\Bigr) \;.
\end{equation}

If we optimize this expression over Alice's alphabet we can
determine the maximum achievable throughput per usage. This is
called the channel's channel capacity and is given by
\begin{equation}
C = \max_{\{p_a\}}\, I(A:B) \;.
\end{equation}

\subsubsection{Classical communication via quantum states}

Ultimately, Alice must use some physical carrier to represent the letters
she sends. For simplicity of notation we shall tie the physical states
she generates together with any modifying effects from the channel. Thus
we shall say that to represent a letter $a$ from her alphabet $A$ that
Alice produces a quantum state $\hat \rho_a$. Since letter $a$ is generated
with probability $p_a$ the mean channel state is simply
\begin{equation}
\hat \rho = \sum_a p_a \hat \rho_a \;.
\end{equation}

Now let us suppose that Bob uses some generalized measurement (POVM)
$\hat E_b$, satisfying Eq.~(\ref{POVMdef}),
to try to extract information about the letter Alice was trying to
send.
When Bob is presented with a state $\hat\rho_a$ representing letter $a$
from Alice's alphabet, he will find instead letter $b$ from his own alphabet
with a conditional probability given by Eq.~(\ref{POVMcondprobdef}),
from which one may compute the mutual information $I(A:B)$ using
Eq.~(\ref{mut_info}).

The famous result from \textcite{ASH98} allows us to place an
upper bound on this mutual information via
Eq.~(\ref{Holevobound}). We note that either bound is independent
of Bob's measurement strategy, so an achievable upper bound will
allow us to determine the channel capacity for transmitting
classical information using quantum states. In fact, when this
strategy works it reduces the calculation of the channel capacity
to one of performing a maximum entropy calculation.

When the states used to send information are from an infinite-dimensional
Hilbert space (such as a single-mode bosonic field) we must place
some constraint on the channel usage in order to get a finite value
for this capacity. The canonical constraint in such circumstances
is to presume that there is a constraint on the mean number of quanta
that may pass down the channel per usage $\langle \hat n\rangle = \bar n$.
For this constraint, the maximum entropy may be interpreted as the channel
capacity achieved when Alice uses an alphabet of number states distributed
according to a thermal distribution \cite{CavesDrummondRMP,HPY93}.
In this case, the
channel capacity is given by
\begin{eqnarray}
C&=&(1+\bar n)\ln (1+\bar n) -\bar n\ln \bar n \label{max_chan} \\
&\simeq&1+\ln \bar n \;, \nonumber
\end{eqnarray}
for large $\bar n$.

\begin{figure}[tb]
  \begin{center}
  \begin{psfrags}
    \psfrag{i}[rb]{\large a)$~~~~~~~~~$}
    \psfrag{A}[c]{\large $|n\rangle~~~~~~~~~~~$}
    \psfrag{B}[c]{\large $~~~~~~|m\rangle\langle m|$}
    \epsfxsize=3.0in
    \epsfbox[-160 0 520 140]{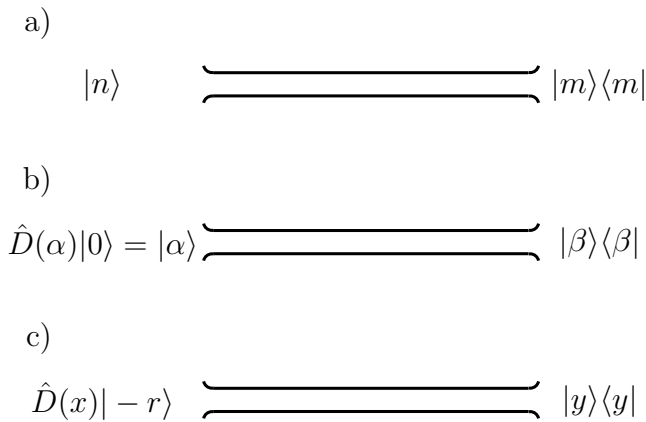}
  \end{psfrags}
  \end{center}
\begin{center}
  \begin{psfrags}
    \psfrag{i}[rb]{\large b)$~~~~~~~~~$}
    \psfrag{A}[c]
{\large $\hat D(\alpha)|0\rangle=|\alpha\rangle~~~~~~~~~~~$}
    \psfrag{B}[c]{\large $~~~~~~|\beta\rangle\langle \beta|$}
    \epsfxsize=3.0in
    \epsfbox[-160 0 520 120]{channela.eps}
  \end{psfrags}
  \end{center}
\begin{center}
  \begin{psfrags}
    \psfrag{i}[rb]{\large c)$~~~~~~~~~$}
    \psfrag{A}[c]{\large $\hat D(x)|-r\rangle~~~~~~~~~~~$}
    \psfrag{B}[c]{\large $~~~~~~|y\rangle\langle y|$}
    \epsfxsize=3.0in
    \epsfbox[-160 40 520 120]{channela.eps}
  \end{psfrags}
  \end{center}
\caption{Communication channel with a) optimal number-state
alphabet; b) coherent-state alphabet; and c) squeezed-state alphabet.}
\label{channel_num_coh_sq}
\end{figure}

Given a constraint on the mean number of photons per channel usage
$\langle \hat n\rangle = \bar n$ we shall now compare three choices
of alphabets for transmission by Alice, see Fig.~\ref{channel_num_coh_sq}.
Fig.~\ref{channel_num_coh_sq}a shows the optimal strategy using an
input alphabet of number states and ideal photon-number detection.
Fig.~\ref{channel_num_coh_sq}b shows the channel with the same constraint
operating with an input alphabet of coherent states and an
heterodyne detection. With this extra constraint on the input alphabet
the maximal throughput is given by \cite{JPG62,CYS68,YY86}
\begin{equation}
C^{\rm coh} = \ln (1+\bar n) \simeq \ln \bar n \;.
\label{C_coh}
\end{equation}
Finally, if Alice uses a strategy involving a squeezed state alphabet
(labeled by the displacements $x$) and homodyne detection, see
Fig.~\ref{channel_num_coh_sq}c,
the maximal throughput is given by \cite{YY86}
\begin{equation}
C^{\rm sq}=\ln(1+2\bar n) \simeq \ln 2 + \ln \bar n \;.
\label{C_sq}
\end{equation}
We can see that for large $\bar n$ each of these schemes has a capacity
which differs by only around one bit per usage!

\subsubsection{Dense coding}

In dense coding, Alice and Bob communicate via two channels, however,
Alice only needs to `modulate' one of them. The second channel
is used to transmit
one half of a standard (entangled) state to Bob; since this channel is
not modulated it may be sent at any time, including prior to its need
for communication. In this way, part of the communication channel may
be run {\it off-peak}. Classically there is no way to achieve this sort
of operation.

In general, Alice's local action is sufficient to span a system with
the square of the Hilbert-space dimension of the piece she holds.
Since information is essentially the logarithm of the number of
distinguishable states,
\begin{equation}
\log(n^2) = 2\log n \;,
\end{equation}
one can generally expect a doubling of the channel capacity. This
accounting assumes that the off-peak usage required to transmit the
shared entanglement, or to otherwise generate it, comes at no cost.

Consider the specific case of EPR beams $(1,2)$ approximated
by the two-mode squeezed state with Wigner function
\begin{eqnarray}
W_{{\rm EPR}}\!\!\!\!\!\!\!&&(\alpha_{1},\alpha_{2}) \\
&&={\frac{4}{\pi ^{2}}}\exp [-e^{-2r}(\alpha
_{1}-\alpha_{2})_{R}^{2}-e^{2r}(\alpha_{1}-\alpha_{2})_{I}^{2}  \nonumber \\
&&\phantom{={\frac{4}{\pi ^{2}}}\exp} -e^{2r}(\alpha_{1}+%
\alpha_{2})_{R}^{2}-e^{-2r}(\alpha _{1}+\alpha _{2})_{I}^{2}] \;,   \nonumber
\end{eqnarray}
where the subscripts $R$ and $I$ refer to real and imaginary parts of the
field amplitude $\alpha $, respectively.

\begin{figure}[tb]
  \begin{center}
  \begin{psfrags}
    \psfrag{EPR}[b]{\large $~~$EPR}
    \psfrag{A}[t]{\large Alice$~~$}
    \psfrag{B}[b]{\large Bob}
    \psfrag{D}[c]{\large $\hat D(\alpha)$}
    \psfrag{c}[c]{\large channel}
    \psfrag{b1r}[c]{\large $~~~~~~\beta_{1R}$}
    \psfrag{b2i}[c]{\large $\beta_{2I}$}
    \epsfxsize=3.0in
    \epsfbox[-90 30 280 314]{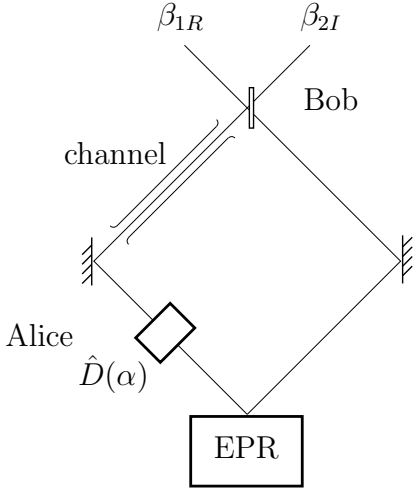}
  \end{psfrags}
  \end{center}
\caption{Schematic of dense coding scheme: An EPR pair is created
and one half sent to Bob. The other half is given to Alice who
modulates it by performing a phase-space translation by the amount $\alpha$.
This half is now passed on to Bob who recombines the two halves and
performs a pair of homodyne measurements on the sum and difference fields.
These signals $\beta_{1R}$ and $\beta_{2I}$ will closely mimic $\alpha$
for a sufficiently strongly entangled state.}
\label{dense_coding}
\end{figure}

As shown in Fig.~\ref{dense_coding}, signal modulation is performed only
on Alice's mode, with the second mode treated as an overall shared
resource by Alice and Bob. The modulation scheme that Alice chooses is
simply to displace her mode by an amount $\alpha$. This leads to a
displaced Wigner function given by $W_{\rm EPR}(\alpha_1-\alpha,\alpha_2)$,
corresponding to the field state that is sent via the quantum channel from
Alice to Bob.

Upon receiving this transmitted state (consisting of Alice's modulated
mode), the final step in the dense-coding protocol is for Bob to combine
it with the shared resource he holds and retrieve the original classical
signal $\alpha$ with as high a fidelity as possible. As indicated in
Fig.~\ref{dense_coding}, this demodulation can be performed with a
simple $50:50$ beam splitter that superposes these modes to yield
output fields that are the sum and difference of the input fields
and which we label as $\beta_1$ and $\beta_2$, respectively. The
resulting state emerging from Bob's beam splitter has Wigner function
\begin{eqnarray}
&&W_{{\rm sum/diff}}(\beta _{1},\beta _{2}) \\
&&~~=W_{{\rm EPR}}\biglb((\beta _{1}+\beta _{2})/\sqrt{2}-\alpha ,(\beta
_{1}-\beta _{2})/\sqrt{2}\bigrb)\;.  \nonumber
\end{eqnarray}
The classical signal that we seek is retrieved by homodyne detection,
which measures the analogues of position and momentum for the sum and
difference fields $(\beta_1,\beta_2)$. For ideal homodyne detection
the resulting outcomes are distributed according to
\begin{equation}
p_{\beta |\alpha}={\frac{2e^{2r}}{\pi }}\exp (-2e^{2r}|\beta -\alpha /
\sqrt{2}|^{2})\;,
\end{equation}
where $\beta =\beta _{1R}+i\beta_{2I}$ and represents a highly peaked
distribution about the complex displacement $\alpha /\sqrt{2}$. For large
squeezing parameter $r$ this allows us to extract the original signal
$\alpha $ which we choose to be distributed as
\begin{equation}
p_{\alpha }={\frac{1}{\pi \sigma ^{2}}}\exp (-|\alpha |^{2}/\sigma ^{2})\;.
\end{equation}
Note that this displaced state has a mean number of photons given by
\begin{equation}
\bar{n}=\sigma ^{2}+\sinh ^{2}r\;.  \label{nbar}
\end{equation}

In order to compute the quantity of information that may be sent through
this dense-coding channel we note the unconditioned probability for the
homodyne statistics is given by
\begin{equation}
p_\beta={\frac{2}{\pi(\sigma^2 +e^{-2r})}} \exp\left(\frac{-2|\beta|^2}
{\sigma^2 +e^{-2r}}\right) \;.
\end{equation}
The mutual information describing the achievable information throughput of
this dense-coding channel is then given by
\begin{eqnarray}
H^{{\rm dense}}(A:B) &=& \int d^2\beta\, d^2\alpha\,
p_{\beta|\alpha}p_\alpha
\ln\left( {\frac{p_{\beta|\alpha}}{p_\beta}}\right)  \nonumber \\
&=& \ln(1+\sigma^2e^{2r}) \;.
\end{eqnarray}
For a fixed $\bar n$ in Eq.~(\ref{nbar}) this information is optimized when
$\bar n = e^r\sinh r$, i.e., when $\sigma^2 = \sinh r\cosh r$ so yielding a
dense-coding capacity of
\begin{equation}
C^{{\rm dense}} = \ln (1+\bar n+\bar n^2) \;,  \label{dc}
\end{equation}
which for large squeezing $r$ becomes
\begin{equation}
C^{{\rm dense}} \sim 4 r \;.  \label{dclarger}
\end{equation}

How efficient is this dense coding in comparison to single channel
coding? Let us place a `common' constraint of having a fixed mean
number of photons $\bar n$ which can be modulated. The maximal channel
capacity as given in Eq.~(\ref{max_chan}) and substituting
$\bar n = e^r \sinh r$ gives
\begin{equation}
C\sim 2 r \;,
\end{equation}
for large squeezing $r$. This is just one-half of the asymptotic
dense coding mutual information, see Eq.~(\ref{dclarger}). Thus
asymptotically, at least, the dense coding scheme allows twice as
much information to be encoded within a given state, although it has
an extra expense (not included within the simple constraint $\bar n$)
of requiring shared entanglement.

It is worth noting that this dense coding scheme does {\it not\/} always
beat the optimal single channel capacity. Indeed, for small squeezing
it is worse. The break-even squeezing required for dense coding to equal
the capacity of the optimal single channel communication is
\begin{equation}
r_{{\rm break-even}} \simeq 0.7809 \;,
\end{equation}
which corresponds to roughly 6.78 dB of two-mode squeezing or to
$\bar n \simeq 1.884$. This break-even point takes into account the
difficulty of making highly squeezed two-mode squeezed states. No
similar difficulty has been factored into making ideal number states
used in the benchmark scheme with which our dense coding scheme is
compared.

A fairer comparison is against single-mode coherent state communication
with heterodyne detection. Here the channel capacity is given
by Eq.~(\ref{C_coh}) and is {\it always\/} beaten by the optimal dense
coding scheme described by Eq.~(\ref{dc}).

An improvement on coherent state communication is squeezed state
communication with a single mode. The channel capacity of this
channel is given by Eq.(\ref{C_sq}) and is beaten by the dense
coding scheme of Eq.~(\ref{dc}) for $\bar n > 1$, i.e., the
break-even squeezing required is
\begin{equation}
r^{{\rm sq}}_{{\rm break-even}} \simeq 0.5493 \;,
\end{equation}
which corresponds to 4.77 dB.

Note that this cv protocol should allow for high efficiency, {\it
unconditional\/} transmission with encoded information sent every
inverse bandwidth time. This situation is in contrast to
implementations that employ weak parametric down conversion, where
transmission is achieved {\it conditionally\/} and relatively
rarely. In fact, \textcite{KM96} obtained rates of only 1 in
$10^7$ per inverse bandwidth time \cite{HW00}. By going to strong
down conversion and using cv entanglement, much higher efficiency
should be achievable.

It should be noted that one feature about dense coding is our assumption
that the shared entanglement may be sent off-peak. This implicitly assumes
that the entanglement can be stored (possibly for long periods of time).
Continuous-variable entanglement has now been shown to be easy to create
in collective atomic states and to be efficiently transformable back and
forth to optical entanglement \cite{BJ01} (see Sec.~\ref{qmemory}).
Thus, although the storage
times are still very small, there is potential for a high-bandwidth
technology here.

\subsection{Quantum error correction}\label{qerrorcorr}

Let us now proceed with a few remarks on
an alternative method for the reliable transmission of
quantum information, which is not based on shared entanglement
such as quantum teleportation combined with
entanglement distillation
(the latter is the subject of Sec.~\ref{entdistillation}).
This method is called quantum error correction \cite{Shor2}.

In a quantum error correction scheme used for communication purposes,
quantum states are sent directly through a potentially noisy
channel after encoding them into a larger system that contains
additional auxiliary subsystems.
When this larger system is subject to errors during its
propagation through the quantum channel, under certain circumstances
these errors can be corrected at the receiving station and the
input quantum state can be retrieved, in principle, with unit fidelity.

For discrete variables, a lot of theoretical work on quantum error
correction has been done, for example in Refs.~\cite{Shor2,Cald2,KnillPRA}.
Shortly after the proposal for the realization of cv
teleportation, the known qubit quantum error correction codes were
translated to continuous variables \cite{SamQEC,Lloyd}.
These schemes appeared to require active nonlinear operations
such as quantum non-demolition coupling for the implementation
of the C-NOT gate \cite{SamQEC}. However, later it turned out that
also cv quantum error correction codes can be implemented using only
linear optics and resources of squeezed light \cite{Sam98c}. This was
shown for the nine-wavepacket code \cite{Sam98c}, the analogue of
Shor's nine-qubit code \cite{Shor2}.  An open question is still how
to implement the five-wavepacket code using only linear optics and
squeezed light.

A more robust set of quantum error-correcting codes over
continuous variables was proposed by \textcite{Gottesman00a}.
These codes protect dv quantum information, i.e., states of a
finite-dimensional system, from decoherence by encoding it into
the infinite-dimensional Hilbert space of a cv system (``an
oscillator''). The advantage of this variation over the codes
described above is that the codes from \textcite{Gottesman00a}
allow the effective protection against small ``diffusive'' errors,
which are closer to typical realistic loss mechanisms. In the
codes discussed above, small errors comparable or smaller than
readout errors cannot be corrected and are additive with each
``protective'' operation.

\subsection{Quantum cryptography}\label{qcryptsec}

In this section, we give an overview
of the various proposals of
continuous-variable quantum cryptography
(or quantum key distribution).
We further discuss the absolute theoretical
security and the verification of experimental
security of cv quantum key distribution.
Finally, we will conclude this section with a few remarks
on quantum secret sharing with continuous variables.

\subsubsection{Entanglement-based versus prepare and measure}

For qubit-based quantum cryptography there have been two basic
schemes. Those involving the sending of states from non-orthogonal
bases, such as the original ``BB84 protocol'' \cite{BB84}, and
those based on sharing entanglement between the sender and
receiver, such as Ekert's scheme \cite{Ekert}. The protocols
without entanglement may be termed ``prepare and measure''
schemes, where Alice randomly prepares a sequence of
non-orthogonal states to be sent to Bob and Bob measures these
states in a randomly chosen basis. In general, the
entanglement-based schemes, without Ekert's approach using Bell
inequalities, can also be interpreted as state preparation at a
distance. Due to the entanglement, the states nonlocally prepared
at the receiving station should be correlated to the sender's
states which are measured in a randomly chosen basis. The
entanglement-based schemes are then equivalent to schemes such as
the BB84 protocol \cite{BennettMermin92}.

Conversely, it has been shown that the presence of entanglement in
the quantum state {\it effectively} distributed between Alice and
Bob is a necessary precondition for any secure quantum key
distribution protocol \cite{Curty}. In this sense, the notion of
entanglement can be recovered in the prepare and measure schemes
which do not appear to rely upon entanglement. The crucial point
is that the correlations given by the classical data of Alice's
and Bob's measurements, described by a joint classical probability
distribution $P(A,B)$, must not be consistent with a separable
state \cite{Curty}. This requirement is independent of the
particular physical implementation, where usually the prepare and
measure schemes seem more practical than those based on the
distribution of entanglement. Thus, a first test of secure quantum
key distribution is to check for (optimal) entanglement witnesses
(observables that detect entanglement), given a set of local
operations and a corresponding classical distribution $P(A,B)$
\cite{Curty}. An example for such an entanglement witness would be
the violation of Bell inequalities as in Ekert's scheme
\cite{Ekert}. However, even if there are no such violations, a
suitable entanglement witness to prove the presence of quantum
correlations may still be found. In the cv case, a particularly
practical witness is given by the Duan criterion in
Eq.~(\ref{Duan3}) or Eq.~(\ref{generalDuan}), based solely upon
efficient homodyne detection. Now bearing in mind that {\it any}
secure quantum key distribution scheme must rely upon the
effective distribution of entanglement, a similar categorization
into prepare and measure schemes and those based on entanglement
can be made for the various proposals of cv quantum cryptography.

\subsubsection{Early ideas and recent progress}

The schemes that do not rely on entanglement are mostly based on
alphabets involving (non-orthogonal) coherent states as the signal
states. For example, \textcite{Mu96} utilize four coherent states
and four specific local oscillator settings for the homodyne
detection, enabling the receiver to conclusively identify a bit
value. \textcite{Huttner95} use generalized measurements (POVM's)
instead, which may sometimes yield inconclusive results for a bit
value encoded in weak coherent states. The scheme of Huttner {\it
et al.} is actually a combination of the BB84 \cite{BB84} and
``B92'' \cite{B92} qubit protocols, the latter of which requires
just two arbitrary non-orthogonal states. The basic idea behind
this combination is to make the two states in each pair of basis
states, which are orthogonal in BB84, non-orthogonal instead as in
B92. By using non-orthogonal states in each pair, one gets the
additional advantage of the B92 protocol, namely, that an
eavesdropper (``Eve'') cannot deterministically distinguish
between the two states in each basis. The usual disadvantage of
not being able to create single-photon states, but rather weak
coherent-state pulses (where pulses on average contain less than
one photon), is then turned into a virtue. How a receiver
optimally distinguishes between two coherent signal states for
these coherent-state schemes was shown by \textcite{Banaszek99} to
be possible using a simple optical arrangement.

The use of squeezed states rather than coherent states was
investigated by \textcite{Hillery00}. His analysis of security,
explicitly including the effects of loss, is in some sense
realistic, though ignoring collective attacks.\footnote{commonly,
a distinction is made between three different classes of attacks
\cite{Norbert}: in an individual attack, each signal is coupled to
a probe, and each probe is measured independently of the others,
whereas in a collective attack, several probes are collected and
measured jointly; the most general attack is the coherent attack
in which many signals are coupled to many probes followed by a
collective measurement of the probes.} In addition, two kinds of
eavesdropper attack are studied: man-in-the-middle (or
intercept-resend) measuring a single quadrature; and quantum-tap
using a beam splitter after which again only a single quadrature
is measured. Hillery found that losses produce a significant
degradation in performance, however, he suggested that this
problem could be ameliorated by pre-amplification.

The entanglement-based quantum cryptographic schemes within the
framework of cv quantum optics rely on the correlations of the
quadratures of two-mode squeezed states. \textcite{Cohen97}
considered the idealized case with an unphysical infinite amount
of squeezing to give perfect correlations. More realistically,
\textcite{Pereira00} (which we note was first circulated as a
preprint in 1993) considered `cryptography' based on finitely
squeezed two-mode light beams (their paper described a scheme more
reminiscent of dense coding than of a standard quantum
cryptographic protocol, since it is based on preshared
entanglement and the transmission of one half of the entangled
state).

\textcite{Ralph00a} has considered cv quantum cryptography in two
variations: first, a scheme where the information is encoded onto
just a single (bright) coherent state, and second, an
entanglement-based scheme, where the bit strings are impressed on
two (bright) beams squeezed orthogonally to each other before
being entangled via a beam splitter (i.e., becoming entangled in a
two-mode squeezed state). In assessing these schemes, Ralph
considered three non-collective attacks by Eve. The first two
involved the eavesdropper acting as man-in-the-middle; in one by
measuring a fixed quadrature via homodyne detection and in the
other by measuring both quadratures via heterodyne detection [or
an Arthurs-Kelly type double homodyne detection \cite{Arth}] and
reproducing the signal based on the measured values. The third
used a highly asymmetric beam splitter as a quantum `tap' on the
communication channel after which simultaneous detection of both
quadratures [again \'{a} la Arthurs-Kelly \cite{Arth}] was used to
maximize information retrieval.

In his former, entanglement-free scheme, Ralph found that the
third of his three eavesdropping strategies allowed Eve to obtain
significant information about the coherent state sent with only
minimal disturbance in the bit-error rate observed between Alice
and Bob. Thus, this first scheme proved inferior in comparison to
normal qubit scenarios. By contrast, his entanglement-based scheme
apparently gave comparable security to qubit schemes when analyzed
against the same three attacks. Indeed, for this
entanglement-based scheme a potential eavesdropper is revealed
through a significant increase in the bit-error rate (for a sample
of data sent between Alice and Bob). Ralph has also considered an
eavesdropping strategy based on quantum teleportation and shows
again that there is a favorable trade-off between the extractable
classical information and the disturbance of the signals passed on
to the receiver \cite{Ralph00b}. We note, however, that enhanced
security in this entanglement-based scheme requires high levels of
squeezing and low levels of loss in the channel. Ralph's latter
work \cite{Ralph00b} includes an analysis of losses.

\textcite{Reid00} has considered a similar scheme, exploiting the
``Heisenberg correlations'' [Eq.~(\ref{EPRcondition})] between the
modes of a two-mode squeezed state. In fact, this scheme is
directly analogous to Ekert's qubit scheme \cite{Ekert}, where the
protection against Eve is provided by Alice and Bob being able to
observe a Bell inequality violation. The security analysis was
limited to studying a quantum-tap based attack using a beam
splitter and measurement of a single quadrature. In addition, like
Hillery and Ralph, Reid includes losses in her analysis.

Finally, we note two other theoretical works by
\textcite{Silberhorn00} (an entanglement-based protocol) and by
\textcite{GrangierQCR} (a protocol solely relying upon the
non-orthogonality of coherent states). Very recently, the latter
scheme based on coherent states was implemented experimentally
\cite{cvcryptoNature03}. It is also demonstrated in
Ref.~\cite{cvcryptoNature03} that this protocol is, in principle,
secure for any value of the line transmission rate. Initially, a
line transmission below $50\%$, corresponding to line loss above 3
dB, was thought to render secure key distribution impossible.
Conversely, a scheme with line loss $\leq 3$ dB was considered
secure (to some extent), because the no-cloning bound for coherent
states (Sec.~\ref{qcloningsec}) prevents Eve from obtaining better
signals than Bob (when Eve replaces the lossy channel by a perfect
one and employs beam-splitter based cloning of the coherent
signals as the supposedly optimal eavesdropping strategy). As for
the existence of secure schemes beyond the 3 dB loss limit, one
should realize that the entanglement of a cv resource (two-mode
squeezed states), though being degraded, never vanishes completely
for any degree of the loss \cite{Duan,Kimblecrit}.\footnote{note
that this statement no longer holds true when excess noise is
present as well \cite{Duan,Kimblecrit}. Correspondingly, in the
presence of any finite excess noise, a loss limit for secure key
distribution does exist for any coherent-state based cv scheme
\cite{Hirano04}.} In other words, the necessary precondition for
secure key distribution according to the theorem of
\textcite{Curty}, namely the presence of quantum correlations, can
be satisfied for any line loss. Let us briefly discuss how this
potential security of schemes beyond 3 dB loss may be exploited,
especially by utilizing classical techniques.

The information-theoretic condition for
secure communication, i.e., for enabling extraction
of a secure key using privacy amplification
\cite{BennettCrepeau95}
and error correction techniques
\cite{BrassardSavail94},
is given by the following relation for the mutual information
[Eq.~(\ref{mutualinfodef})] between the participants,
\begin{eqnarray}\label{seccommcond}
I(A:B) > {\rm max}\{I(A:E),I(E:B)\}\;.
\end{eqnarray}
In other words, the mutual information between Alice and Bob,
$I(A:B)$, must exceed the information that either of them shares
with Eve. For losses beyond 3 dB, the condition $I(A:B) > I(A:E)$
is always violated using the classical standard techniques.
However, there are methods to beat this 3 dB loss limit. One such
method is using, in addition to the classical techniques,
entanglement purification and quantum memories, which are both
presently not available in a feasible form (see
Sec.~\ref{entdistillation} and Sec.~\ref{qmemory}). Alternatively,
one may use a ``reverse reconciliation'' protocol, which is the
method advertised by \textcite{cvcryptoNature03}. It enables, in
principle, security of the scheme in Ref.~\cite{cvcryptoNature03}
for arbitrarily small line transmission rate. Reverse
reconciliation basically means that Alice tries to guess what was
received by Bob instead of a protocol where Bob guesses what was
sent by Alice. Another promising method to beat the 3 dB loss
limit is based on a post-selection procedure
\cite{SilberhornNorbert02}. The implementation of error correction
techniques in this scheme might be less demanding than in the
scheme of \textcite{cvcryptoNature03}. As for the signals in the
post-selection based scheme of \textcite{SilberhornNorbert02},
like in the scheme of \textcite{cvcryptoNature03}, simple coherent
states suffice. Note that for the above schemes, although it seems
likely that the cloning-based beam splitting attack is the optimal
attack by Eve, an absolute proof of security would require
analyzing more general attacks by Eve [including non-Gaussian
attacks; see Ref.~\cite{GrosshansCerf}, where it is shown that for
the scheme of \textcite{cvcryptoNature03}, under certain
reasonable assumptions, the individual Gaussian attack is optimal,
being superior to any non-Gaussian coherent attack; note that this
analysis excludes the alternative protocol of
\textcite{SilberhornNorbert02} based on post-selection].

\subsubsection{Absolute theoretical security}

 From single-wavepacket non-collective attacks considered above there has been
great progress recently for cv quantum cryptography in the
detailed proof of absolute theoretical security for one scheme
\cite{Gottesman00a,Gottesman00b}. This scheme is the cv analogue
of the original BB84 scheme. Following the Shor and Preskill
\cite{Shor00} proof for absolute security for the original qubit
proposal, \textcite{Gottesman00b} have generalized the proof.

The key theoretical construct is to embed the communication into
the context of quantum error correction codes. These are not actually
needed to run the protocol, but greatly simplify the proof.
Then given provable bounds to the quantity of information the eavesdropper
can have about the key,
classical error correction codes and classical privacy
amplification are used to reduce this quantity by any desired amount.
This works within some bounds of information
captured by the eavesdropper.
Further, imperfect resources may be treated as a channel defect (or as
an effect of eavesdropping) and so are also easily included.

In the protocol considered, a signal is sent as a squeezed state
with either positive or negative squeezing (which corresponds to
squeezing around conjugate quadratures). It is proved that if the
noise in the quantum channel is weak, squeezed signal states of
just 2.51 dB are sufficient in principle to ensure the protocol's
security \cite{Gottesman00b}. For non-squeezing based
coherent-state schemes, such a proof of unconditional security is
not available yet [see \textcite{GrosshansCerf}].

Heuristically, it appears that the original rough and ready reasoning
of security based on single-shot non-collective attacks really does
impart absolute security. This suggests strongly that the protocols
discussed previously will be found to be similarly absolutely
secure when enhanced or supplemented by classical error correction and
privacy amplification.

Remaining issues appear to be:

\noindent
1.) Re-analysis of this proof in a broadband context. In particular, can
the protocol be run in a $cw$ manner or do complications occur which
necessitate pulsed operation. For example, in $cw$ operation the signal
switching limitations must be accounted for in addition to limitations
in the detection process. The answers to this would have a sizable
impact on the potential bit-rates available.

\noindent
2.) Attempts to use the Shor-Preskill and Gottesman-Preskill approach
to try to complete the proofs of absolute theoretical security for the
various schemes considered previously. Detailed
criteria could then be established for each protocol. This analysis could
be of potential benefit by providing significant flexibility and hence
allow for resolution of various implementation-related design tensions.

\noindent
3.) Experimental verifiability of the claims of absolute security.
This last point will be considered now.

\subsubsection{Verifying experimental security}

So, we have seen that there are already approaches to theoretical proofs
for absolute security. Unfortunately, such theoretical proofs must be treated
somewhat skeptically. Questions must still be asked about how the
theoretical ideas were implemented. Were extra Hilbert-space dimensions
`written' into during the sending or receiving processes by Alice and
Bob. It appears that the only acceptable approach to truly resolve this
problem is through experimental criteria. One way of thinking about this
is in terms of an arms race. We have been hurriedly building the defenses,
but perhaps neglected some subtle loopholes because of unintended
mismatches between ideal conceptualization and actual realization. The
question remains: can an eavesdropper find a way through our defenses?
To find out, it makes most sense to take seriously the position of
devil's advocate, but in the laboratory, and work towards serious
eavesdropping scenarios in order to put the intended ideally secure
schemes through their paces.

To that end a natural first approach for the eavesdropper (in the absence
of a full quantum computer) would be to consider an asymmetric cloning
strategy, whereby as little or as much information gain versus
disturbance could be produced.
It should be noted that Ralph suggests using teleportation as an
eavesdropping strategy \cite{Ralph00a}.
This strategy deserves more consideration,
but it unnecessarily limits the eavesdropper to non-collective attacks.
By contrast, general cloning strategies should encompass the
same performance, but without imposing this restriction.

\textcite{Cerf00c} have applied the work on optimal asymmetric
cloning to the question of eavesdropping on Gaussian channels. For
an individual attack based on measuring a random quadrature, the
quantum information gain versus disturbance was investigated. They
showed that the information gained by the eavesdropper was, in
this case, equal to that lost by the receiver. This sort of
analysis forms a basis for experimentally implementable
verification schemes. The immediate further work here is to
convert the quantum circuits into realizable quantum optics
hardware. This translation is considered in Sec.~\ref{qcloningsec}
on the cloning of cv quantum states.

In summary,
there is now one theoretically proven secure quantum cryptographic
scheme involving quantum continuous variables \cite{Gottesman00b}.
It seems likely that those
schemes which appear secure based on individual attacks will be shown to be
generally secure in a similar manner. If true this would give
freedom in the approaches taken to implement any final scheme. Questions
still remain about the translation of theoretical protocols to real
implementations and whether new loopholes will not be created during this
phase.

\subsubsection{Quantum secret sharing}\label{qsecretsharing}

Quantum secret sharing can be thought of as
a multi-party generalization of quantum cryptography
where a message is not only protected
against potential eavesdroppers.
In addition, the relevant information can only be retrieved
from several people who collaborate.
The first quantum secret sharing scheme was proposed
for qubits using GHZ states as an entanglement resource
\cite{Hillqsecret}.
The GHZ states are used to split information in such a way
that if one is in possession of all of the subsystems,
the information can be recovered, but if one has only some
of the subsystems, it cannot.
This statement applies both to classical and to quantum
information \cite{Hillqsecret}. In the former case,
a key can be established between all participants
and using the key requires all participants working together.
An eavesdropper would introduce errors and could be detected.
In the latter case, for example, a qubit can be recovered
after its quantum information has been split into two or more
parts (obviously, only the former scenario is a multi-party extension
of what is known as quantum key distribution).

In the context of continuous variables, the analogue of the qubit
GHZ state, the maximally entangled $N$-mode state $\int
dx\,|x,x,\ldots ,x\rangle$ is indeed suitable for quantum secret
sharing. We know that this state is an eigenstate with total
momentum zero and all relative positions $x_i-x_j=0$
$(i,j=1,2,\ldots ,N)$. These correlations may be similarly
exploited as the two-mode correlations in the entanglement-based
quantum cryptography schemes. In fact, their exploitation is
equivalent to the two-party sender-receiver scenario when all
participants except for the sender team up and share the
information about local momentum measurements to yield a total
``receiver momentum''. This would enable one to secretly share
classical information protected against eavesdropping. Of course,
in a more realistic scenario, the finite squeezing that affects
the quantum correlations of the cv multi-party entangled states
[Eq.~(\ref{PVLcorrfamily})] must be taken into account.

Continuous-variable secret sharing of quantum information was
proposed by \textcite{Tyc}. In their scheme, as opposed to the
communication scenario considered by \textcite{Hillqsecret}, the
quantum information is to be shared {\it locally} and only
sufficiently large (but arbitrary) subgroups of all the
participants can have access to the secret quantum information.
The multi-mode entangled states used in the scheme of
\textcite{Tyc} are also producible with squeezed light and beam
splitters. It was already demonstrated experimentally how a
coherent state can be shared using cv tripartite entanglement,
where any two of three parties can reconstruct the state to some
extent, but a single party cannot \cite{Lance}.

\subsection{Entanglement distillation}\label{entdistillation}

In order to transfer quantum information reliably
in the presence of loss, quantum teleportation
must be combined with entanglement distillation protocols.
The common security proofs for quantum cryptography
are based on entanglement distillation too.

In general, the term ``entanglement distillation'' refers to any
procedure that aims at distilling from a particular number of
imperfectly entangled states a smaller number of better entangled
states using local operations and classical communication.
Commonly, a distinction is made between distillation schemes for
purifying {\it mixed} entangled states after their two halves have
been distributed through noisy channels [``entanglement
purification'' \cite{Benn2}] and those schemes which concentrate a
number of {\it pure} nonmaximally entangled states into a smaller
number of better entangled pure states [``entanglement
concentration'' \cite{Benn3}].

As for the concentration of pure nonmaximally entangled states,
there are various approaches \cite{Benn3,Nielsen2,Nielsen,Bose1}.
For example, entanglement swapping may serve as an entanglement
concentration protocol capable of turning two copies of a
nonmaximally entangled state into one maximally entangled copy
with nonzero probability \cite{Bose1}. The original proposal of
entanglement concentration included the ``Schmidt projection'' and
the ``Procrustean'' methods \cite{Benn3}.

The Schmidt projection method requires at least two nonmaximally
entangled pairs and becomes efficient for large numbers of pairs.
It is based on a collective measurement [of the ``Hamming weight''
\cite{Mosca}] of all qubits at one side, projecting all pairs onto
a subspace spanned by states having a common Schmidt coefficient.
The measurement result is then classically communicated to the
other side (alternatively, the same collective measurement
performed at the other side would yield the same result and make
classical communication dispensable). This method also works for
dimensions $d>2$ \cite{Benn3}. In the asymptotic limit, turning
the total state vector of $n$ nonmaximally entangled input pairs
into that of $m$ maximally entangled output pairs, can be
described via the majorization criterion for {\it deterministic}
entanglement transformations \cite{Nielsen2,Nielsen}. In
entanglement concentration, we have $m<n$.

Another method for entanglement concentration is the so-called
``Procrustean'' method which represents a filter operation applied
to just a single copy of a nonmaximally entangled state
\cite{Benn3}. With some nonzero probability (even for higher
finite dimensions) a successful filter operation leads to maximum
entanglement.

So what is known about the distillation of continuous-variable
entangled states? As for the concentration of entanglement, we
have seen in Sec.~\ref{entswappsec} that cv entanglement swapping
based on Gaussian (quadrature) Bell measurements does not lead to
an enhancement of the initial entanglement \cite{Parker}. In fact,
the output entanglement becomes even worse than that of the
inputs. As opposed to the entanglement swapping with qubits
\cite{Bose1}, there is no probabilistic element in cv entanglement
swapping which may sometimes lead to better and sometimes to worse
entanglement. An example of probabilistic entanglement
concentration of a single copy of a pure Gaussian two-mode
squeezed state into a more highly entangled non-Gaussian (but
still infinite-dimensional) state via non-Gaussian operations
(subtracting photons) was presented by \textcite{Opatr}.

As for the general distillation of entanglement including
purification, a ``continuous-variable'' protocol was proposed by
Duan {\it et al.} based on local photon number quantum
non-demolition measurements \cite{Duanpurif1}. The entanglement of
bipartite Gaussian states can be both concentrated and purified
using this scheme \cite{Duanpurif1}. Its applicability to all
bipartite Gaussian states was proven by \textcite{Gezapurif} using
the reduction criterion. In the pure-state case, the scheme
corresponds to the Schmidt projection method. However, though
feasible \cite{Duanpurif2}, its experimental realization is very
difficult. Moreover, in this scheme, the distilled entangled
states end up in a finite-dimensional Hilbert space. Another cv
entanglement concentration protocol based on a cross Kerr
interaction was proposed by \textcite{Fiurasekdist2}.

How to distill Gaussian entangled states in a feasible way to
obtain better Gaussian or {\it continuous-variable} entanglement
is a very subtle question. In particular, recently, it was proven
using Gaussian CPTP maps that entanglement distillation within the
class of Gaussian states by applying Gaussian operations (which
are those particularly feasible such as beam splitting, homodyne
detection, etc.) is impossible
\cite{Eisertdist,Fiurasekdist,Giedkedist}. In a
``continuous-variable'' distillation scheme, at some stage
non-Gaussian states, and hence non-Gaussian operations, must be
part of the protocol. A possible approach to this is to apply a
first non-Gaussian step, for instance, via a measurement that
discriminates between the vacuum state and states containing
photons. Subsequently, further Gaussian operations may be used.
Through this kind of protocol, proposed by \textcite{Browne} and
\textcite{Eisertdist2}, cv entangled states can be distilled into
highly entangled approximately Gaussian states.

\subsection{Quantum memory}\label{qmemory}

A way of storing continuous quantum information is a crucial component
of a fully integrated technology. In order to be able to store such
information for extended periods, it seems clear that any suitable scheme
will require a way of storing it in atomic states. A natural way of
achieving this is via the collective spin of an atomic ensemble,
as discussed in Sec.~\ref{polspinrep}. If the
ensemble is highly polarized, then small excursions of the collective spin
away from some fixed axis will mimic the phase-space structure of a
harmonic oscillator. As the size of the ensemble increases, the patch
approximating an ideal `flat' phase-space increases also.

A beam-splitter like coupling between optical Stokes
operators and the collective atomic spin can be achieved
for strongly polarized off-resonant
coupling. In particular, for light propagating along the $z$-axis, the
coupling is well described by the Jaynes-Cummings model. For a sufficiently
off-resonant interaction, no population transfer will occur. Thus, only
second-order transitions can produce any effect, leading to an effective
Hamiltonian \cite{Brune92}
\begin{equation}
\hat H_{\rm eff}\propto \hat S_z \hat F_z \;.
\end{equation}
It has been noted that this yields a QND probe of $\hat F_z$ of the
atomic sample \cite{Happer67}.

In the limit of small interaction times, the equations of motion are
\begin{eqnarray}\label{atomicQND}
\hat F_x^{\rm (out)} &\simeq& \hat F_x^{\rm (in)}
-a \hat S_z^{\rm (in)}\hat F_y^{\rm (in)} \nonumber \\
\hat F_y^{\rm (out)} &\simeq& \hat F_y^{\rm (in)}
+a \hat S_z^{\rm (in)}\hat F_x^{\rm (in)} \\
\hat F_z^{\rm (out)} &\simeq& \hat F_z^{\rm (in)} \nonumber \;,
\end{eqnarray}
and similar equations with $\hat S_j$ interchanged with $\hat F_j$
throughout. The constant $a$ depends on several parameters
among which are, for example,
the spontaneous emission rate of the atoms and the transverse cross
section and the detuning of the light beam.
A beam-splitter like coupling may be obtained, for example,
between $\hat F_y$ and $\hat S_z$ when the atomic sample is highly polarized
along the $x$-axis and for a strongly $x$-polarized optical beam.
By having the
beam propagate through the sample along different directions, any suitable
beam-splitter coupling between components may be made. This is
the basis for a series of beautiful cv experiments
involving spin-squeezed atomic samples \cite{Polzik02}.
We will briefly discuss these experiments in
Sec.~\ref{expwcvgenforatoms} and Sec.~\ref{expwcvqmemory}.

\section{Quantum Cloning with Continuous Variables}
\label{qcloningsec}

In this section, we investigate the consequences of the famous
quantum no-cloning theorem, independently found by
\textcite{Woott} and by \textcite{Dieks}, for continuous quantum
variables. As mentioned in Sec.~\ref{qcryptsec}, a potential
application of cv quantum cloning is to implement eavesdropping
strategies for cv quantum cryptography.

\subsection{Local universal cloning}

We now consider the possibility of approximately
copying an unknown quantum state at a given location using
a particular sequence of unitary transformations (a quantum circuit).
Entanglement as a potentially nonlocal resource is therefore
not necessarily needed, but it might be an ingredient at the intermediate
steps of the cloning circuit.

\subsubsection{Beyond no-cloning}

The no-cloning theorem, originally derived for qubits, in general
forbids exact copying of unknown nonorthogonal (or simply arbitrary)
quantum states \cite{Woott,Dieks}.
The first papers that went ``beyond the no-cloning theorem''
and considered the possibility of approximately copying nonorthogonal
quantum states initially referred to qubits and later, more generally, to
finite-dimensional systems \cite{Buzek,gisinmassar,Bruss1,Bruss2,Werner2}.
Based on these results, a cloning experiment has been proposed
for qubits encoded as single-photon states
\cite{Simonclon}, and two other optical qubit
cloning experiments have been realized already \cite{Martini,Wan}.

What about the situation when we have $N$ quantum systems of
arbitrary dimension each prepared in the same, but arbitrary input
state and we want to convert them into $M$ ($M> N$) systems that
are each in a quantum state as similar as possible to the input
state? By using an axiomatic approach,\footnote{It is pointed out
by \textcite{Werner2} that the ``constructive'' approach (the
coupling of the input system with an apparatus or ``ancilla''
described by a unitary transformation, and then tracing out the
ancilla) consists of completely positive trace-preserving (CPTP)
operations. Therefore, any constructively derived quantum cloner
is in accordance with the axiomatic definition that an admissible
cloning machine must be given by a linear CPTP map. Conversely,
any linear CPTP map can be constructed via the constructive
approach.} Werner was able to derive the cloning map that yields
the optimal $N$ to $M$ cloning fidelities for $d$-dimensional
states \cite{Werner2}
\begin{eqnarray}\label{Wernersresult}
F=\frac{N(d-1)+M(N+1)}{M(N+d)}\equiv F_{{\rm clon},N,M}^{{\rm univ},d}\;.
\end{eqnarray}
For this optimum cloning fidelity, we use the superscript ``univ''
to indicate that any $d$-dimensional quantum state is universally copied
with the same fidelity.
Let us now further investigate universal cloning machines,
for both discrete and continuous variables.

\subsubsection{Universal cloners}

A universal cloner is capable of optimally copying arbitrary
quantum states with the same fidelity independent of the
particular input state. Bu\u{z}ek and Hillery's universal 1 to 2
qubit cloner \cite{Buzek} leads to two identical copies
$\hat\rho_a$ and $\hat\rho_b$. It is a symmetric universal cloner.
An asymmetric universal 1 to 2 cloner would distribute the quantum
information of the input state {\it unequally} among the two
output states. The fidelity of one output state is then better
than the optimum value for symmetric cloning, whereas the fidelity
of the other output state has become worse. Such a potentially
asymmetric cloning device represents a quantum information
distributor that generates output states of the form \cite{Sam}
\begin{eqnarray}\label{1to2distributor}
\hat\rho_a&=&(1-A^2)|s\rangle_{a\,a} \langle s|+
\frac{A^2}{d}\,\mbox{1$\!\!${\large 1}}_a\;,\nonumber\\
\hat\rho_b&=&(1-B^2)|s\rangle_{b\,b} \langle s|+
\frac{B^2}{d}\,\mbox{1$\!\!${\large 1}}_b\;,
\end{eqnarray}
for an arbitrary $d$-dimensional input state
$|s\rangle_a=\sum_{n=0}^{d-1}c_n|n\rangle_a$.
The parameters $A$ and $B$ are related via $A^2+B^2+2AB/d=1$ \cite{Sam}.
The two extreme cases are when the entire quantum information is kept
by the original system ($A=0$) and when it is completely transferred
to the other system ($B=0$).
It follows directly from the {\it covariant} form of the above density
operators that the fidelity of the information transfer is
input-state independent. The second term proportional to
$\mbox{1$\!\!${\large 1}}/d$ in each density operator represents
``noise'' added by the information transfer process \cite{Sam}.

It was shown by \textcite{Sam} that the above quantum information
distributor can be constructed from a single family of quantum
circuits. This kind of quantum circuit was previously used as a
quantum computational network for universal qubit cloning, in
which case it consists of four C-NOT gates pairwise acting on the
input qubit $a$ and two qubits $b$ and $c$ in an entangled state
\cite{Buzek1997}. For arbitrary dimensions, the analogous circuit
can be used with C-NOT operations generalized to $d$ dimensions,
$|n\rangle |m\rangle \rightarrow |n\rangle |n\oplus m\rangle$, and
a corresponding $d$-dimensional entangled state of systems $b$ and
$c$ \cite{Sam}. In a discretized phase space $(x_k,p_k)$
\cite{Buzekdiscrete1,Buzekdiscrete2,Buzekdiscrete3}, the entangled
state has the form $|\chi\rangle_{bc}=A |x_0\rangle_b
|p_0\rangle_c + B (\sum_{k=0}^{d-1} |x_k\rangle_b
|x_k\rangle_c)/\sqrt{d}$, where $|x_0\rangle$ and $|p_0\rangle$
are ``zero-position'' and ``zero-momentum'' eigenstates
respectively. The continuous limit for this state is then obvious,
and its regularized form consists of quadrature squeezed vacuum
states and a two-mode squeezed vacuum state of squeezing $r$
\cite{Sam}. The parameters $A$ and $B$ are then related as
$A^2+B^2+4AB/\sqrt{4+2\sinh^2 2r}=1$ and the C-NOT operations
become conditional shifts in phase space, $|x\rangle |y\rangle
\rightarrow |x\rangle |x+y\rangle$ \cite{SamQEC}. Expressed in
terms of position and momentum operators, the sequence of four
generalized C-NOT operations\footnote{Note that the C-NOT
operation $|n\rangle |m\rangle \rightarrow |n\rangle |n\oplus
m\rangle$ is its own inverse only for qubits ($d=2$). For higher
dimensions, $\hat U_{ab}=\sum_{n,m=0}^{d-1}|n\rangle_{a\,a}\langle
n|\otimes |n\oplus m\rangle_{b\,b} \langle m|$ and $\hat
U^{\dagger}_{ab}$ differ describing conditional shifts in opposite
directions. The same applies to the continuous-variable C-NOT
operation $|x\rangle |y\rangle \rightarrow |x\rangle |x+y\rangle$
\cite{SamQEC}. Therefore, there is a slight modification in the
sequence of four C-NOT's from $d=2$ to $d>2$: $\hat U_{ca}\hat
U^{\dagger}_{ba}\hat U_{ac}\hat U_{ab}$. Making the C-NOT its own
inverse $\hat U=\hat U^{\dagger}$ could be achieved by defining
$|x\rangle |y\rangle \rightarrow |x\rangle |x-y\rangle$
\cite{Alber}.} acting on modes $a$ (the original), $b$, and $c$,
can be written as \cite{Sam,Cerf1}
\begin{eqnarray}\label{4CNOT}
\hat U_{abc}=\exp[-2i(\hat{x}_c-\hat{x}_b)\hat{p}_a]\,
\exp[-2i\hat{x}_a(\hat{p}_b+\hat{p}_c)]\;.
\end{eqnarray}
Here, $\exp(-2i\hat x_k \hat p_l)$ corresponds to a single C-NOT operation
with ``control'' mode $k$ and ``target'' mode $l$ ($l$ shifted
conditioned upon $k$). After applying $\hat U_{abc}$
to mode $a$ and the regularized state of modes $b$ and $c$,
the resulting fidelities of the universal continuous-variable
quantum information distributor in the limit of large squeezing turn out
to be $F=B^2$ for mode $a$ and $F=A^2$ for mode $b$.
Symmetric cloning with $A=B$ then means $A^2=B^2=1/2$ for infinite squeezing
and hence a duplication fidelity of $1/2$ \cite{Sam}.

Similarly, for universal symmetric $N$ to $M$ cloning of arbitrary
continuous-variable states, one obtains the optimum cloning fidelity
\cite{Sam}
\begin{eqnarray}
F_{{\rm clon},N,M}^{{\rm univ},\infty}=\frac{N}{M}\;,
\end{eqnarray}
which is exactly the infinite-dimensional limit $d\to\infty$ of
Werner's result in Eq.~(\ref{Wernersresult}).
This result looks suspiciously classical. In fact, in the continuous limit,
the universal cloner simply reduces to a {\it classical probability
distributor}. For example, the optimum 1 to 2 cloner can be mimicked
by a completely classical device that relies on a coin toss.
 From the two input states of that device, the original input state
and an entirely random state (ideally an infinite-temperature thermal state),
either state is sent to output $a$ and the other one to output $b$ or
vice-versa depending on the result of the coin toss. Then,
on average, with a small overlap between the original input state and the
random state, the two output clones have a cloning fidelity of $1/2$
\cite{Sam}. These observations are further confirmed by the fact that
there is no entanglement between systems $a$ and $b$ at the output
of the universal continuous-variable cloner, as opposed to any universal
finite-dimensional cloner \cite{Sam}.

Let us summarize at this point: we have discussed fidelity boundaries
for universal $N$ to $M$ cloners. These boundaries, the optimal cloning
fidelities, can in fact be attained by means of a single family of
quantum circuits. There is a universal design for these quantum circuits
in any Hilbert space dimension and for a given dimension these circuits
represent universal cloning machines copying arbitrary input states with
the same optimal fidelity. Furthermore, we have seen that the
{\it universal} continuous-variable cloner
is not very interesting, since it
is a purely classical device. Does a continuous-variable cloning machine
possibly become nonclassical and hence more interesting when it is designed
to copy quantum states drawn from a limited alphabet?
We will now turn to this question.

\subsection{Local cloning of Gaussian states}

\subsubsection{Fidelity bounds for Gaussian cloners}

In the first papers that considered continuous-variable cloning,
the set of input states to be copied was restricted to Gaussian states
\cite{Cerf1,Cerf2}. The optimal cloning fidelity for turning $N$ identical
but arbitrary coherent states into $M$ identical approximate copies,
\begin{eqnarray}\label{NMcohstclonfid}
F_{{\rm clon},N,M}^{{\rm coh\,st},\infty}= MN/(MN+M-N)\;,
\end{eqnarray}
was derived by \textcite{Cerf2}. The approach there was to reduce
the optimality problem of the $N\rightarrow M$ cloner to the task
of finding the optimal $1\rightarrow\infty$ cloner, an approach
previously applied to universal qubit cloning \cite{Bruss2}. Let
us briefly outline the derivation for qubits in order to reveal
the analogy with that for coherent states.

The operation of the universal $N\rightarrow M$ qubit cloner can be
characterized by a shrinking factor $\eta_{\rm clon}(N,M)$, shrinking the
Bloch vector of the original input state \cite{Preskillectures}
\begin{eqnarray}
\hat\rho_a^{\rm in}=\frac{1}{2}(\mbox{1$\!\!${\large 1}}_a+
\vec s_a^{\,{\rm in}}\cdot\vec\sigma)\;,
\end{eqnarray}
so that the output density operator of each copy becomes
(for example for $a$)
\begin{eqnarray}
\hat\rho_a^{\rm out}=\frac{1}{2}[\mbox{1$\!\!${\large 1}}_a+
\eta_{\rm clon}(N,M)\,
\vec s_a^{\,{\rm in}}\cdot\vec\sigma]\;.
\end{eqnarray}
The optimal cloners are those with maximum $\eta_{\rm clon}(N,M)\equiv
\bar{\eta}_{\rm clon}(N,M)$.
The derivation of the fidelity boundaries then relies on two facts:
the shrinking factors for concatenated cloners multiply and the optimum
cloning shrinking factor for infinitely many copies
$\bar{\eta}_{\rm clon}(N,\infty)$ equals the shrinking factor for the
optimal quantum state estimation through measurements
$\bar{\eta}_{\rm meas}(N)$ given $N$ identical input states.
This leads to the inequality
$\eta_{\rm clon}(N,M)\,\eta_{\rm clon}(M,L)\leq
\bar{\eta}_{\rm clon}(N,L)$
and also (with $L\to\infty$)
$\eta_{\rm clon}(N,M)\,\bar{\eta}_{\rm clon}(M,\infty)\leq
\bar{\eta}_{\rm clon}(N,\infty)$, which gives the lowest upper bound
\begin{eqnarray}
\eta_{\rm clon}(N,M)\leq \frac{\bar{\eta}_{\rm clon}(N,\infty)}
{\bar{\eta}_{\rm clon}(M,\infty)}=\frac{\bar{\eta}_{\rm meas}(N)}
{\bar{\eta}_{\rm meas}(M)}\;.
\end{eqnarray}
Because of the optimal shrinking factor
$\bar{\eta}_{\rm meas}(N)=N/(N+2)$ due to a measurement
\cite{MassarPopescu}, we obtain
\begin{eqnarray}
\bar{\eta}_{\rm clon}(N,M)=\frac{N}{M}
\frac{M+2}{N+2}\;.
\end{eqnarray}
This result for qubits
yields the correct optimum $N\rightarrow M$ cloning
fidelity given by Eq.~(\ref{Wernersresult}) for dimension $d=2$,
when inserted into
\begin{eqnarray}\label{fidelitywithNinputs}
F&=&\langle\psi_{\theta_0,\phi_0}|\hat\rho_{\rm out}|
\psi_{\theta_0,\phi_0}\rangle=\frac{1}{2}+
\frac{\bar{\eta}_{\rm clon}(N,M)}{2}\nonumber\\
&=&F_{{\rm clon},N,M}^{{\rm univ},2}\;,
\end{eqnarray}
for arbitrary qubit states \cite{Preskillectures}
\begin{eqnarray}
|\psi_{\theta_0,\phi_0}\rangle=
\cos \frac{\theta_0}{2}\, e^{-i\phi_0/2}|0\rangle+
\sin \frac{\theta_0}{2}\, e^{+i\phi_0/2}|1\rangle\;.
\end{eqnarray}
In fact, the resulting fidelities do not
depend on the particular values of $\theta_0$ and $\phi_0$.

An analogous approach for the derivation of the optimum
coherent-state cloning fidelities is based on
the fact that the excess noise variances in the quadratures
due to the cloning process {\it sum up} when an $N\rightarrow L$
cloner is described by two cloning machines,
an $N\rightarrow M$ and an $M\rightarrow L$,
operating in sequence,
$\lambda_{\rm clon}(N,L)=\lambda_{\rm clon}(N,M)+
\lambda_{\rm clon}(M,L)$.
With the optimum (minimal) excess noise variances
defined by $\bar{\lambda}_{\rm clon}(N,L)$,
we find now the largest lower bound
\begin{eqnarray}\label{lambdaineq}
\lambda_{\rm clon}(N,M)\geq \bar{\lambda}_{\rm clon}(N,\infty)-
\bar{\lambda}_{\rm clon}(M,\infty)\;.
\end{eqnarray}
The quantity $\bar{\lambda}_{\rm clon}(N,\infty)$ can be inferred
from quantum estimation theory \cite{Holevo}, because
it equals the quadrature variance of an optimal joint measurement
of $\hat x$ and $\hat p$ on $N$ identically prepared systems,
$\bar{\lambda}_{\rm clon}(N,\infty)=\bar{\lambda}_{\rm meas}(N)
=1/2N$ \cite{Cerf2}.
For instance, the optimal simultaneous
measurement of $\hat x$ and $\hat p$ on a single system $N=1$
yields for each quadrature a variance of
$\bar{\lambda}_{\rm meas}(1)=1/2=1/4+1/4$ \cite{Arth},
corresponding to the intrinsic minimum-uncertainty noise (one unit
of vacuum) of the input state plus one extra unit of vacuum due
to the simultaneous measurement. Reconstructing a coherent state
based on that measurement gives the correct coherent-state input
plus two extra units of vacuum [this is exactly the procedure
Alice and Bob follow in classical teleportation with
an optimal average fidelity of $1/2$ for
arbitrary coherent states, Eq.~(\ref{fid3})].
Since infinitely many copies can be made this way, the optimal
measurement can be viewed as a potential $1\rightarrow \infty$ or,
in general, an $N\rightarrow \infty$ cloner.
In fact, analogously to the qubit case \cite{Bruss2},
the optimal measurement (optimal state estimation) turns out
to be the optimal
$N\rightarrow \infty$ cloner, and  hence $\bar{\lambda}_{\rm
clon}(N,\infty)=\bar{\lambda}_{\rm meas}(N) =1/2N$. This result
combined with the inequality of Eq.~(\ref{lambdaineq})
gives the optimum (minimal) excess
noise induced by an $N\rightarrow M$ cloning process \cite{Cerf2},
\begin{eqnarray}\label{optimumexcess}
\bar{\lambda}_{\rm clon}(N,M)=\frac{M-N}{2MN}\;.
\end{eqnarray}
Inserting this excess noise into Eq.~(\ref{fid4}) with $g=1$ and a
coherent-state input [where $\sigma_x=\sigma_p=1/2+
\bar{\lambda}_{\rm clon}(N,M)$] leads to the correct fidelity
in Eq.~(\ref{NMcohstclonfid}).
Note that this optimal fidelity does not depend on the particular
coherent amplitude of the
input states. Any ensemble of $N$ identical coherent states is cloned
with the same fidelity. The $M$ output clones are in covariant form.
This means the cloning machine can be
considered state-independent with respect to the limited alphabet of
arbitrary coherent states (``it treats all coherent states equally well'').
Of course, this does not hold when the cloner is
applied to arbitrary infinite-dimensional states without any restriction
to the alphabet.
In this sense, the optimal covariant coherent-state cloner is nonuniversal.

When the coherent-state
alphabet is extended to squeezed-state inputs, optimality is provided
only if the excess cloning
noise is squeezed by the same amount as the input state.
However, this requires knowledge about the input state's
squeezing, making the cloner {\it state-dependent when applied to all
Gaussian states}. Yet Gaussian input states with fixed and known
squeezing $r$, of which the coherent-state alphabet is just the special
case $r=0$, are optimally cloned in a covariant
fashion.

To summarize, for arbitrary qubits,
the optimal cloner shrinks the input state's Bloch vector
by a factor $\bar{\eta}_{\rm clon}(N,M)$ without changing its orientation;
the output clones all end up in the same mixed state.
For arbitrary coherent states, the optimal cloner adds an excess noise
$\bar{\lambda}_{\rm clon}(N,M)$ to the input state without changing
its mean amplitude; the coherent-state copies are all in the same mixed
state. {\it In both cases, this ensures covariance and optimality}.

What kind of transformation do we need to achieve optimal coherent-state
cloning? In fact, the Four-C-NOT transformation
in Eq.~(\ref{4CNOT}) can be used
to construct an optimal $1\rightarrow 2$ coherent-state cloner,
covariant under displacement and rotation in phase space \cite{Cerf1}.
The entangled state of modes $b$ and $c$ then has to be
(in our units) \cite{Cerf1}
\begin{eqnarray}\label{4CNOTentstate}
|\chi\rangle_{bc}\propto\int dx\, dy\, \exp(-x^2-y^2)
\,|x\rangle|x+y\rangle\;.
\end{eqnarray}
An alternative, non-entanglement based, optical circuit for the
optimal local $N\rightarrow M$ cloning of coherent states
will be discussed in the next section.

\subsubsection{An optical cloning circuit for coherent states}

So far, we have only discussed the fidelity boundaries
for the $N\rightarrow M$ coherent-state cloner. In general,
finding an optimal cloning transformation and proving
that it achieves the maximum fidelities is a fundamental issue
in quantum information theory.
In the case of coherent states, implementing an
$N\rightarrow M$ symmetric cloning transformation
that attains Eq.~(\ref{NMcohstclonfid})
only requires a phase-insensitive linear amplifier and
a series of beam splitters \cite{localcl,Jaromirlocalcl}.

As the simplest example,
let us focus on coherent-state duplication ($N=1$, $M=2$).
The coherent state to be cloned is given by the
annihilation operator $\hat a_0$, and
an additional ancilla mode is similarly represented
by $\hat a_z$.
The optimal duplication can be implemented in two steps via the
two canonical transformations,
\begin{eqnarray}\label{dupli}
\hat a'_0  =  \sqrt{2} \hat a_0 + \hat a_z^{\dagger}, \qquad
\hat a'_z  =  \hat a_0^{\dagger} + \sqrt{2} \hat a_z, \nonumber\\
\hat a''_0 =  \frac{1}{\sqrt{2}}(\hat a'_0+\hat a_1), \quad
\hat a''_1 =  \frac{1}{\sqrt{2}}(\hat a'_0-\hat a_1),
\end{eqnarray}
where the mode described by $\hat a_1$ is another ``blank''
mode assumed to be in the vacuum state.
These transformations preserve the bosonic commutation rules
for the two clones, modes $0''$ and $1''$.

%Fig20
\begin{figure}[htb]
\vspace{1cm}
\begin{center}
\epsfxsize=0.88 \hsize\leavevmode\epsffile{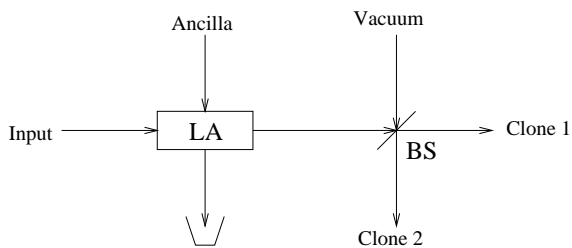}
\end{center}
\caption{Implementation of a $1 \to 2$ continuous-variable cloning machine.
LA stands for linear amplifier, and BS represents a phase-free
50:50 beam splitter.}
\label{fig20}
\end{figure}

As shown in Fig.~\ref{fig20}, the interpretation of these
transformations is straightforward: the first step (which
transforms $\hat a_0$ and $\hat a_z$ into $\hat a_0'$ and $\hat
a_z'$) corresponds to a phase-insensitive amplifier whose (power)
gain $G$ is equal to 2, while the second step (which transforms
$\hat a_0'$ and $\hat a_1$ into $\hat a_0''$ and $\hat a_1''$) is
a phase-free 50:50 beam splitter \cite{capri,Ralphpriv}. As
discussed by \textcite{cave}, the ancilla $z$ involved in linear
amplification can always be chosen such that $\langle\hat
a_z\rangle=0$, so that we have $\langle\hat
a''_0\rangle=\langle\hat a''_1\rangle= \langle\hat a_0\rangle$ as
required. Finally, the optimality of our cloner can be confirmed
from known results on linear amplifiers. For an amplifier of
(power) gain $G$, each quadrature's excess noise variance is
bounded by \cite{cave}:
\begin{equation}\label{amplibound}
\lambda_{LA} \geq (G-1)/4.
\end{equation}
Hence, the optimal amplifier of gain $G=2$ yields
$\lambda_{LA} =1/4$. This leads to
quadrature variances of both clones equal to $1/2$,
corresponding to one extra unit of vacuum due to the cloning
procedure. This one extra unit is indeed the optimal
(minimal) amount according to Eq.~(\ref{optimumexcess})
for $N=1$ and $M=2$. The corresponding optimal fidelity is
$2/3$ [Eq.~(\ref{NMcohstclonfid})].
Let us now turn from local cloning of cv quantum states
to cloning at a distance.

\subsection{Telecloning}\label{telecloningsection}

What about conveying quantum
information via a ``multiuser quantum channel'' (MQC) simultaneously to
several receivers? The no-cloning theorem that generally forbids {\it
perfect} cloning of unknown nonorthogonal quantum states
then also disallows cloning
over a distance. This prevents the MQC from being
able to produce {\it exact} clones of the sender's input state at all
receiving stations. The MQC, however, can provide
each receiver with at least a part of the input quantum information and
distribute {\it approximate} clones with non-unit fidelity \cite{Buzek}.
This cloning at a distance or ``telecloning'' may be
seen as the ``natural generalization of teleportation to the many-recipient
case'' \cite{Murao}.

For qubits, telecloning has
been studied theoretically, first with one input sent to two
receivers \cite{Bruss1}, and more generally, with one input
\cite{Murao} and $N$ identical inputs \cite{Duertele} distributed among
$M$ receivers. The telecloning scenario with one input copy and
$M$ receivers has been extended to $d$-level systems \cite{Murao2}.

Clearly a telecloner needs entanglement as soon as its fidelity
is greater than the maximum fidelity attainable by classical teleportation
$F_{\rm class}$. In fact, for {\it universal} $1\rightarrow M$ qubit cloning
we have $F_{{\rm clon},1,M}^{{\rm univ},2}> F_{\rm class}=2/3$
[Eq.~(\ref{Wernersresult})], whereas
for $1\rightarrow M$ cloning of {\it coherent states}
we have $F_{{\rm clon},1,M}^{{\rm coh\,st},\infty}> F_{\rm class}=1/2$
[Eq.~(\ref{NMcohstclonfid})](for the bounds on classical teleportation,
see Sec.~\ref{telepcritsec}).
Therefore, optimal telecloning cannot be achieved by
``classical telecloning'', i.e., by simply measuring
the input state and sending copies of the classical
result to all receivers.
On the other hand, in the limit $M\to\infty$, both
$F_{{\rm clon},1,M}^{{\rm univ},2}\to F_{\rm class}=2/3$ and
$F_{{\rm clon},1,M}^{{\rm coh\,st},\infty}\to F_{\rm class}=1/2$
which implies that {\it no} entanglement is needed for
infinitely many copies [this observation reflects the previously
discussed relations
$\bar{\eta}_{\rm clon}(N,\infty)=\bar{\eta}_{\rm meas}(N)$ and
$\bar{\lambda}_{\rm clon}(N,\infty)=\bar{\lambda}_{\rm meas}(N)$ with
$N=1$]. Thus, only the optimal telecloning to an infinite number
of receivers is achievable via classical telecloning. {\it Otherwise,
for a finite number of receivers, entanglement is needed.}

The most wasteful scheme
would be a protocol in which the sender locally creates $M$ optimum
clones and perfectly teleports one clone to each receiver using $M$
maximally entangled two-party states \cite{Murao,Murao2}.
A much more economical
strategy is that all participants share an appropriate multipartite
entangled state as a quantum channel.
Such states can be found both for discrete variables
\cite{Bruss1,Murao,Duertele,Murao2} and for continuous variables
\cite{telecl}.

The recipe for building such an MQC for continuous variables
is as follows \cite{telecl}:
first, make a bipartite entangled state by combining two
squeezed vacua with squeezing parameter $r$, where one is
squeezed in $x$ and the other in $p$, at a
phase-free $50:50$ beam splitter.
Then keep one half (say mode 1) and send the other half
together with $M-1$ vacuum modes through an $M$-splitter,
Eq.~(\ref{PVLNsplitt}).
Mode 1 is now given to the sending station and the $M$ output modes
of the $M$-splitter are distributed among the $M$ receivers.
The symmetric $1\rightarrow M$ telecloning protocol
then works similar to the $1\rightarrow 1$ teleportation protocol.
The sender performs a cv Bell measurement
on mode 1 and the input mode to be transferred and sends the
results via classical channels to all the receivers.
Eventually, each receiver can produce an optimal clone
by applying the corresponding phase-space displacements
to his mode.
For coherent-state inputs, the optimal cloning
fidelities $F_{{\rm clon},1,M}^{{\rm coh\,st},\infty}$,
Eq.~(\ref{NMcohstclonfid}) with $N=1$,
are attained by adjusting the squeezing parameter
according to
\begin{eqnarray}\label{PVLconditions}
e^{-2r}=\frac{\sqrt{M}-1}{\sqrt{M}+1}\;.
\end{eqnarray}
Hence, the generation of the MQC requires no more than
two $|10\log_{10}[(\sqrt{M}-1)/(\sqrt{M}+1)]|$ dB squeezed states and
$M$ beam splitters. This is about 7.7 dB
for $M=2$, 5.7 dB for $M=3$, 4.8 dB for $M=4$, and 4.2 dB for $M=5$.
That the squeezing and hence the entanglement approaches zero as $M$
increases is consistent with the convergence of the optimum cloning
fidelity $F_{{\rm clon},1,M}^{{\rm coh\,st},\infty}=M/(2M-1)$
to $F_{\rm class}=1/2$. Conversely,
for optimal $1\rightarrow 1$ quantum teleportation
attaining unit fidelity, infinite squeezing is needed.

More generally, using the above MQC with the squeezing
parameter given by Eq.~(\ref{PVLconditions}),
arbitrary quantum states can be transferred
from a sender to $M$ receivers with
equal minimum excess noise in each output state.
This can enable one, for instance,
to teleport entanglement to all receivers \cite{telecl}.
Further, the protocol based on the MQC forms a cloning circuit
(an optimal one for coherent states)
with no need to amplify the input.

Let us finally emphasize that the cv telecloning scheme discussed
above works without maximum bipartite entanglement (corresponding
to the unphysical case of infinite squeezing), whereas the
existing dv schemes rely on maximum two-party entanglement
\cite{Murao,Duertele,Murao2}. The only known exception is the
$1\rightarrow 2$ qubit telecloner of \textcite{Bruss1} which uses
nonmaximum entanglement.

\section{Quantum Computation with Continuous Variables}
\label{qcompsec}

We now consider the necessary and sufficient conditions for constructing
a universal quantum computer using continuous variables. As an example, it
is shown how a universal quantum computer for the amplitudes of the
electromagnetic field might be constructed using linear optics, squeezers
and at least one further non-linear optical element such as the Kerr effect.

\subsection{Universal quantum computation}

Ordinarily, a universal quantum computer applies ``local''
operations that effect only a few variables at a time (such
operations are called quantum logic gates): by repeated
application of such local operations it can effect any unitary
transformation over a finite number of those variables to any
desired degree of precision \cite{seth1,seth2}.\footnote{This
definition of quantum computation corresponds to the normal
``circuit'' definition of quantum computation as in, e.g.,
\textcite{Deutsch}; \textcite{Yao}.}

However, since an arbitrary unitary transformation over even a
single continuous variable requires an infinite number of
parameters to define, it typically cannot be approximated by any
finite number of quantum operations, each of which would be
specified by a finite number of parameters. At first sight,
therefore, it might seem that quantum computation over continuous
variables would be an ill-defined concept. Despite this
difficulty, it is nonetheless possible to define a notion of
universal quantum computation over continuous variables for
various subclasses of transformations, such as those that
correspond to Hamiltonians that are polynomial functions of the
operators corresponding to the continuous variables: A set of
continuous quantum operations will be termed universal for a
particular set of transformations if one can by a finite number of
applications of the operations approach arbitrarily closely to any
transformation in the set.

Consider a single continuous variable corresponding to the dimensionless
operator $\hat x$, with conjugate variable $\hat p$ satisfying
$[\hat x,\hat p]=i/2$.
We first investigate the problem of constructing Hamiltonians that
correspond to arbitrary polynomials of $\hat x$ and $\hat p$.
% For definiteness, we shall choose the operator ordering so that these
% polynomials are linear combinations of terms like $\hat x^n \hat p^m$.
It is clearly necessary that one be able to apply the Hamiltonians
$\pm \hat x$ and $\pm \hat p$ themselves. In the Heisenberg picture,
applying a Hamiltonian $\hat H$ gives a time evolution for an
operator $\hat A$ as $\hat A(t)=e^{i\hat Ht}A(0)e^{-i\hat Ht}$.
Accordingly, applying the Hamiltonian $\hat x$ for time $t$ takes
$\hat x \rightarrow \hat x$, $\hat p \rightarrow \hat p - \frac{t}{2}$,
and applying $\hat p$ for time $t$ takes
$\hat x\rightarrow \hat x + \frac{t}{2}$, $\hat p\rightarrow \hat p$:
the Hamiltonians $\hat x$ and $\hat p$ have the effect of shifting
the conjugate variable by a constant.

A simple geometric construction allows one to determine what Hamiltonian
transformations can be constructed by the repeated application of
operations from some set. Apply the Hamiltonian $\hat B$ for time
$\delta t$, followed by $\hat A$, $-\hat B$, $-\hat A$, each for the
same time.  Since
\begin{equation}
e^{-i\hat A\delta t} e^{-i\hat B\delta t} e^{i\hat A\delta t}
e^{i\hat B\delta t}
= e^{[\hat A,\hat B]\,{\delta t}^2}
+ O({\delta t}^3)\;, \label{seth_eq1}
\end{equation}
in the limit that $\delta t \rightarrow 0$, the result is the same as
if one had applied the Hamiltonian $-i[\hat A,\hat B]$ for time
${\delta t}^2$.
Similarly, since
\begin{equation}
e^{i\hat A\delta t/2} e^{i\hat B\delta t/2} e^{i\hat B\delta t/2}
e^{i\hat A\delta t/2}
= e^{i(\hat A+\hat B)\,\delta t} + O({\delta t}^3)\;, \label{sum_ham}
\end{equation}
in the limit that $\delta t \rightarrow 0$, the result is the same as
if one had applied the Hamiltonian $\hat A+\hat B$ for time
$\delta t$. In general then, if one can apply a set of Hamiltonians
$\{\pm \hat H_i\}$, one can construct any Hamiltonian that is a linear
combination of Hamiltonians of the form $\pm i[\hat H_i,\hat H_j]$,
$\pm [\hat H_i,[\hat H_j,\hat H_k]]$, etc.
\cite{seth11,seth12,seth13}, and no other Hamiltonians.
That is, one can construct the Hamiltonians in the algebra generated
from the original set by commutation.  This result makes it relatively
straightforward to determine the set of Hamiltonians that can be
constructed from simpler operations.

Now apply this result to the continuous variables introduced above.
The application of the boost $\pm \hat x$ and translation $\pm \hat p$
for short periods of time clearly allows the construction of any
Hamiltonian $a\hat x+b\hat p+c$ that is linear in $\hat x$ and $\hat p$;
this is all that it allows. To construct more complicated Hamiltonians
one must also be able to perform operations that are higher-order
polynomials in $\hat x$ and $\hat p$.
Suppose now that one can apply the quadratic Hamiltonian
\begin{equation}
\hat H = \hat x^2+\hat p^2\;.
\end{equation}
Application of this Hamiltonian for time
$t$ takes
\begin{eqnarray}
\hat x&\rightarrow& \cos t\, \hat x - \sin t \,\hat p \nonumber \\
\hat p&\rightarrow& \cos t \,\hat p + \sin t \,\hat x \;.
\end{eqnarray}
For an electromagnetic field, such an operation corresponds to a simple
phase shift. Note that since $e^{i\hat Ht}$ is periodic with period
$1/4\pi$, one can effectively apply $-\hat H$ for a time $\delta t$
by applying $\hat H$ for a time $4\pi -\delta t$.  The simple commutation
relations between $\hat H$, $\hat x$ and $\hat p$ imply that the addition
of $\pm \hat H$ to the set of operations that can be applied allows the
construction of Hamiltonians of the form $a\hat H+b\hat x+c\hat p+d$.

Suppose that in addition to translations and phase shifts one can
apply the quadratic Hamiltonian $\pm\hat S$ with
\begin{equation}
\hat S = \hat x\hat p+\hat p\hat x \;.
\end{equation}
Applying Hamiltonian $\hat S$ takes
\begin{eqnarray}
\hat x&\rightarrow& e^t \hat x \nonumber \\
\hat p&\rightarrow& e^{-t}\hat p \;.
\end{eqnarray}
Colloquially, $\hat S$ ``stretches'' $\hat x$ and
``squeezes'' $\hat p$ by some amount. In the case of the electromagnetic
field, $\hat S$ corresponds to a squeezer operating in the parametric
approximation. It is easily verified that
\begin{equation}
[\hat H,\hat S]=2i(\hat x^2-\hat p^2)\;.
\end{equation}
Looking at the algebra generated from $\hat x$, $\hat p$, $\hat H$ and
$\hat S$ by commutation, one sees that translations, phase shifts, and
squeezers allow the construction of any Hamiltonian that is quadratic
in $\hat x$ and $\hat p$, and of no Hamiltonian of higher order.

To construct higher-order Hamiltonians, nonlinear operations are required.
One such operation is the ``Kerr'' Hamiltonian
\begin{equation}\label{Kerrhamiltonian}
\hat H^2=(\hat x^2+\hat p^2)^2\;,
\end{equation}
corresponding to a $\chi^{(3)}$ process
in nonlinear optics. This higher-order Hamiltonian has the key
feature that commuting it with a polynomial in $\hat x$ and $\hat p$
typically {\it increases\/} its order. By evaluating a few commutators, e.g.,
\begin{eqnarray}
[\hat {H^2},\hat x ]
&=& -2i(\hat x^2\hat p+ \hat p^3)+\mbox{lower-order terms} \nonumber \\
\phantom{}[ \hat H^2,\hat p ]
&=& 2i(\hat x \hat p^2 + \hat x^3)+\mbox{lower-order terms} \nonumber \\
\phantom{}[ \hat x,[\hat H^2,\hat S]]
&=& -8\hat p^3 +\mbox{lower-order terms}\nonumber \\
\phantom{}[ \hat p,[\hat H^2,\hat S]]
&=& 8\hat x^3 +\mbox{lower-order terms} \;,
\end{eqnarray}
one sees that the algebra generated by $\hat x$, $\hat p$, $\hat H$,
$\hat S$ and $\hat H^2$ by commutation includes all third order polynomials
in $\hat x$ and $\hat p$. A simple inductive proof now shows that one
can construct Hamiltonians that are arbitrary Hermitian polynomials in
any order of $\hat x$ and $\hat p$. Suppose that one can construct a
polynomial of order $M$ consisting of any specific term
\begin{equation}
\hat x^{M-n}\hat p^n \;, \qquad n\le M \;.
\end{equation}
Since we may already create any quadratic Hermitian Hamiltonian and since
\begin{eqnarray}
[\hat x^2, \hat x^{M-n}\hat p^n] &=&
\frac{ni}{2} x^{M-(n-1)}\hat p^{(n-1)} \nonumber \\
&&+\mbox{lower-order terms} \nonumber \\
\phantom{}[\hat p^2, \hat x^{M-n}\hat p^n] &=&
\frac{-(M-n)i}{2} x^{M-(n+1)}\hat p^{(n+1)} \nonumber \\
&&+\mbox{lower-order terms}\;,
\end{eqnarray}
then it is easy to see that we may create {\it all\/} polynomials of order
$M$. Further, since we may create the Hamiltonians $\hat x^3$ or
$\hat p^3$ and since
\begin{eqnarray}
[\hat x^3 , \hat x^n \hat p^m]
&=& \frac{3mi}{2} \hat x^{n+2} \hat p^{m-1}
+\mbox{lower-order terms} \nonumber \\
\phantom{}[\hat x^3 , \hat x^n \hat p^m]
&=&\frac{-3ni}{2} \hat x^{n-1} \hat p^{m+2}
+\mbox{lower-order terms}\nonumber \\
\end{eqnarray}
one can by judicious commutation of $\hat x^3$ and $\hat p^3$ with
monomials of order $M$ construct any monomial of order $M+1$. Since
any polynomial of order $M+1$ can be constructed from monomials of order
$M+1$ and lower, by applying linear operations and a single nonlinear
operation a finite number of times one can construct polynomials of
arbitrary order in $\hat x$ and $\hat p$ to any desired degree of accuracy.
Comparison with similar results for the discrete case \cite{seth14}
shows that the number of operations required grows as a small polynomial
in the order of the polynomial to be created, the accuracy to which
that polynomial is to be enacted, and the time over which it is to be
applied.

The use of the Kerr Hamiltonian $\hat H^2$ was not essential: any higher
order Hamiltonian would be satisfactory. Note that commutation of a
polynomial in $\hat x$ and $\hat p$ with $\hat x$ and $\hat p$ themselves
(which have order one) always reduces the order of the polynomial by at
least one, commutation with $\hat H$ and $\hat S$ (which have order two) never
increases the order, and commutation with a polynomial of order three or
higher typically increases the order by at least one. Judicious commutation
of $\hat x$, $\hat p$, $\hat H$ and $\hat S$ with an applied Hamiltonian
of order three or higher therefore allows the construction of arbitrary
Hermitian polynomials of any order in $\hat x$ and $\hat p$. Alternatively,
it has recently been proposed that a measurement-induced nonlinearity
(using ideal photodetection) could be used in an optical scheme without
the need for nonlinear materials \cite{Gottesman00a,Bar01}.
The physical
realization of such nonlinearities is an important quest for quantum
information theory over continuous variables.

The above set of results shows that simple linear operations, together
with a single nonlinear operation, allow one to construct arbitrary
polynomial Hamiltonian transformations of a single quantum variable.
Let us now turn to more than one variable, e.g., the case of an
interferometer in which many modes of the electromagnetic field interact.
Suppose now that there are many variables, $\{\hat x_i,\hat p_i\}$, on each
of which the simple single-variable operations described above can be
performed. Now let the variables interact with each other. For simplicity,
we assume that we can apply interaction Hamiltonians $\pm \hat B_{ij}$ with
$\hat B_{ij}= (\hat p_i\hat x_j-\hat x_i\hat p_j)$: a more complicated
interaction Hamiltonian can always be used to generate interactions of
this form by combining it with single-variable operations.
This operation has the effect of taking
\begin{eqnarray}
\hat A_i&\rightarrow& \cos t \,\hat A_i + \sin t \,\hat A_j \nonumber \\
\hat A_j&\rightarrow& \cos t \,\hat A_j - \sin t \,\hat A_i \;,
\end{eqnarray}
where $\hat A_i=\{\hat x_i,\hat p_i\}$ and $\hat A_j=\{\hat x_j,\hat p_j\}$.
For the electromagnetic field, $\hat B_{ij}$ functions as a beam splitter,
linearly mixing together the two modes $i$ and $j$. By repeatedly taking
commutators of $\hat B_{ij}$ with polynomials in $\hat x_i$ and $\hat p_i$,
for different $i$, it can be easily seen by the same algebraic arguments
as above that it is possible to build up arbitrary Hermitian polynomials
in $\{\hat x_i, \hat p_i \}$.

This concludes the derivation of the main result: simple linear
operations on continuous variables, together with any nonlinear
operation and any interaction between variables suffice to
enact to an arbitrary degree of accuracy Hamiltonian operators
that are arbitrary Hermitian polynomials of the set of continuous variables.
In the case of modes of the electromagnetic field, linear operations
such as translations, phase shifts, squeezers, and beam splitters,
combined with some nonlinear operation such as a Kerr nonlinearity, allow
one to perform arbitrary polynomial transformations on those modes.
Note that in contrast to the case of qubits, in which a nonlinear
coupling between qubits is required to perform universal quantum
computation, in the continuous case only {\it single variable\/}
nonlinearities are required, along with linear couplings between
the variables.

In analogy with information over classical continuous variables,
which is measured in units of ``nats'' (1 nat = ${\rm log}_2 e$
bits), the unit of continuous quantum information will be called
the ``qunat.'' Two continuous variables in the pure state
$|\psi\rangle_{12}$ possess $-{\rm Tr} \hat\rho_1 \ln \hat\rho_1$
qunats of entanglement, where $\hat\rho_1 = {\rm Tr}_2
|\psi\rangle_{12}\langle\psi|$. For two squeezed vacua (squeezed
by an amount $e^{-r}$) entangled using a beam splitter the entropy
so computed from the approximate EPR state is given by
\begin{equation}
S(\hat\rho)=(1+\bar n)\ln (1+\bar n) - \bar n \ln \bar n \, ~~
{\rm qunats} \label{seth_eq2}
\end{equation}
with $\bar n = e^r \sinh r$. For example, $e^{2r} = 10$ gives 10 dB of
squeezing in power, corresponding to $r = 1.151$. By Eq.~(\ref{seth_eq2}),
two continuous variables entangled using a 10 dB squeezer then possess
$2.607$ qunats of shared, continuous quantum information, equivalent
to $3.762$ qubits of discrete quantum information. This is comparable
to the degree of entanglement currently available using ion-trap
quantum computers.

Quantum computation over cv can be thought of as
the systematic creation and manipulation of qunats. Universal quantum
computation for polynomial transformations of continuous variables
effectively allows one to perform quantum floating point manipulations
on those variables. For example, it is clearly possible using linear
operations alone to take the inputs $\hat x_1$, $\hat x_2$ and to map
them to $\hat x_1$, $a\hat x_1+b\hat x_2 +c$. Similarly, application of the
three-variable Hamiltonian $2\hat x_1\hat x_2\hat p_3$ takes
\begin{eqnarray}
\hat x_1 &\rightarrow& \hat x_1 \nonumber \\
\hat x_2 &\rightarrow& \hat x_2 \\
\hat x_3 &\rightarrow& \hat x_3 + \hat x_1 \hat x_2 t \nonumber \;,
\end{eqnarray}
that is, this operation  allows one to multiply $\hat x_1$ and $\hat x_2$
and place the result in the ``register'' $\hat x_3$.  A wide variety of
quantum floating point operations are possible.

The ability to create and manipulate qunats depends crucially on the
strength of squeezing and of the nonlinearities that one can apply.
Ten dB squeezers (6 dB after attenuation in the measurement apparatus)
currently exist \cite{seth16}. High-$Q$ cavity quantum electrodynamics
can supply a strong Kerr effect in a relatively lossless context, and
quantum logic gates constructed for qubits could be used to provide
the nonlinearity for continuous quantum variables as well \cite{seth17}.
Here the fact that only single-mode nonlinearities are required for
universal quantum computation simplifies the problem of effecting
continuous quantum logic. Nonetheless, the difficulty of performing
repeated nonlinear operations in a coherent and loss-free manner is
likely to limit the possibilities for quantum computation over the
amplitudes of the electromagnetic field. Vibrational modes of ions in
traps or excitations of a Bose-Einstein condensate might provide the
long-lived, lossless states required for quantum computation over
continuous variables.

\subsection{Extension of the Gottesman-Knill theorem}

Quantum mechanics allows for information processing that could not be
performed classically.  In particular, it may be possible to perform
an algorithm efficiently on a quantum computer that cannot be performed
efficiently on a classical one. The Gottesman-Knill theorem~\cite{Got99}
for discrete-variable (qubit) quantum information provides a valuable tool
for assessing the classical complexity of a given process. Essentially, it
states that any quantum algorithm that starts in the computational basis
and employs only a restricted class of gates (Hadamard, phase, C-NOT,
and Pauli gates), along with projective measurements in the computational
basis, can be efficiently simulated on a classical computer \cite{Nielsen}.
The Gottesman-Knill theorem reveals that a large class of quantum
algorithms do not provide a speedup over classical processes.

Here we develop the continuous-variable extension to the Gottesman-Knill
theorem. This result helps us understand what algorithms performed by a
cv quantum computer may be efficiently simulated by a
conventional classical computer. By contrast, it is exactly the algorithms
for which such an efficient simulation is impossible which is the major
subject of attention of quantum computation.

We note that the issue of efficient classical
simulation of a cv process is more involved than for the discrete case.
Continuous-variable
quantum states will typically only be defined for some limited
precision. For example, the states used in cv experiments are
approximations to the idealized computational basis. These basis states
are infinitely squeezed states whereas any experimental implementation
will involve only finitely squeezed states \cite{Llo99}.  Furthermore,
measurements are part of the quantum computation and, even in the
computational basis, are subject to experimental constraints (such as
detection efficiency).  A good classical simulation must be robust
against such imperfections.

Despite these complications, we shall present a set of sufficient
conditions for a cv quantum information process to ensure that it can
be efficiently simulated classically. To prove this theorem, we employ
the techniques of stabilizers \cite{Nielsen} that are used for qubits.
Using this formalism, it is possible to simulate a quantum algorithm
by following the evolution of the set of stabilizers, rather than the
evolution of quantum states. For a non-trivial set of algorithms this
procedure requires only a short description at each step and so may
be simulated efficiently without recourse to the exponential overhead of
explicitly recording all the terms of a quantum superposition
(which even for a single continuous
variable could require an infinite number
of terms). For cv processes the stabilizer formalism is particularly
simple when expressed in terms of their generators.

We first must identify the cv analogue
to the qubit computational basis. Here
we take it to be the set of position eigenstates $|x\rangle$
\cite{Llo99,SamQEC,Sam98c}. Next, given this choice we must identify the
cv analogues to each of the operations which are considered in the
qubit-version of the Gottesman-Knill theorem. For qubits the so-called
Pauli gates perform bit flips, phase flips or a combination of the two. For
cv states, the natural analogue would be to perform position
translations, momentum kicks or combinations thereof. Thus, for a
single cv, the Pauli operator analogs are
\begin{equation}
  \label{eq:PauliOperators}
  \hat X(x) \equiv e^{-2 i x \hat{p}} \; , \quad
  \hat Z(p) \equiv e^{2 i p \hat{x}} \; ,
\end{equation}
for $x$, $p$ both real. These operators are non-commutative obeying
\begin{equation}
  \label{eq:PauliOperatorCommutation}
  \hat X(x) \hat Z(p) = e^{-2 ixp} \hat Z(p) \hat X(x) \; .
\end{equation}
On the computational basis these operators act as
\begin{equation}
  \label{eq:ActionPauliOnCompBasis}
  \hat X(x')|x\rangle = |x + x'\rangle \; , \quad
  \hat Z(p)|x\rangle = e^{2 ipx}|x\rangle \, .
\end{equation}

We define the SUM gate as the cv analogue of the C-NOT gate. It provides
the basic interaction gate for a pair of cv systems $i$ and $j$ via
\begin{equation}
  \label{eq:DefSUM}
  \widehat {\rm SUM}_{ij} \equiv e^{ -2i \hat{x}_i \otimes \hat{p}_j} \; .
\end{equation}
Referring to Eq.~(\ref{eq:PauliOperators}) the action of this gate
on the Pauli operators is given by
\begin{eqnarray}
  \label{eq:ActionSUM}
\widehat{\rm SUM}_{ij}:\;
\hat X_i(x) \otimes \hat {\openone}_j
&\to&\hat X_i(x) \otimes \hat X_j(x)\nonumber \\
\hat Z_i(p) \otimes \hat {\openone}_j
&\to& \hat Z_i(p) \otimes \hat {\openone}_j  \nonumber \\
\hat {\openone}_i \otimes \hat X_j(x)
&\to& \hat {\openone}_i \otimes \hat X_j(x)  \nonumber \\
\hat {\openone}_i \otimes \hat Z_j(p)
&\to& \hat Z_i(p)^{-1} \otimes \hat Z_j(p) \; .
\end{eqnarray}
This gate describes the unitary transformation used in a back-action
evading or quantum nondemolition process.

The Fourier transform $\hat\mathcal{F}$ is the cv analogue of the Hadamard
transformation [Eq.~(\ref{fourier})].
It can also be defined as
\begin{equation}
  \label{eq:DefFourier}
\hat\mathcal{F} \equiv e^{ i \pi(\hat{x}^2 + \hat{p}^2)/2 } \; ,
\end{equation}
and its action on the Pauli operators is
\begin{eqnarray}
  \label{eq:ActionFourier}
  \hat\mathcal{F}:\; \hat X(x) &\to& \hat Z(x) \nonumber \\
  \hat Z(p) &\to& \hat X(p)^{-1} \; .
\end{eqnarray}
The ``phase gate'' $\hat P(\eta)$ is a squeezing operation for cv,
defined by
\begin{equation}
  \label{eq:DefPhase}
\hat P(\eta) \equiv e^{ i \eta \hat{x}^2 } \; ,
\end{equation}
and its action on the Pauli operators is given by
\begin{eqnarray}
  \label{eq:ActionPhase}
\hat P(\eta):\; \hat X(x)
&\to& e^{i\eta x^2} \hat X(x)\hat Z(\eta x) \nonumber\\
\hat Z(p) &\to& \hat Z(p) \;,
\end{eqnarray}
which is analogous to that of the discrete-variable phase gate $\hat P$
\cite{Gottesman00a}.

For the cv operators defined above, SUM, $\hat\mathcal{F}$, $\hat P(\eta)$,
$\hat X(x)$ and $\hat Z(p)$ are sufficient to simulate all possible
quadratic Hermitian Hamiltonians, as we saw in the last section.

We now have the necessary components to prove our main result.  We
employ the stabilizer formalism used for discrete variables and follow
the evolution of generators of these stabilizers rather than the states.
To start with, let us consider the ideal case of a system with an
initial state in the computational basis of the form
$|x_1,\,x_2,\,\ldots,\,x_n\rangle$. This state may be fully characterized
by the eigenvalues of the generators of $n$ Pauli operators
$\{\hat x_1,\,\hat x_2,\ldots,\,\hat x_n\}$.  Any cv
process or algorithm can then be modeled by following the evolution
of the generators of these $n$ Pauli operators, rather than by following
the evolution of the states in the infinite-dimensional Hilbert
space $\mathcal{L}^2(\mathbb{R}^n)$.  If we restrict ourselves to the
gate operations SUM, $\hat\mathcal{F}$,
$\hat P(\eta)$, $\hat X(x)$ and $\hat Z(p)$
our job is straightforward, since each of the stabilizers evolves only
to a simple tensor product of Pauli operators. In other words, these
operations map linear combinations of Pauli operator generators to
linear combinations of Pauli operator generators (each $\hat{x}_i$ and
$\hat{p}_i$ is mapped to sums of $\hat{x}_j$, $\hat{p}_j$, $j=1,\ldots,n$
in the Heisenberg picture).  For each of the $n$ generators describing
the initial state, one must keep track of $2n$ real coefficients describing
this linear combination.  To simulate such a system, then, requires
following the evolution of $2n^2$ real numbers.

In the simplest case, measurements (in the computational basis) are
performed at the end of the computation. An efficient classical
simulation involves simulating the statistics of linear combinations
of Pauli operator generators.  In terms of the Heisenberg evolution,
the $\hat{x}_j$ are described by their initial eigenvalues, and the
$\hat{p}_j$ in the sum by a uniform random number. This prescription
reproduces the statistics of all multi-mode correlations for
measurements of these operators.

Measurement in the computational basis plus feed-forward {\it during\/}
the computation may also be easily simulated for a sufficiently
restricted class of feed-forward operations; in particular, operations
corresponding to feed-forward displacement (not rotation or squeezing,
though this restriction will be dropped below) by an amount
proportional to the measurement result.  Such feed-forward operations
may be simulated by the Hamiltonian that generates the SUM gate with
measurement in the computational basis delayed until the end of the
computation. In other words, feed-forward from measurement can be
treated by employing conditional unitary operations with delayed
measurement \cite{Nielsen}, thus reducing feed-forward to the case
already treated.

In practice, infinitely-squeezed input states are not available.
Instead, the initial states will be of the form
\begin{equation}
  \hat S_1(r_1)\otimes\hat S_2(r_2)\otimes\cdots\otimes \hat S_n(r_n)
  |0, 0, \ldots, 0\rangle \,,
\end{equation}
where here $|0\rangle$ is a vacuum state and $\hat S(r)$, with
$r\in \mathbb{R}$, is the squeezing operation.  Now the vacuum states
may {\it also\/} be described by stabilizers `generated' by
$\{\hat x_1+i\hat p_1,\,\hat x_2+i\hat p_2,\,\ldots,\,\hat x_n+i\hat p_n\}$
which are now complex linear combinations of the earlier generators.
Although these `generators' are non-Hermitian, the operators obtained
by their exponentiation do indeed behave as stabilizers. Combining the
initial squeezing operators into the computation, a classical
simulation of a gate array consisting of operations from the set
SUM, $\hat\mathcal{F}$, $\hat P(\eta)$, $\hat X(x)$ and $\hat Z(p)$ requires
following the evolution of $4n^2$ numbers (twice that of infinitely
squeezed inputs due to the real and imaginary parts).
Measurements in the computational basis are again easily simulated in
terms of this Heisenberg evolution, by treating each of the $x_i$ and
$p_i$ as random numbers independently sampled from a Gaussian
distribution with widths described by the vacuum state.  Simulation of
measurement plus feed-forward follows exactly the same prescription as
before.

Furthermore, the condition for ideal measurements can be relaxed.
Finite-efficiency detection can be modeled by a linear-loss mechanism
\cite{Yue80}.  Such a mechanism may be described by quadratic
Hamiltonians and hence simulated by quadratic Hamiltonians and hence
the allowed gate elements.  Note that these allowed gate elements are
precisely those that preserve Gaussian states; i.e., they transform
Gaussians to Gaussians; this observation allows us to remove our earlier
restriction on feed-forward gates and allow for classical feed-forward
of any allowed gate operation.  Note that non-Gaussian components to the
states cannot be modeled in this manner.

Finally, it should be noted that modeling the evolution requires
operations on real-valued (continuous) variables, and thus must be
discretized when the simulation is done on a discrete (as opposed to
analog) classical computer. The discretization assumes a finite error,
which will be bounded by the smaller of the initial squeezing or the
final ``resolution'' due to detector efficiency, and this error
must remain bounded throughout the simulation.  As only the
operations of addition and multiplication are required, the
discretization error can be kept bounded with a polynomial cost
to efficiency. This completes our demonstration of the extension of
the Gottesman-Knill theorem to continuous variables.

As with the discrete-variable case, these conditions do not mean that
entanglement between the $n$ oscillator systems is not allowed; for
example, starting with (separable) position eigenstates, the
Fourier-transform gate combined with the SUM gate leads to entanglement.
Thus, algorithms that produce entanglement between systems may still
satisfy the conditions of the theorem and thus may be simulated
efficiently on a classical computer; included are those used for cv
quantum teleportation \cite{SamKimble,Fur98}, quantum cryptography
\cite{Ralph00a,Hillery00,Reid00,Gottesman00b}, and error correction
\cite{SamQEC,Sam98c}.  Although these processes are of a
fundamentally quantum nature and involve entanglement between
systems, the extended Gottesman-Knill theorem demonstrates that they
do not provide any speedup over a classical simulation.  This theorem,
therefore, provides a valuable tool in assessing the classical
complexity of simulating these quantum processes.

As shown in the previous section, in order to generate all unitary
transformations given by an arbitrary polynomial Hermitian Hamiltonian
(as is necessary to perform universal cv quantum computation), one must
include a gate described by a Hamiltonian other than an inhomogeneous
quadratic in the canonical operators, such as a cubic or higher-order
polynomial.  Transformations generated by these Hamiltonians do not
preserve the linear structure of the generators of the stabilizers, and
thus cannot be described efficiently by the stabilizer formalism.
These nonlinear transformations can be used in cv algorithms and
may provide a significant speedup over any classical process.

\section{Experiments with Continuous Quantum Variables}
\label{expwcv}

In this section, we discuss some experiments
based on continuous quantum variables.
These include the generation of squeezed-state EPR entanglement
via optical parametric amplification and via the Kerr
effect. Qualitatively different manifestations of cv entanglement
are that between two atomic ensembles, created
in an experiment in Copenhagen \cite{BJ01},
and that between more than two optical modes,
experimentally generated and verified
in Tokyo for three modes \cite{Aoki03}.
Quantum teleportation of
coherent states has been achieved already in Pasadena
at Caltech \cite{Fur98,Zhang03}
and in Canberra \cite{BowenLam03}.
We will briefly show how to describe
these experiments in a realistic broadband fashion.
Further important cv experiments include
the dense coding experiment of Li {\it et al.}
utilizing bright EPR beams \cite{LiXY02}, the coherent-state
based quantum key distribution experiment by Grangier
and his group \cite{cvcryptoNature03}, and the demonstration of a
quantum memory effect by the Polzik group \cite{Schori02}.

\subsection{Generation of squeezed-state EPR entanglement}
\label{expwcvgen}

The generation of dv qubit entanglement can be achieved
experimentally via weak down-conversion producing
polarization-entangled single photons.
The resulting maximum entanglement is then
`polluted' by a large vacuum contribution
\cite{SamJeffNature}.
The consequence of this is that the entanglement never
emerges from these optical devices in an event-ready fashion.
Since successful (post-selected)
events occur very rarely, one has to cope with very low efficiency
in these single-photon schemes.
However, there are also some advantages of the
single-photon based approaches to entanglement generation
and quantum communication, as we discussed in Sec.~\ref{Intro}.
Great progress has been made in generating
single-photon entanglement, both for the
case of two qubits \cite{Bou} and three qubits \cite{Bouw}.

In the cv setting, the generation
of entanglement, for instance, occurring every inverse bandwidth
time at the output of an optical parametric
amplifier is more efficient than in the single-photon schemes.
When making an entangled two-mode squeezed state,
the vacuum contribution that originates from the
down-conversion source need not be excluded via post-selection.
It is still contained in the resulting
nonmaximally entangled output state,
as expressed by Eq.~(\ref{twin}) in an idealized
discrete-mode description. We will now discuss
a more realistic description of the resulting
broadband entangled state that emerges from
an optical parametric amplifier.
This type of quadrature entanglement was used
in the recent cv quantum communication experiments
\cite{Fur98,Zhang03,BowenLam03,LiXY02}.
The first experiment to produce cv broadband
EPR correlations of this kind was performed in 1992
\cite{Ou92,Ou92applphys}.
Recently, this ``conventional'' quadrature entanglement
was also transformed into cv polarization entanglement
exhibiting correlations in the Stokes operators
of two beams \cite{BowenTreps02}.
Another recent scheme to create
cv broadband entanglement
was based on a different nonlinear optical interaction,
namely the Kerr effect in an optical fiber \cite{Silberhorn}.

\subsubsection{Broadband entanglement via optical parametric
amplification}\label{bbentvianopa}

A broadband entangled state is generated either directly by
nondegenerate parametric amplification in a cavity
(NOPA, also called
``nondegenerate parametric down conversion'') or by combining
at a beam splitter two independently squeezed fields produced
via degenerate down conversion.
This observation is the broadband extension
of the fact that a two-mode squeezed state
is equivalent to two single-mode squeezed states combined
at a $50:50$ beam splitter. Just as for the two discrete modes
in Eqs.~(\ref{TWOMODESQfromBS}) and (\ref{2modeHeis}),
this can be easily seen in the ``continuum''
representation \cite{Schum}
of the quadrature operators,
\begin{eqnarray}\label{1.14before}
\hat{x}(\Omega)&=&\frac{1}{2}
\Big[\sqrt{1+\frac{\Omega}{\omega_0}}\,
\hat{b}(\omega_0+\Omega)+\nonumber\\
&&\quad\quad\sqrt{1-\frac{\Omega}{\omega_0}}\,
\hat{b}^{\dagger}(\omega_0-\Omega)\Big],\nonumber\\
\hat{p}(\Omega)&=&\frac{1}{2i}
\Big[\sqrt{1+\frac{\Omega}{\omega_0}}\,
\hat{b}(\omega_0+\Omega)-\nonumber\\
&&\quad\quad\sqrt{1-\frac{\Omega}{\omega_0}}\,
\hat{b}^{\dagger}(\omega_0-\Omega)\Big],
\end{eqnarray}
where $\omega_0$ is the optical central frequency
and $\Omega>0$ some small modulation frequency.
Here, the annihilation and creation operators
(now no longer dimensionless, but each in
units of root time, $\sqrt{\rm s}$)
satisfy the commutation relation
$[\hat{b}(\omega),\hat{b}^{\dagger}(\omega')]=
\delta(\omega - \omega')$.
The commutators for the quadratures are \cite{Schum}
\begin{eqnarray}\label{bbquadrcommut}
&&[\hat{x}(\Omega),\hat{x}(\Omega')]=
[\hat{x}(\Omega),\hat{p}(\Omega')]=
[\hat{p}(\Omega),\hat{p}(\Omega')]=0,
\nonumber\\
&&[\hat{x}(\Omega),\hat{x}^{\dagger}(\Omega')]=
[\hat{p}(\Omega),\hat{p}^{\dagger}(\Omega')]=
\frac{\Omega}{2\omega_0}\,\delta(\Omega - \Omega'),
\nonumber\\
&&[\hat{x}(\Omega),\hat{p}^\dagger(\Omega')]=
[\hat{x}^\dagger(\Omega),\hat{p}(\Omega')]=
\frac{i}{2}\,\delta(\Omega - \Omega').
\nonumber\\
\end{eqnarray}
This is a suitable formalism for analyzing
two-photon devices such as the parametric amplifier.
Due to the nonlinear optical interaction,
a pump photon at frequency $2\omega_0$ can be annihilated
to create two photons at the frequencies
$\omega_0\pm\Omega$ and, conversely, two photons
can be annihilated to create a pump photon.
Thus, the light produced by the amplifier always consists
of pairs of modes at frequencies $\omega_0\pm\Omega$.

Now it is convenient to define the upper-case operators
in the rotating frame about
the optical central frequency $\omega_0$ (for the NOPA,
half the pump frequency),
\begin{eqnarray}\label{1.9}
\hat{B}(t)=\hat{b}(t)e^{i\omega_0t}.
\end{eqnarray}
Via the Fourier transform
\begin{eqnarray}\label{1.10}
\hat{B}(\Omega)=\frac{1}{\sqrt{2\pi}}\int dt\,\hat{B}(t)e^{i\Omega t},
\end{eqnarray}
the fields may then
be described as functions of the modulation frequency
$\Omega$ with the commutation relation
\begin{eqnarray}\label{bbcommutator}
[\hat{B}(\Omega),\hat{B}^{\dagger}(\Omega')]=\delta(\Omega - \Omega').
\end{eqnarray}
In the rotating frame, using $\hat{b}(\omega_0\pm\Omega)=
\hat{B}(\pm\Omega)$ and the approximation $\Omega\ll \omega_0$,
the frequency resolved
``broadband'' quadrature amplitudes
of Eq.~(\ref{1.14before}) for a mode $k$
(for instance, a spatial or a polarization mode)
may be written in a form more reminiscent of the ``discrete'' quadratures
in Eq.~(\ref{quadraturesdef}) as \cite{Ou92applphys}
\begin{eqnarray}\label{1.14}
\hat{X}_k(\Omega)&=&\frac{1}{2}[\hat{B}_k(\Omega)+\hat{B}^{\dagger}_k
(-\Omega)],\nonumber\\
\hat{P}_k(\Omega)&=&\frac{1}{2i}[\hat{B}_k(\Omega)-\hat{B}^{\dagger}_k
(-\Omega)],
\end{eqnarray}
with the broadband annihilation operators $\hat{B}_k(\Omega)$.
The only nontrivial commutation relations for the
non-Hermitian quadratures are now
\begin{eqnarray}\label{bbquadrcommutinrotating}
[\hat{X}(\Omega),\hat{P}^\dagger(\Omega')]=
[\hat{X}^\dagger(\Omega),\hat{P}(\Omega')]=
\frac{i}{2}\,\delta(\Omega - \Omega').
\nonumber\\
\end{eqnarray}
The actually measurable quantities are the Hermitian
real and imaginary parts of these quadratures, satisfying
the usual commutation relations
\begin{eqnarray}\label{bbquadrcommutinrotatinghermit}
[{\rm Re}\hat{X}(\Omega),{\rm Re}\hat{P}(\Omega')]&=&
[{\rm Im}\hat{X}(\Omega),{\rm Im}\hat{P}(\Omega')]
\nonumber\\
&=&\frac{i}{4}\,\delta(\Omega - \Omega')\,.
\end{eqnarray}
Using the quadrature squeezing spectra,
\begin{eqnarray}\label{spectra}
\langle\Delta\hat{X}_1^{\dagger}(\Omega)\Delta\hat{X}_1
(\Omega')\rangle&=&\langle\Delta\hat{P}_2^{\dagger}(\Omega)
\Delta\hat{P}_2(\Omega')\rangle\nonumber\\
&=&
\delta(\Omega-\Omega')|S_+(\Omega)|^2/4,\nonumber\\
\langle\Delta\hat{X}_2^{\dagger}(\Omega)\Delta\hat{X}_2
(\Omega')\rangle&=&\langle\Delta\hat{P}_1^{\dagger}(\Omega)
\Delta\hat{P}_1(\Omega')\rangle\nonumber\\
&=&
\delta(\Omega-\Omega')|S_-(\Omega)|^2/4,\nonumber\\
\end{eqnarray}
two independently squeezed fields coming
from two degenerate optical parametric oscillators (OPO's)
can be described as
\begin{eqnarray}\label{general}
&&\hat{X}_1(\Omega)=S_+(\Omega) \hat{X}^{(0)}_1
(\Omega),\;\;\;\hat{P}_1(\Omega)=S_-(\Omega)
\hat{P}^{(0)}_1(\Omega),
\nonumber\\
&&\hat{X}_2(\Omega)=S_-(\Omega) \hat{X}^{(0)}_2
(\Omega),\;\;\;\hat{P}_2(\Omega)=S_+(\Omega)
\hat{P}^{(0)}_2(\Omega),\nonumber\\
\end{eqnarray}
where $|S_-(\Omega)|<1$ refers to the quiet quadratures
and $|S_+(\Omega)|>1$
to the noisy ones, and the superscript `$(0)$' denotes vacuum modes.
These fields
can be used as a broadband EPR source when they are
combined at a beam splitter \cite{PvLbroad},
\begin{eqnarray}\label{general2}
\hat{X}_1'(\Omega)&=&\frac{1}{\sqrt{2}}S_+(\Omega)
\hat{X}^{(0)}_1(\Omega)+\frac{1}{\sqrt{2}}S_-(\Omega)
\hat{X}^{(0)}_2(\Omega),\nonumber\\
\hat{P}_1'(\Omega)&=&\frac{1}{\sqrt{2}}S_-(\Omega)
\hat{P}^{(0)}_1(\Omega)+\frac{1}{\sqrt{2}}S_+(\Omega)
\hat{P}^{(0)}_2(\Omega),\nonumber\\
\hat{X}_2'(\Omega)&=&\frac{1}{\sqrt{2}}S_+(\Omega)
\hat{X}^{(0)}_1(\Omega)-\frac{1}{\sqrt{2}}S_-(\Omega)
\hat{X}^{(0)}_2(\Omega),\nonumber\\
\hat{P}_2'(\Omega)&=&\frac{1}{\sqrt{2}}S_-(\Omega)
\hat{P}^{(0)}_1(\Omega)-\frac{1}{\sqrt{2}}S_+(\Omega)
\hat{P}^{(0)}_2(\Omega).\nonumber\\
\end{eqnarray}
In this state, the upper and lower sidebands
around the central frequency exhibit EPR-type correlations
similar to those in Eq.~(\ref{2moderelpostotmom}),
\begin{eqnarray}\label{bbsqcorrelations}
\hat{U}(\Omega)&\equiv&
\hat{X}_1'(\Omega)-\hat{X}_2'(\Omega)=\sqrt{2}\,S_-(\Omega)
\hat{X}^{(0)}_2(\Omega)\,,\nonumber\\
\hat{V}(\Omega)&\equiv&
\hat{P}_1'(\Omega)+\hat{P}_2'(\Omega)=\sqrt{2}\,S_-(\Omega)
\hat{P}^{(0)}_1(\Omega)\,,
\end{eqnarray}
and therefore
\begin{eqnarray}\label{broadbandcorrelations}
\langle\Delta\hat{U}^{\dagger}(\Omega)\Delta\hat{U}
(\Omega')\rangle&=&
\delta(\Omega-\Omega')|S_-(\Omega)|^2/2,\nonumber\\
\langle\Delta\hat{V}^{\dagger}(\Omega)\Delta\hat{V}
(\Omega')\rangle&=&
\delta(\Omega-\Omega')|S_-(\Omega)|^2/2.
\end{eqnarray}
The corresponding sum condition Eq.~(\ref{Duan3}) with $\bar a=1$,
necessarily satisfied by any separable state, is now violated for
that range of modulation frequencies for which the resources are
squeezed,\footnote{There is actually no rigorous broadband
derivation of the inseparability criteria in the literature,
including the corresponding broadband analogue to the sum
condition Eq.~(\ref{Duan3}). Here we also only give the
``continuum'' equations for the broadband EPR state and apply it
directly to the discrete sum condition at the squeezing
frequencies.}
\begin{eqnarray}\label{broadbandcorrelationsDuan}
\langle\Delta\hat{U}^{\dagger}(\Omega)\Delta\hat{U}
(\Omega')\rangle &+&
\langle\Delta\hat{V}^{\dagger}(\Omega)\Delta\hat{V}
(\Omega')\rangle \\
&&\,= \delta(\Omega-\Omega')|S_-(\Omega)|^2.\nonumber
\end{eqnarray}
In recent experiments,
such violations at some squeezing frequency were detected
for the verification of cv quadrature entanglement
\cite{Furucvbook,Silberhorn} or, similarly, of
cv polarization entanglement \cite{BowenTreps02}.
Exactly these violations were also needed for accomplishing
the recent quantum communication protocols
\cite{Fur98,Zhang03,BowenLam03,LiXY02}.

If the squeezed fields for entanglement generation
come from two OPO's,
the nonlinear optical interaction is due to a
$\chi^{(2)}$ medium.
In general, Eq.~(\ref{spectra}) may define arbitrary
squeezing spectra of two statistically identical but
independent broadband squeezed states.
Before obtaining the ``broadband EPR state'',
the squeezing of the two initial fields may be generated by
any suitable nonlinear interaction.
The optical Kerr effect, based on a $\chi^{(3)}$ interaction
may also serve as such a suitable interaction.

\subsubsection{Kerr effect and linear interference}
\label{Kerrinterfer}

The first light-squeezing experiment was published in 1985
\cite{Slush}. In this experiment, squeezed light
was generated via four-wave mixing.
Though involving the production of photon pairs
as in parametric down-conversion, the process
of four-wave mixing is based on a $\chi^{(3)}$ interaction.

Initially, the main conceptual difficulty in creating a detectable
squeezing effect via a  $\chi^{(3)}$ interaction was that such a
process is very weak in all transparent media. In particular, in
order to achieve measurable quantum noise reduction against
additional classical (thermal) noise, large light energy density
and long interaction lengths are required. These requirements led
to the proposal to use an optical fiber for nondegenerate
four-wave mixing \cite{Levenson}. The proposal referred to a
dispersionless $cw$ type of four-wave mixing. In the response of
the fiber material to an external field, the dominant nonlinear
contribution corresponds to a $\chi^{(3)}$ interaction (``Kerr
effect''), because the $\chi^{(2)}$ susceptibility vanishes in a
glass fiber \cite{Agrawal}. The Kerr effect is equivalent to an
intensity dependent refractive index. A squeezing experiment
confirming the $cw$ theory of four-wave mixing in a single-mode
fiber \cite{Levenson} was successfully conducted by
\textcite{Shelby}. Soon after this experiment, the quantum theory
of light propagation and squeezing in an optical fiber was
extended to include pulsed pump fields and group-velocity
dispersion \cite{Carter}.

By using stochastic equations for describing the classical
propagation plus the evolution of the quantum noise in the fiber,
Carter {\it et al.} proposed the squeezing of quantum
fiber-solitons \cite{Carter}. This theory was then experimentally
confirmed by \textcite{Rosenbluh}.

What is the potential advantage of using optical fibers and light
pulses with respect to applications in quantum communication? At
the communication wavelength of $1.55$ $\mu$m, glass fibers have
an absorption minimum with very low losses and negative dispersion
which enables one to use stable soliton pulses
\cite{Agrawal,Drummond2}. A fiber-based quantum communication
system can be potentially integrated into existing fiber-optics
communication networks. Moreover, an optical fiber naturally
offers long interaction times for producing squeezed light. Short
light pulses and solitons have large peak power and photon number
density which enhances the effective $\chi^{(3)}$ nonlinearity in
the fiber and hence the potential squeezing.

The Kerr interaction Hamiltonian,
\begin{eqnarray}\label{KerrHam}
\hat{H}_{\rm int}=
\hbar\kappa\,
\hat{a}^{\dagger 2}\hat a^2 =
\hbar\kappa\,\hat n (\hat n -1)\;,
\end{eqnarray}
with $\kappa$ proportional to $\chi^{(3)}$, is quartic rather than
quadratic [see Eq.~(\ref{OPOHam})] as for optical parametric
amplification or for conventional four-wave mixing. For the
quartic Hamiltonian, the Kerr interaction would turn a coherent
state into a ``banana-shaped'' state which after a suitable
phase-space displacement has reduced number and increased phase
uncertainty though essentially still a number-phase minimum
uncertainty state \cite{Kitagawa}. This state corresponds to a
photon number squeezed state with sub-Poissonian statistics, as
opposed to the ordinary quadrature squeezed state. It is closer to
a Fock state than to a quadrature eigenstate. However, in the
regime of large photon number and small nonlinearity [which for
example applies to quantum solitons for sufficiently small
propagation distance \cite{Kaertner}] quantum fluctuations higher
than those of second order can be neglected. The quartic
Hamiltonian is then effectively reduced to a quadratic one (note
that squeezing due to the former preserves the photon number,
whereas that due to the latter does not). In fact, the fiber Kerr
nonlinearity is so small that the radius of curvature of the
``banana'' is far larger than its length. The difference between
such a state and an ordinary squeezed state with an ``elliptic''
phase-space distribution is therefore negligible.

Recently, bipartite cv entanglement was created
through an optical fiber with optical pulses squeezed via the Kerr
$\chi^{(3)}$ nonlinearity \cite{Silberhorn}.
The entanglement-generating mechanism in this experiment
was indeed similar to the creation of the broadband EPR state
in Eq.~(\ref{general2}) by combining the two squeezed fields
in Eq.~(\ref{general}): first, the Kerr nonlinearity
in the fiber was exploited to produce two independent squeezed beams
(more precisely, an asymmetric fiber Sagnac interferometer was used
to make two amplitude or photon number squeezed beams
of orthogonal polarization).
The squeezed fields were then combined {\it outside the fiber}
at a beam splitter. As described,
in order to obtain Kerr-induced squeezing,
the beams must have non-zero intensity, they must be {\it bright},
as opposed to the squeezed vacuum states in Eq.~(\ref{general}).

\subsection{Generation of long-lived atomic entanglement}
\label{expwcvgenforatoms}

In Sec.~\ref{cventanglement}, in particular
Sec.~\ref{genentsection}, we discussed how to make entanglement
from sources of nonclassical light such as squeezed states using a
network of beam splitters. In fact, in order to create cv
entanglement using linear optics, at least one of the input modes
must be in a nonclassical state. Otherwise, if all the input modes
are in a vacuum or coherent state, the output state that emerges
from the beam splitters will always remain separable. Similarly,
entanglement-based quantum communication schemes utilizing
atom-light interactions also seem to rely upon nonclassical light
as a resource [see, for instance, the protocol of
\textcite{KuzPolzik00}]. Moreover, other atom-light protocols
require the atoms be trapped in high-$Q$ optical cavities [for
example, in Ref.~\cite{ParkinsKimble99}]. However, remarkably,
entanglement between free-space atomic ensembles may also be
created using only coherent light, as it was proposed by
\textcite{DuanPolzik00}. The QND coupling given in
Eq.~(\ref{atomicQND}) is a suitable interaction to achieve this.
After the successive transmission of a light beam through {\it
two} separate atomic ensembles, the light is detected such that
the atomic states are reduced to an entangled state, corresponding
to a nonlocal Bell measurement \cite{DuanPolzik00}. Such an
experiment, along the lines of the proposal of
\textcite{DuanPolzik00}, was performed in Copenhagen \cite{BJ01}.
The long-lived entanglement generated in this experiment was
between two clouds of atoms, each consisting of a caesium gas
containing about $10^{12}$ atoms. The creation of this
entanglement between material objects is an important step towards
storing quantum information in an (optical) communication protocol
and proves the feasibility for the implementation of light-atom
quantum interfaces using a similar approach. Further experimental
investigations towards a quantum memory, also based on this kind
of approach, will be briefly described in
Sec.~\ref{expwcvqmemory}.

The experiment for creating atomic entanglement \cite{BJ01}
is based on the polarization and spin representation
for cv quantum information, as discussed in Sec.~\ref{polspinrep}.
Hence, the light and the atoms are described via the Stokes
operators and the operators for the collective spin, respectively.
More precisely, if the atomic samples are spin-polarized along
the $x$-axis with a large classical value, and similarly
for the light,
the only quantum variables used for the entanglement
generation are the atomic operators $\hat F_y$ and $\hat F_z$
and the light operators $\hat S_y$ and $\hat S_z$.
In other words, the $y$ and $z$ components of spin and polarization
play the roles of the effective phase-space variables both for the
atomic and the light system, respectively.

Now when an off-resonant light pulse classically polarized along
the $x$-axis ($\hat S_x\simeq \langle \hat S_x\rangle\equiv S$)
is transmitted along the $z$-axis through two atomic samples
with opposite classical spins along the $x$-axis,
$\hat F_{xj}\simeq \langle \hat F_{xj}\rangle$, $j=1,2$,
$\langle \hat F_{x1}\rangle
=-\langle \hat F_{x2}\rangle\equiv F$,
the input-output relations are given by [see Eq.~(\ref{atomicQND})]
\begin{eqnarray}
\hat S_y^{\rm (out)} &=& \hat S_y^{\rm (in)}
+a S\,(\hat F_{z1}^{\rm (in)} + \hat F_{z2}^{\rm (in)})\,,
\nonumber\\
\hat S_z^{\rm (out)} &=& \hat S_z^{\rm (in)},
\nonumber\\
\hat F_{y1}^{\rm (out)} &=& \hat F_{y1}^{\rm (in)}
+a F\,\hat S_z^{\rm (in)},\quad
\hat F_{y2}^{\rm (out)} = \hat F_{y2}^{\rm (in)}
-a F\,\hat S_z^{\rm (in)},
\nonumber\\
\hat F_{z1}^{\rm (out)} &=& \hat F_{z1}^{\rm (in)},
\quad
\hat F_{z2}^{\rm (out)} = \hat F_{z2}^{\rm (in)}.
\end{eqnarray}
These equations show that for a sufficiently large
value of the quantity $a\,S$, a measurement of
$\hat S_y^{\rm (out)}$ reveals the value of
the total $z$-spin in a QND fashion,
$\hat F_{z1}^{\rm (in)} + \hat F_{z2}^{\rm (in)}=
\hat F_{z1}^{\rm (out)} + \hat F_{z2}^{\rm (out)}$.
At the same time, the total $y$-spin is conserved
as well, neither being changed by the interaction,
$\hat F_{y1}^{\rm (in)} + \hat F_{y2}^{\rm (in)}=
\hat F_{y1}^{\rm (out)} + \hat F_{y2}^{\rm (out)}$, nor
affected by the measurement thanks to the vanishing commutator
$[\hat F_{y1} + \hat F_{y2},\hat F_{z1} + \hat F_{z2}]=0$.
Upon repeating this procedure with a different light pulse,
but now measuring the total $y$-spin in a QND fashion
(which will not change the previously measured value of the total
$z$-spin), both the total $z$-spin and the total $y$-spin
may be precisely determined.
Thus, the resulting state of the two atomic samples
has arbitrarily small variances for both $y$ and $z$ components
of the total spin,
\begin{eqnarray}\label{Duanforatoms3}
\langle[\Delta(\hat F_{y1}+\hat F_{y2})]^2\rangle+
\langle[\Delta(\hat F_{z1}+\hat F_{z2})]^2\rangle
\rightarrow 0\;.
\end{eqnarray}
This would, in the ideal case, lead to a maximal violation
of the necessary separability condition in
Eq.~(\ref{Duanforatoms2}) (for $x$ and $z$ components
interchanged).
Under realistic experimental conditions, however, with imperfections
caused by, for instance, losses of the light on the way from
one sample to the other and spin-state decay between the
two measurements, the resulting
atomic state does not become perfectly entangled.
Moreover, the vacuum noise of the incoming light pulse
prevents the creation of a maximally entangled state.
The outgoing state, prepared after the measurements, is then
similar to a nonmaximally entangled two-mode squeezed state.

In the experiment in Copenhagen, the protocol described above
was slightly modified by adding a magnetic field oriented along
the $x$-axis. Using only a single entangling light pulse,
both the $y$ and the $z$ spin projections can be
measured this way. The generated entangled state
was maintained for more than $0.5$ ms. This relatively long
lifetime is due to the high symmetry of the state.
The entanglement is based on the collective properties
of the two atomic ensembles such that
the coherence of the entangled superposition state is not
destroyed when only a few atoms interact with the environment.
This kind of robustness would not be obtainable in a maximally
entangled multi-particle state.
The degree of entanglement verified in the Copenhagen experiment
corresponds to a fidelity of $F\approx 0.55$ when using
the entangled state for teleporting an atomic sample in a coherent
spin state. This clearly exceeds the classical boundary of $F=0.5$.

Although the entanglement produced in Copenhagen was bipartite,
i.e., between two atomic clouds, one can easily think of
an extension to more atomic samples.
As for an experiment in which the creation of such a genuine
multipartite entanglement has been accomplished already,
we will now return to the all-optical regime
of squeezed-light resources and linear-optics transformations.

\subsection{Generation of genuine multipartite entanglement}
\label{expwcvgenformanyparties}

We have seen that
a particularly efficient way to generate entanglement
between electromagnetic modes is to let squeezed light beams
interfere using linear optics. For instance,
the generation of tripartite entanglement,
the entanglement between three optical modes,
only requires combining
three input modes at two beam splitters, where at least
one of these input modes is in a squeezed state
(see Sec.~\ref{genentsection}).
The resulting entangled modes, even when spatially separated,
exhibit quantum correlations,
as described by Eq.~(\ref{PVLcorrfamily}) for $N=3$.
However, due to experimental imperfections,
the three-mode states generated in the laboratory
become noisy mixed states that might be partially or even
fully separable.
Therefore, one has to verify experimentally that the
generated state is indeed {\it fully inseparable},
exhibiting {\it genuine tripartite entanglement}.
Such an unambiguous verification can be achieved
even without determining the entire correlation matrix
of the generated Gaussian three-mode state.
It is sufficient to detect a set of suitable linear
combinations of the quadratures (see Sec.~\ref{verentanglement}).
However, these must contain
the positions and momenta of all modes involved.

In an experiment in Tokyo \cite{Aoki03}, such a cv tripartite
entangled state was created by combining three independent
squeezed vacuum states at two beam splitters.
For verification, the variances of the entangled state's
relative positions and total momentum were measured.
The following total variances were obtained \cite{Aoki03},
\begin{eqnarray}\label{ineq}
{\rm I}. \quad
\langle [\Delta (\hat{x}_1 - \hat{x}_2)]^2 \rangle
+
\langle [\Delta (\hat{p}_1 + \hat{p}_2 + \hat{p}_3)]^2 \rangle
\nonumber \\
= 0.851 \pm 0.062 < 1,
\nonumber \\
{\rm II}. \quad
\langle [\Delta (\hat{x}_2 - \hat{x}_3)]^2 \rangle
+
\langle [\Delta (\hat{p}_1 + \hat{p}_2 + \hat{p}_3)]^2 \rangle
\nonumber \\
= 0.840 \pm 0.065 < 1,
\nonumber \\
{\rm III}. \quad
\langle [\Delta (\hat{x}_3 - \hat{x}_1)]^2 \rangle
+
\langle [\Delta (\hat{p}_1 + \hat{p}_2 + \hat{p}_3)]^2 \rangle
\nonumber \\
= 0.867 \pm 0.062 < 1.
\label{ineqresult}
\end{eqnarray}
These results clearly show the nonclassical correlations among the
three modes. Moreover, according to \textcite{PvLAkira03}, the
above inequalities unambiguously prove the full inseparability of
the generated tripartite entangled state. In fact, the measured
variances correspond to violations of the conditions in
Eq.~(\ref{3partycritgenansatz}) with Eq.~(\ref{threemodecombin})
and Eq.~(\ref{3partyassumption}). Thus, any partially separable
form is ruled out and the generated state can only be fully
inseparable.

\subsection{Quantum teleportation of coherent states}\label{telepexperim}

In the dv teleportation experiments
in Innsbruck \cite{Bou} and in Rome \cite{PVLBoschi},
the teleported states were single-photon polarization states.
Continuous-variable quantum teleportation of
coherent states has been achieved in Pasadena
at Caltech \cite{Fur98,Zhang03}
and in Canberra \cite{BowenLam03}.
A realistic broadband description of these experiments
can be obtained from the Heisenberg equations for cv quantum
teleportation \cite{PvLbroad}, as given in Sec.~\ref{telepprotocol}.

For the teleportation of an electromagnetic field with finite bandwidth,
the EPR state shared by Alice and Bob is a broadband two-mode
squeezed state as in Eq.~(\ref{general2}).
The incoming electromagnetic field
to be teleported, $\hat{E}_{\rm in}(z,t)=\hat{E}_{\rm in}^{(+)}(z,t)+
\hat{E}_{\rm in}^{(-)}(z,t)$, traveling in the positive-$z$ direction and
having a single unspecified polarization,
can be described by its positive-frequency part
\begin{eqnarray}\label{field}
\hat{E}_{\rm in}^{(+)}(z,t)&=&[\hat{E}_{\rm in}^{(-)}(z,t)]^{\dagger}
\nonumber\\
&=&
\int_{\rm W} d\omega\frac{1}{\sqrt{2\pi}}\left(\frac{u\hbar\omega}
{2cA_{\rm tr}}\right)^{1/2}\hat{b}_{\rm in}(\omega)e^{-i\omega(t-z/c)}.
\nonumber\\
\end{eqnarray}
The integral runs over a relevant bandwidth W centered on $\omega_0$
and $A_{\rm tr}$ represents the transverse structure of the field.
The parameter $u$ is a units-dependent constant.
By Fourier transforming the incoming field
in the rotating frame, we obtain the input modes
as a function of the modulation frequency $\Omega$,
$\hat{B}_{\rm in}(\Omega)$.
As for the transverse structure and the polarization
of the input field, we assume that both are known to Alice and Bob.

Using the broadband EPR state of Eq.~(\ref{general2}),
for her Bell detection, Alice combines mode 1 with
the unknown input field at a 50:50 beam splitter.
She obtains the quadratures
$\hat{X}_{\rm u}(\Omega)=\frac{1}{\sqrt{2}}\hat{X}_{\rm in}(\Omega)
-\frac{1}{\sqrt{2}}\hat{X}_1(\Omega)$ and
$\hat{P}_{\rm v}(\Omega)=\frac{1}{\sqrt{2}}\hat{P}_{\rm in}(\Omega)+
\frac{1}{\sqrt{2}}\hat{P}_1(\Omega)$ to be measured.
The photocurrent operators for the two homodyne detections,
$\hat{i}_{\rm u}(t)\propto \hat{X}_{\rm u}(t)$ and
$\hat{i}_{\rm v}(t)\propto \hat{P}_{\rm v}(t)$,
can be written (without loss of generality we assume $\Omega>0$) as
%note that these operators are indeed Hermitian as they must be
\begin{eqnarray}\label{currents}
\hat{i}_{\rm u}(t)&\propto&\int_{\rm W} d\Omega\,
h_{\rm el}(\Omega)\left[\hat{X}_{\rm u}(\Omega)e^{-i\Omega t}+
\hat{X}^{\dagger}_{\rm u}(\Omega)e^{i\Omega t}\right],\nonumber\\
\hat{i}_{\rm v}(t)&\propto&\int_{\rm W} d\Omega\,
h_{\rm el}(\Omega)\left[\hat{P}_{\rm v}(\Omega)e^{-i\Omega t}+
\hat{P}^{\dagger}_{\rm v}(\Omega)e^{i\Omega t}\right],
\nonumber\\
\end{eqnarray}
assuming a noiseless, classical local oscillator and
with $h_{\rm el}(\Omega)$
representing the detectors' responses within their electronic bandwidths
$\Delta\Omega_{\rm el}$: $h_{\rm el}(\Omega)=1$ for $\Omega\leq
\Delta\Omega_{\rm el}$ and zero otherwise. We assume that the relevant
bandwidth W ($\sim$ MHz) is fully covered by the electronic bandwidth
of the detectors ($\sim$ GHz). Therefore, $h_{\rm el}(\Omega)\approx 1$
in Eq.~(\ref{currents}) is a good approximation.
The two photocurrents are measured and fed forward
to Bob via a classical channel with sufficient RF bandwidth.
They can be viewed as complex quantities in order to respect
the RF phase.
Any {\it relative} delays between the classical information conveyed by
Alice and Bob's EPR beam must be such that $\Delta t\ll 1/\Delta\Omega$
with the inverse bandwidth of the EPR source $1/\Delta\Omega$.
Bob's final amplitude and phase modulations correspond to
\begin{eqnarray}
\hat{X}_2(\Omega)\longrightarrow\hat{X}_{\rm tel}(\Omega)&=&\hat{X}_2
(\Omega)+g(\Omega)\sqrt{2}\hat X_{\rm u}(\Omega),\nonumber\\
\hat{P}_2(\Omega)\longrightarrow\hat{P}_{\rm tel}(\Omega)&=&\hat{P}_2
(\Omega)+g(\Omega)\sqrt{2}\hat P_{\rm v}(\Omega),\nonumber\\
\end{eqnarray}
with a frequency-depending gain $g(\Omega)$.

For unit gain, $g(\Omega)\equiv 1$, the teleported field is
\begin{eqnarray}\label{1.30}
\hat{X}_{\rm tel}(\Omega)&=&\hat{X}_{\rm in}(\Omega)-
\sqrt{2}S_-(\Omega)\hat{\bar{X}}^{(0)}_2(\Omega),\nonumber\\
\hat{P}_{\rm tel}(\Omega)&=&\hat{P}_{\rm in}(\Omega)+
\sqrt{2}S_-(\Omega)\hat{\bar{P}}^{(0)}_1(\Omega).
\end{eqnarray}
Obviously, for unit-gain teleportation at all relevant
frequencies, it turns out that
{\it the variance of each teleported quadrature is given
by the variance of the input quadrature plus
twice the squeezing spectrum of the quiet quadrature of a decoupled
mode in a ``broadband squeezed state''} as in Eq.~(\ref{general}).
Thus,
the excess noise in each teleported quadrature due the teleportation
process is, relative to the vacuum noise, {\it twice} the
squeezing spectrum $|S_-(\Omega)|^2$ of Eq.~(\ref{spectra}).

In the teleportation experiment of \textcite{Fur98}, the
teleported states described fields at modulation frequency
$\Omega/2\pi=2.9$ MHz within a bandwidth $\pm\Delta\Omega/2\pi=30$
kHz. Due to technical noise at low modulation frequencies, the
resulting nonclassical fidelity, $F=0.58\pm 0.02$ (exceeding the
limit of 1/2 for classical teleportation of coherent states), was
achieved at these higher frequencies $\Omega$. The amount of
squeezing was about 3 dB, where the broadband EPR source as
described by Eq.~(\ref{general2}) was generated via interference
of two independent OPO's. In a second coherent-state teleportation
experiment at Caltech \cite{Zhang03}, a fidelity of $F=0.61\pm
0.02$ was achieved, which is a slight improvement compared to the
first experiment. Finally, in the most recent cv quantum
teleportation in Canberra \cite{BowenLam03}, the best fidelity
observed was $F=0.64\pm 0.02$.

\subsection{Experimental dense coding}
\label{expwcvdense}

As described in Sec.~\ref{qdensecod}, as opposed to the reliable
transfer of quantum information through a classical channel via
quantum teleportation, dense coding aims at transmitting classical
information more efficiently using a quantum channel. Thus, the
roles of the classical channel and the quantum channel are
interchanged. However, like quantum teleportation, dense coding
also relies upon preshared entanglement. In a dense coding scheme,
the amount of classical information transmitted from Alice to Bob
is increased when Alice sends her half of a preshared entangled
state through a quantum channel to Bob. In order to accomplish
this, the local operations performed by Alice and Bob are
interchanged compared to those in quantum teleportation: Alice
encodes the classical information by unitarily transforming her
half of the entangled state; Bob eventually retrieves this
information through a Bell measurement on his part of the
entangled state and the other part obtained from Alice. In a dv
qubit-based implementation, two bits of classical information can
be conveyed by sending just one qubit. As discussed in
Sec.~\ref{qdensecod}, comparing a cv implementation based on
squeezed-state entanglement against heterodyne-detection based
single-mode coherent state communication, the channel capacity of
the latter, given by Eq.~(\ref{C_coh}), is always (for any nonzero
squeezing of the entangled state) beaten by the optimal dense
coding scheme described by Eq.~(\ref{dc}). In order to double the
capacity of single-mode coherent state communication, the dense
coding requires infinite squeezing. This is similar to the ideal
cv quantum teleportation where the fidelity limit of classical
coherent-state teleportation is exceeded for any nonzero squeezing
of the entanglement resource, and unit fidelity is achieved in the
limit of infinite squeezing.

In the dense coding experiment of \textcite{LiXY02}, bright EPR
beams were employed, similar to the entanglement created by
\textcite{Silberhorn}. However, as mentioned in
Sec.~\ref{Kerrinterfer}, the squeezed states in the experiment by
Silberhorn {\it et al.} were produced via the Kerr $\chi^{(3)}$
nonlinearity. The bright squeezed beams were then combined at a
beam splitter to build the entanglement. By contrast, the bright
EPR entanglement in the dense coding experiment of Li {\it et al.}
was directly generated from a nondegenerate parametric amplifier
(NOPA, based on a $\chi^{(2)}$ interaction, see
Sec.~\ref{bbentvianopa}). In order to accomplish the dense coding
protocol on Bob's side, the ``usual'' cv Bell measurement as
described in Sec.~\ref{qdensecod} (using a 50:50 beam splitter and
two homodyne detectors with strong local oscillator fields) must
be replaced by a ``direct Bell measurement'' \cite{Zhang00} due to
the nonzero intensity of the entangled beams. This direct Bell
measurement corresponds to the direct detection of the two bright
outputs from a 50:50 beam splitter. Eventually, the sum and the
difference photocurrents yield
\begin{eqnarray}\label{densecurrents}
\hat{i}_+(\Omega)&\propto&\hat{X}_1'(\Omega)-\hat{X}_2'(\Omega)
+X_s(\Omega),\nonumber\\
\hat{i}_-(\Omega)&\propto&\hat{P}_1'(\Omega)+\hat{P}_2'(\Omega)
+P_s(\Omega),
\end{eqnarray}
where, using the same notation as in
Sec.~\ref{bbentvianopa} and Sec.~\ref{telepexperim},
$X_s(\Omega)$ and $P_s(\Omega)$ are the quadratures
corresponding to Alice's classical signal modulations,
and $\hat{X}_j'(\Omega)$ and $\hat{P}_j'(\Omega)$ are those
belonging to the entangled NOPA beams.
As described by Eq.~(\ref{bbsqcorrelations}),
due to the EPR-type correlations at those frequencies
where squeezing occurs, Bob can simultaneously retrieve
Alice's classical modulations $X_s(\Omega)$ and $P_s(\Omega)$
with a better accuracy than that given by the vacuum noise
limit of two uncorrelated beams.
In the dense coding experiment of Li {\it et al.},
the NOPA's position quadratures $\hat{X}_j'(\Omega)$ and
momentum quadratures $\hat{P}_j'(\Omega)$ were actually
anticorrelated and correlated, respectively,
corresponding to interchanged signs in the expressions
of Eq.~(\ref{densecurrents}), i.e.,
$\hat{X}_1'(\Omega)+\hat{X}_2'(\Omega)$ and
$\hat{P}_1'(\Omega)-\hat{P}_2'(\Omega)$.
The measured variances were up to 4 dB below the vacuum noise limit.

On the other hand, the individual NOPA beams are very noisy. The
noise background in the signal channel, measured by Li {\it et
al.} without exploiting the correlations with the other EPR beam,
was about $4.4$ dB above the corresponding vacuum limit. As a
result, the signal is to some extent protected against
eavesdropping; only the authorized receiver who holds the other
half of the EPR beam can retrieve the transmitted signal. This
potential application of cv dense coding to the secure
transmission of classical information was first realized by
\textcite{Pereira00}. Other quantum cryptography protocols
utilizing EPR-type cv entanglement were discussed in
Sec.~\ref{qcryptsec}. In the next section, we will turn to a
non-entanglement based cv quantum key distribution protocol,
experimentally demonstrated by the Grangier group
\cite{cvcryptoNature03}.

As a final remark of the current section on cv dense coding,
let us mention that the experiment of Li {\it et al.}
demonstrates the potential of cv dense coding
for unconditional signal transmission with high efficiency only
when the distribution of the preshared entanglement
is not counted as part of the communication
(see Sec.~\ref{qdensecod}).
In other words, the entanglement distribution must be
accomplished ``off-peak''. Otherwise, no advantage can be gained
via the dense coding protocol, in agreement with
Holevo's bound in Eq.~(\ref{Holevobound}).
On the other hand, non-entanglement based ``true''
quantum coding schemes \cite{Schumacher95} can be considered
as well. These schemes may indeed outperform their classical
counterparts, but would require a quantum computational step
for the information decoding at Bob's side.
An optical proof-of-principle experiment of this type has been
performed in the single-photon based dv regime
demonstrating a superadditive capacity unattainable without
quantum coding \cite{Sasakiexp}. The conditional quantum gate
required for the decoding was achieved in this experiment by encoding
the quantum information into the spatial and the polarization
modes of a single photon. So far, no quantum coding experiment
of this type has been performed in the cv domain using the
more ``practical'' cv signals. The main difficulty of such an
experiment would be the cv quantum gate in Bob's decoding
procedure. It would require a non-Gaussian operation
based on nonlinearities beyond those described by
a quadratic interaction Hamiltonian and the LUBO
transformation in Eq.~(\ref{LUBO}) (see Sec.~\ref{qcompsec}).

\subsection{Experimental quantum key distribution}
\label{expwcvkey}

In Sec.~\ref{qcryptsec}, we gave an overview of the
various proposals of cv quantum key distribution.
Some of these proposals are based on the use of entanglement
and others are ``prepare and measure'' schemes without
directly utilizing entangled states.

As for the experimental progress, a BB84-like (entanglement-free)
quantum cryptography scheme was implemented by \textcite{Hirano00}
at telecommunication wavelengths using four non-orthogonal
coherent states. A somewhat more genuine continuous-variable
protocol, also based on coherent states, was recently implemented
by \textcite{cvcryptoNature03}. This scheme, proposed by
\textcite{GrangierQCR}, relies upon the distribution of a Gaussian
key \cite{Cerf00c}. Alice continuously modulates the phase and
amplitude of coherent light pulses and Bob eventually measures
these pulses via homodyne detection. The continuous data obtained
must then be converted into a binary key using a particular
reconciliation algorithm \cite{CerfvanAsche}. Complete secret key
extraction can be achieved, for instance, via a reverse
reconciliation technique (followed by privacy amplification)
\cite{cvcryptoNature03}. This method, experimentally implemented
by \textcite{cvcryptoNature03}, provides security against
arbitrarily high losses, even beyond the 3 dB loss limit of direct
reconciliation protocols, as discussed briefly in
Sec.~\ref{qcryptsec}.

In the experiment of \textcite{cvcryptoNature03}, the mutual
information between all participants, Alice, Bob, and Eve, was
experimentally determined for different values of the line
transmission, in particular, including losses of 3.1 dB. The
measured values confirmed the potential security of the scheme
according to the information-theoretic condition in
Eq.~(\ref{seccommcond}), which is sufficient for secure key
extraction using privacy amplification and error correction
techniques. Eventually, net key transmission rates of about $1.7$
megabits per second for a loss-free line and $75$ kilobits per
second for losses of $3.1$ dB were obtained. The signal pulses in
the experiment of \textcite{cvcryptoNature03} contained up to 250
photons and were emitted at a wavelength of 780 nm. The
limitations of this experiment were considered to be essentially
technical, allowing for further improvement on the present scheme.
Therefore, implementing this or related schemes at
telecommunication wavelengths could lead to efficient,
high-bit-rate quantum key distribution over long distances.

\subsection{Demonstration of a quantum memory effect}
\label{expwcvqmemory}

The creation of long-lived atomic entanglement, as described in
Sec.~\ref{expwcvgenforatoms}, is a first step for storing optical
quantum information in atomic states for extended periods and
hence implementing light-atom quantum interfaces. Using a similar
approach, a proof-of-principle demonstration of such a quantum
memory effect was achieved in a further experiment by the Polzik
group \cite{Schori02}. In this experiment, the quantum properties
of a light beam were (partially) recorded in a long-lived atomic
spin state; thus, this experiment goes beyond a previous one where
only a short-lived squeezed spin state of an atomic ensemble was
generated via complete absorption of nonclassical light
\cite{JHJLS99}.

Similar to the experiment for the creation of atomic entanglement
\cite{BJ01}, the experiment by \textcite{Schori02} also relies
upon the polarization and spin representation for cv quantum
information, as discussed in Sec.~\ref{polspinrep}. Thus, the
light and the atoms are described via the Stokes operators and the
operators for the collective spin, respectively. The atom-light
interaction employed in the experiment is again based on the QND
type coupling between the atomic spin and the polarization state
of light, as described in Sec.~\ref{qmemory}.

For describing the experiment by \textcite{Schori02}, we again
consider an atomic sample classically spin-polarized along the
$x$-axis, $\hat F_{x}\simeq\langle \hat F_{x}\rangle\equiv F$, and
similarly for the light, $\hat S_x\simeq \langle \hat
S_x\rangle\equiv S$. Hence again the only quantum variables
involved in the protocol are the atomic operators $\hat F_y$ and
$\hat F_z$ and the light operators $\hat S_y$ and $\hat S_z$,
i.e., the $y$ and $z$ components of spin and polarization are the
effective phase-space variables. The quantum properties of light
to be transferred to the atoms are those of a vacuum or a squeezed
optical field, i.e., those of a (pure) Gaussian state of light. An
off-resonant light pulse prepared in such a state propagates
through the atomic sample along the $z$-axis and leaves its trace
on the sample. As for the relevant input-output relations of this
interaction, we may now only write [see Eq.~(\ref{atomicQND})]
\begin{eqnarray}\label{qmemoryinteraction}
\hat F_{y}^{\rm (out)} &=& \hat F_{y}^{\rm (in)} +a F\,\hat
S_z^{\rm (in)}\,,\\
\label{qmemoryinteraction2}
\hat S_y^{\rm
(out)} &=& \hat S_y^{\rm (in)} +a S\,\hat F_{z}^{\rm (in)}.
\end{eqnarray}
In the experiment by \textcite{Schori02}, the optical input state
was squeezed in the Stokes operator $\hat S_y$ and correspondingly
antisqueezed in $\hat S_z$. This antisqueezing was mapped onto the
atomic state and eventually read out through a detection of the
outgoing light. The protocol consists of the following steps:
first, $\hat S_z$ is mapped onto the atomic variable $\hat F_y$,
as can be seen in Eq.~(\ref{qmemoryinteraction}). However, like in
the experiment for creating atomic entanglement, the protocol is
slightly modified by applying a constant magnetic field oriented
along the $x$-axis. This gives rise to Larmor precession where the
value of the Larmor frequency determines the frequency component
of light to be stored in the atomic sample. Including the magnetic
field, the actual evolution of the atomic spin is more complicated
than it is described by Eq.~(\ref{qmemoryinteraction}). However,
the coupling term responsible for the back action of light onto
atoms per time interval $dt$ is still given by $a F\,\hat
S_z(t)\,dt$, leading to the corresponding change $d\hat F_{y}(t)$
depending on the value of $\hat S_z(t)$. Moreover, due to the
external magnetic field, $\hat F_{y}(t)$ and $\hat F_{z}(t)$ get
linked with each other such that $\hat F_{z}$ and hence $\hat
F_{y}$ can be read out via $\hat S_y$ according to
Eq.~(\ref{qmemoryinteraction2}). For this last step, one can
exploit that $\hat S_y^{\rm (out)}$ in
Eq.~(\ref{qmemoryinteraction2}) is more sensitive to $\hat
F_{z}^{\rm (in)}$ due to the squeezing of $\hat S_y^{\rm (in)}$.

In the experiment by \textcite{Schori02}, the power spectrum of
$\hat S_y^{\rm (out)}$ was measured, yielding clear evidence that
the antisqueezed variable $\hat S_z^{\rm (in)}$ was stored in the
atomic sample. Hence it was shown that partial information about
an optical Gaussian quantum state, i.e., the value of one
quadrature variable can be recorded in an atomic sample. This
storage of quantum information was achieved over a duration of
approximately 2 ms. However, \textcite{Schori02} did not
demonstrate full quantum memory of an optical Gaussian state. In
order to accomplish this, two conjugate variables must be recorded
and for verification, the fidelity between the input and the
reproduced output state has to be determined. For Gaussian signal
states, this corresponds to reproducing values of the output
variances sufficiently close to those of the input variances,
similar to the verification of high-fidelity quantum
teleportation. The experiment by \textcite{Schori02}, however, is
a significant step towards full quantum memory, because it was
shown that long-lived atomic spin ensembles may serve as storage
for optical quantum information sensitive enough to store fields
containing just a few photons.

\section{Concluding remarks}\label{concludremarks}

The field of quantum information has typically concerned itself with the
manipulation of discrete systems such as quantum bits, or ``qubits.''
However, many quantum variables, such as position, momentum or the
quadrature amplitudes of electromagnetic fields, are continuous, leading
to the concept of continuous quantum information.

Initially, quantum information processing with continuous variables
seemed daunting at best, ill-defined at worst. Nonetheless, the first
real success came with the experimental realization of quantum
teleportation for optical fields. This was soon followed by a flood of
activity, to understand the strengths and weaknesses of this type of
quantum information and how it may be processed. The next major
breakthrough was the successful definition of a notion of universal
quantum computation over continuous variables, suggesting that such
variables are as powerful as conventional qubits
for any class of computation.

In some ways continuous-variable computation may not be so different from
qubit-based computation. In particular, limitations due to finite precision
make quantum floating-point operations, like their classical counterparts,
effectively discrete.  Thus we might expect a continuous-variable quantum
computer to perform no better than a discrete quantum computer. However, for
some tasks continuous-variable quantum computers are nonetheless more
efficient. Indeed, in many protocols, especially those relating to
communication, they only require {\it linear\/} operations together with
classical feed-forward and detection. This together with the large
bandwidths naturally available to continuous (optical) variables appears
to give them the potential for a significant advantage.

However, notwithstanding these successes, the very practical optical cv
approach, when solely based upon Gaussian transformations
such as beam-splitter and squeezing transformations, feed-forward and
homodyne detections, is not sufficient for implementing more advanced
or ``genuine'' quantum information protocols.
Any more sophisticated quantum protocol that is truly superior
to its classical counterpart requires a non-Gaussian element.
This may be included on the level of the
measurements, for example, via state preparation conditioned upon
the number of photons detected in a subset of the Gaussian modes.
Alternatively, one may directly apply a non-Gaussian
operation which involves a highly nonlinear optical
interaction described by a Hamiltonian at least cubic
in the mode operators.

Though being a significant first step, communication protocols in
which this non-Gaussian element is missing cannot fully exploit
the advantages offered by quantum mechanics. For example, the
goals in the Gaussian protocols of cv quantum teleportation and
dense coding are reliable transfer of quantum information and
increase of classical capacity, respectively. However, in both
cases, preshared entanglement is required. Using this resource,
via teleportation, fragile quantum information can be conveyed
through a classical communication channel without being subject to
decoherence in a noisy quantum channel. In entanglement-based
dense coding, using an ideal quantum channel, more classical
information can be transmitted than directly through a classical
channel. For transferring quantum information over long distances,
however, entanglement must be distributed through increasingly
noisy quantum channels. Hence entanglement distillation is needed,
and for this, Gaussian resources and Gaussian operations alone do
not suffice. Similarly, ``true'' quantum coding would require a
non-Gaussian decoding step at the receiving end. In general, any
cv quantum computation that is genuinely quantum and hence not
efficiently simulatible by a classical computer must contain a
non-Gaussian element. Among the communication protocols, cv
quantum key distribution appears in some sense exceptional,
because even in a purely Gaussian implementation it may well
enhance the security compared to classical key distribution
schemes.

The experiments accomplished so far in cv quantum information
reflect the observations of the preceding paragraphs. Gaussian
state preparation, including (multi-party) entangled states, and
Gaussian state manipulation are techniques well understood and
implemented in many laboratories around the globe. However, in
order to come closer to real applications, both for long-distance
quantum communication and for quantum computation, a new
generation of experiments is needed, crossing the border between
the Gaussian and non-Gaussian worlds. Beyond this border,
techniques from the more ``traditional'' single-photon based
discrete-variable domain will have to be incorporated into the cv
approaches. In fact, a real-world application of optical quantum
communication and computation, possibly including atom-light
quantum interfaces and atomic quantum memories, will most likely
combine the assets of both approaches, the continuous-variable one
and that based on discrete variables.

\section*{Acknowledgments}

SLB currently holds a Wolfson-Royal Society Research Merit Award.
This work is funded in part under project QUICOV as part of the
IST-FET-QJPC program. PvL is grateful to M. Curty, J. Eisert, J.
Fiur\'{a}\u{s}ek, A. Furusawa, G. Giedke, N. L\"{u}tkenhaus, and
T.~C. Ralph for very useful discussions. He acknowledges the
financial support of the DFG under the Emmy-Noether programme.

\bibliographystyle{apsrmp}
%\bibliography{nc,cds,sw,connes}
\bibliography{rmp}

\end{document}